\documentclass[11pt,a4paper]{article}
\pdfoutput=1
\usepackage{subfigure}

\usepackage{color}
\usepackage{graphicx}
\usepackage{rotating}
\usepackage{tikz}
\usetikzlibrary{shapes,arrows}
\usepackage{array}
\usepackage{tabularx}
\usepackage{graphics,amsfonts,amssymb,amsmath}
\usepackage{epstopdf}
\usepackage{multirow}
\usepackage{gensymb}
 \PassOptionsToPackage{round, sort, numbers, compress}{natbib}
\usepackage[affil-it]{authblk}

\begin{document}

% \begin{frontmatter}

% \usepackage[textwidth=16cm,textheight=22cm]{geometry}

\title{Tuning the bulk properties of bidisperse granular mixtures by small amount of fines}

\author{Nishant Kumar, Vanessa Magnanimo, Marco Ramaioli and Stefan Luding}
%   \thanks{Electronic address: \texttt{yfarjoun@math.mit.edu}; Corresponding author}}
\affil{Multi Scale Mechanics (MSM), 
Faculty of Engineering Technology, MESA+, 
P.O.Box 217, 7500 AE Enschede, The Netherlands}

\date{Dated: \today}

\maketitle
% \author{}
% 
% \address{}

\begin{abstract}
We study the bulk properties of isotropic bidisperse granular mixtures using discrete element simulations. 
The focus is on the influence of the size (radius) ratio of the two constituents and volume fraction on the 
mixture properties. We show that the effective bulk modulus of a dense granular (base) assembly can be enhanced by up 
to 20\% by substituting as little as 5\% of its volume with smaller sized particles. 
Particles of similar sizes barely affect the macroscopic properties of the mixture. 
On the other extreme, when a huge number of fine particles are included, most of them 
lie in the voids of the base material, acting as rattlers, leading to an overall weakening effect. 
In between the limits, an optimum size ratio that maximizes the bulk modulus of the mixture is found. 
For loose systems, the bulk modulus decreases monotonically with addition of fines regardless of the size ratio. 
Finally, we relate the mixture properties to the ‘typical’ pore size in a disordered structure as induced by the 
combined effect of operating volume fraction (consolidation) and size ratio.
\end{abstract}


\newcommand{\Fv} { F_\mathrm{v} }
\newcommand{\strnrate} {{\mathrm{\dot{\boldsymbol{\mathrm E}}}}}
\newcommand{\epsisodot} { \dot\epsilon_\mathrm{v} }
\newcommand{\epsiso} { \epsilon_\mathrm{v} }

% \begin{linenumbers}

\section{Introduction}

Granular materials are widely used as raw materials in various industries, 
including pharmaceutical, mining, chemical, agricultural, household products and food industries.
In many of these applications, processes involving milling, segregation, agglomeration, 
filtration and sieving are common and often lead to the generation of granular systems with 
large size ratios. Dealing with highly polydisperse systems is exceptionally challenging and 
often requires heuristic assumptions to be made, as prediction/control of the behavior is still un unsolved issue.

On the other hand, it is well known in the geomechanical community that the presence of 
small particles (fines) strongly influences the mechanical behavior of granular soils.
Scientific work on the topic is extensive. Several models have been proposed to 
describe the variation of stiffness and strength of granular mixtures as a function 
of the volume of fines in the system 
(see \cite{thevanayagam2002undrained, rahman2011equivalent, rahman2012initial, ni2004contribution,lade1998effects,salgado2000shear, Yin2014micromechanics, belkhatir2012experimental, chang2011micromechanical} among others). 
Thevenayagam \textit{et al.\ }\cite{thevanayagam2002undrained} proposed the concept of intergranular void ratio (distinct from the measured, apparent voids ratio), 
assuming that up to a certain fines content (dependent on density) the finer grains do not actively participate in the transfer of contact frictional forces. 

The problem has been recently approached also numerically \cite{yan2011effect, kumar2014effects, martin2004isostatic, minh2014strong, ogarko2012equation, langroudi2010transmission},
by using the Discrete Element Method (DEM). 
Ogarko and Luding \cite{ogarko2012equation} found numerically that any polydispersity can be replaced by an equivalent
bidisperse mixture when the size distribution moments are matched, in the case of isotropic compression.
Recent works by Ueda \textit{et al.\ }\cite{ueda2011effect} have explored ranges 
of size and volume ratios of bidisperse granular mixtures to evaluate the shear strength in the quasi static regime. 
Minh \textit{et al.\ }\cite{minh2014strong} studied the strong force network in a bimodal 
granular mixture under uniaxial compression, depending on the quantity of fines in the system, detecting an optimum value of fine content. 
Shaebani \textit{et al.\ }\cite{shaebani2012influence} used a mean field approximation and found a direct relation between the 
mean packing properties of the stiffness components in the case of (uniformly distributed) polydisperse 
aggregates. The micro-macro scaling is realized through a combination of moments of the particle size distribution.

However, all cited works refer to either systems with an homogeneous size distribution or to bidisperse mixtures 
of constant size ratio, where the relative volume of particles is varied. 
To the best of our knowledge, no systematic study has been done to study the effect of including 
a small volume of fines in a granular aggregate. 
The interest first rises from geophysical hazards, like earthquakes, where the material volume remains constant, 
but the size of particles quickly decreases due to breakage. The change in size distribution, even limited 
to a very small volume ratio, have been shown to play an important role in soil stability.
Applications are uncountable in industrial processes, where the focus is optimizing the performances 
of a given (granular) material with minimum modification, i.e.,\ minimum costs. 
Hence, the presence of small particles in a granular mass, which is often associated with weakness, is here turned into an asset for functionality.

We use the DEM to study the effect on micro--and macroscopic quantities of a 
monodisperse granular assembly by substituting only 5/105=4.76\% of its 
volume with particles of different size (and same characteristics), 
thus generating a bidisperse mixture, with the main focus on the bulk stiffness. 
We analyze the properties of the granular mixture on two phase spaces: (i) by varying the size ratio of fines to coarse and 
(ii) by spanning a wide range of volume fraction, and find that at each volume fraction corresponds at optimum size ratio 
that maximizes the bulk modulus. 

This paper is organized as follows:
The simulation method and parameters used 
and the averaging definitions for scalar and tensorial quantities
are given in section\ \ref{sec:simmeth}. 
The preparation test procedure for creating the granular mixture is explained in section\ \ref{sec:prepproc}. 
Section\ \ref{sec:micro} is devoted to the rattlers (that do not contribute to the mechanical stabilities), and the effect of the size of fines on microscopic quantities like the coordination number. 
In section\ \ref{sec:macro}, we discuss the effect of the size of fines on macroscopic quantities like pressure and the jamming volume fraction. 
The effect of the size of fines on the contact network, quantified by the isotropic fabric, is also discussed there.
Finally, section\ \ref{sec:bulk} is devoted to the bulk modulus and its variation with the size of fines for different volume fractions.

\section{Numerical simulation and Material properties}
\label{sec:simmeth}

In this section, the procedure of creating the granular mixture is presented. 
Later, we discuss the contact model and the simulation parameters. 

The reference sample consists of $N_A^0=1050$ monodisperse particles A with radius $r_A = 1.5$[mm]. 
Starting from this base sample, many mixtures are created by substituting a given number of $\left(N_A^0 - N_A^T \right) = 50$ particles,
with particles of species B of different radius $r_B \le r_A$, such that 
the same volume $\left(N_A^0 - N_A^T \right)\left( 4\pi/3 \right) r_A^3 = {N_B^T} \left( 4\pi/3 \right) r_B^3=V_B^T $ of material A is replaced by B. 
The volume ratio of the two components in the final mixture is thus:
\begin{equation}
\label{eq:phidefn}
\Phi =  \frac{V_B^T}{V_A^T} =  \frac{\left(N_A^0 - N_A^T \right)}{N_A^T} = 5\% = \frac{N_B^T}{N_A^T}  \left(\frac{r_B}{r_A}\right)^3,
\end{equation}
and is much smaller than the pore space of the base material.
The size ratio is varied systematically from the base case $r_B/r_A =1$ down to $r_B/r_A = 0.13$; the number of B particles $N_B$ varies together with $r_B$, while the volume ratio 
$\Phi$ is kept constant, as well as the volume of the individual species. 
The total volume of particles is $V_T = V_A^T + V_B^T = 1.05 V_A^T $, so that 
volume of the box $V$ is same for different granular mixtures with different number of B particles $N_B^T$ and at a given volume fraction $\nu = V_T/V$. 
Note that the substitution can be thought of addition when a 
system containing $N_A^T$ particles of A is mixed with B with volume fraction $\Phi = 5\%$ of that of A. 

\begin{figure}[ht]
\centering
\subfigure[]{\includegraphics[width=0.53\textwidth, origin=c,angle=0]{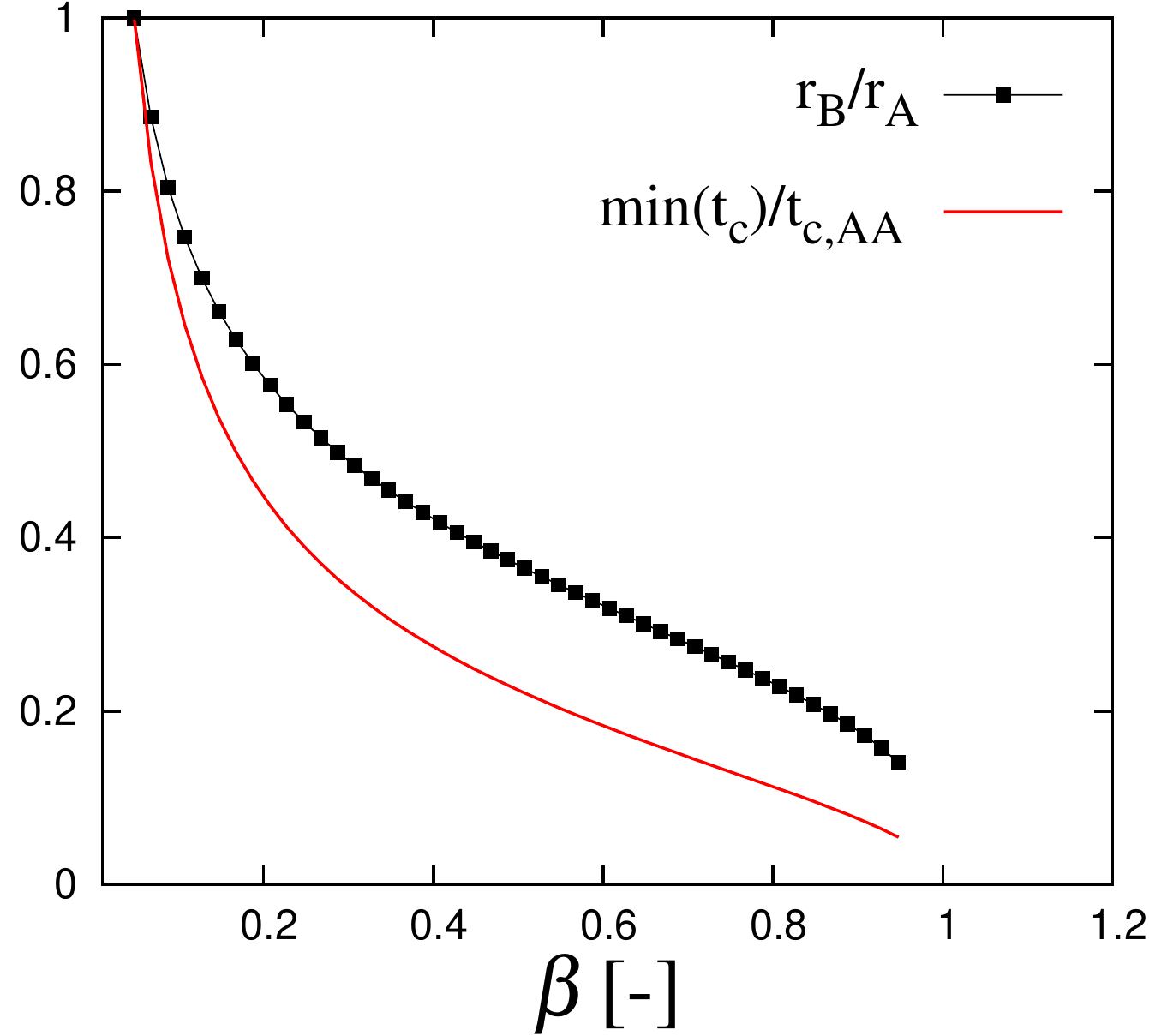}\label{tcvskappa_mu0p0_noscale}}
\subfigure[]{\includegraphics[width=0.53\textwidth, origin=c,angle=0]{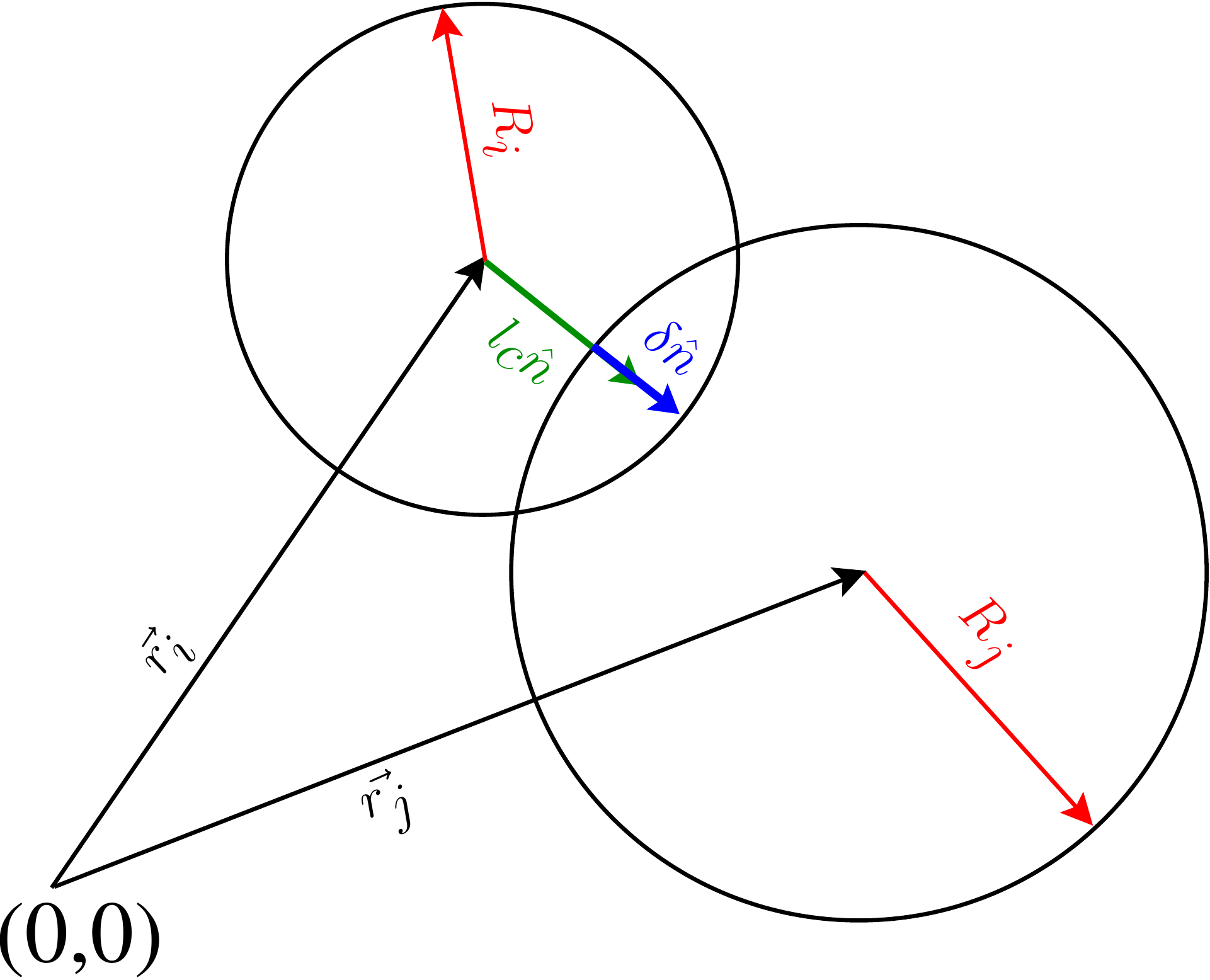}\label{overlap}}
\caption{(a) Variation of the radius ratio $r_B/r_A$ and smallest collision duration $t_c$ with $\beta= N_B^T/\left(N_A^T + N_B^T \right)$.
(b) Sketch of two particles in contact and the direction of the force and branch vectors.}
\label{tcoverlap}
\end{figure}

In order to characterize the mixtures with different $N_B^T$, we define a dimensionless quantity $\beta$ as:
\begin{equation}
\label{eq:betadefn}
\beta = \frac{N_B^T}{ N_A^T + N_B^T },
\end{equation}
which is the ratio of small particles B to the total number of particle in the system. 
$\beta$ is the input parameter of the simulation and is systematically varied to study its effect on the measured micro-macroscopic quantities. 
For small $\beta$, few big particles B are present in the system, while for large $\beta$, many smaller particles B are present. 
The ratio of $N_B^T$ to $N_A^T$ in terms of $\beta$ is given as
\begin{equation}
\label{eq:NBTratio}
 \frac{N_B^T}{N_A^T} =  \frac{\beta}{1-\beta},
\end{equation}
and the size (radius) ratio is  
\begin{equation}
\label{eq:radiusratio}
\frac{r_B}{r_A} = \left( \Phi \frac{N_A^T}{N_B^T} \right)^{1/3} = \Phi^{1/3} \left(\frac{1-\beta}{\beta} \right)^{1/3}.
\end{equation}
The sample made of only A particles is always used as reference case and corresponds to the case ${r_B}/{r_A}=1$. 
It provides the minimum $\beta$, $\beta_\mathrm{min} = \Phi/\left( 1+ \Phi \right)=0.5/1.05=0.0476$, and hence the minimum $N_B^T = \Phi N_A^T$. 
The variation of the radius ratio ${r_B}/{r_A}$ is reported in Fig.\ \ref{tcvskappa_mu0p0_noscale} and shows a monotonic decrease with $\beta$.

The Discrete Element Method (DEM) 
\cite{cundall1979discrete} has been used extensively to study granular materials in biaxial and triaxial geometries
\cite{luding2005shear,kruyt2010micromechanical,duran2010micromechanical,sun2011constitutive, alonsomarroquin2005role,thornton2010quasistatic,thornton2010evolution} under general deformation paths involving 
advanced contact models for fine powders \cite{luding2008cohesive, singh2014effect}. 
In this work, however, we restrict ourselves to the simplest isotropic deformation test and to the linear contact model without any friction between the particles. 
Since DEM is a standard
method, only the contact model parameters relevant for our simulation are briefly discussed.

The simplest linear normal contact force model when two particles $i$ and $j$ interact, as shown in Fig.\ \ref{overlap},
is given as ${\it \bf f}^n_{ij}=f^n_{ij} \hat{\bf n} = ( k_{ij} \delta + \gamma_{ij} \dot\delta) \hat{\bf n}$, where $k_{ij}$ is the contact spring 
stiffness, $\gamma_{ij}$ is the contact viscosity parameter, $\delta$ is the overlap and $\dot\delta$ is 
the relative velocity in the normal direction $\hat{\bf n}$. An artificial background dissipation force, ${\it \bf f}_b=-\gamma_b {\bf v}_i$, proportional to the velocity ${\bf v}_i$ 
of particle $i$ is added (similarly of particle $j$), resembling the damping due to a background medium, as e.g.\ a fluid. 
Note that apart than the radius, materials A and B have the same interacting properties, i.e.\, stiffness, viscosity and density, see Table \ref{parametertable}. 

For a pair of particles $i$ and $j$ with masses $m_i$ and $m_j$, a typical response time is the collision duration 
$t_c^{ij} = \pi / \sqrt {k_{ij}/m_{ij}-(\gamma_{ij}/2m_{ij})^2}$, where $m_{ij} = m_{i}m_{j}/(m_{i}+m_{j})$ is the reduced mass \cite{imole2013hydrostatic}.
In DEM, the integration time-step is chosen to be about 50 times smaller
than the shortest time-scale $t_c = \mathrm{min}\left(t_c^{ij}\right)$ \cite{luding2008cohesive}. The parameters
used in DEM simulations are presented in Table \ref{parametertable}. 
For our system, material B sets the DEM time-step, as $r_B$ and hence $m_{BB}$ is smallest, leading to the smallest $t_c^{BB}$. 
The variation of $t_c$ with $\beta$ can be seen in Fig.\ \ref{tcvskappa_mu0p0_noscale}. 
$t_c$ decreases with increasing $\beta$, meaning that the smaller particles in the mixture lead to a reduction in the collision time and hence to a finer time-step. 
Due to computational limitations, the simulations were performed up to $\beta = 0.957$.

\begin{table}
 \begin{tabular}{l@{\hskip 0.3in}l@{\hskip 0.3in}l@{\hskip 0.5in}l}
    \hline\hline
     Parameter & Symbol & Material A & Material B \\ [1ex]
    \hline\hline \\[-1ex]
    Number of Particles & $N^T$ & $N_A^T =1000$ & $N_B^T$ varied [50--22500] \\ [1ex]
    Radius & $r$ & $r_A = 1.5$ mm & $r_B/r_A = \Phi^{1/3} \left(\frac{1-\beta}{\beta} \right)^{1/3}$  \\ [1ex]
    Particle density & $\rho$ & $\rho_A = 2000$ [kg/m\textsuperscript{3}]  & $\rho_B = \rho_A$ [kg/m\textsuperscript{3}]\\ [1ex]
    Normal stiffness & $k^n$ & $k_A^n = 5.10$\textsuperscript{8} [kg/s\textsuperscript{2}] &  $k_B^n = k_A^n$ \\ [1ex]
    Normal Viscosity & $\gamma$ & 1 [kg/s] &  1 [kg/s] \\ [1ex]
    Background viscosity & $\gamma_{b}$ & 0.1 [kg/s] & 0.1 [kg/s] \\ [1ex]    
%     Inter-particle friction coefficient & $\mu$ & $\mu_A$ = 0, 0.1  & $\mu_B$ = 0, 0.1 \\ [1ex]    
%     Tangential stiffness & $k^t$ & $k_A^t/k_A^n$ = 2/7  & $k_B^t/k_B^n$ = 2/7 \\ [1ex]
  \end{tabular}
  \caption{Summary and numerical values of particle parameters used in the DEM simulations. 
  $\beta$ is the ratio of particles of B to the total number of particles, defined in Eq.\ (\ref{eq:betadefn}). $\Phi=0.05$ is the ratio of volume of B to that of A in the final mixture.}
  \label{parametertable}
\end{table}

\section{Preparation and test procedure}
\label{sec:prepproc}

\begin{figure}[!ht]
\centering
\subfigure[]{\includegraphics[width=0.45\textwidth, angle=0]{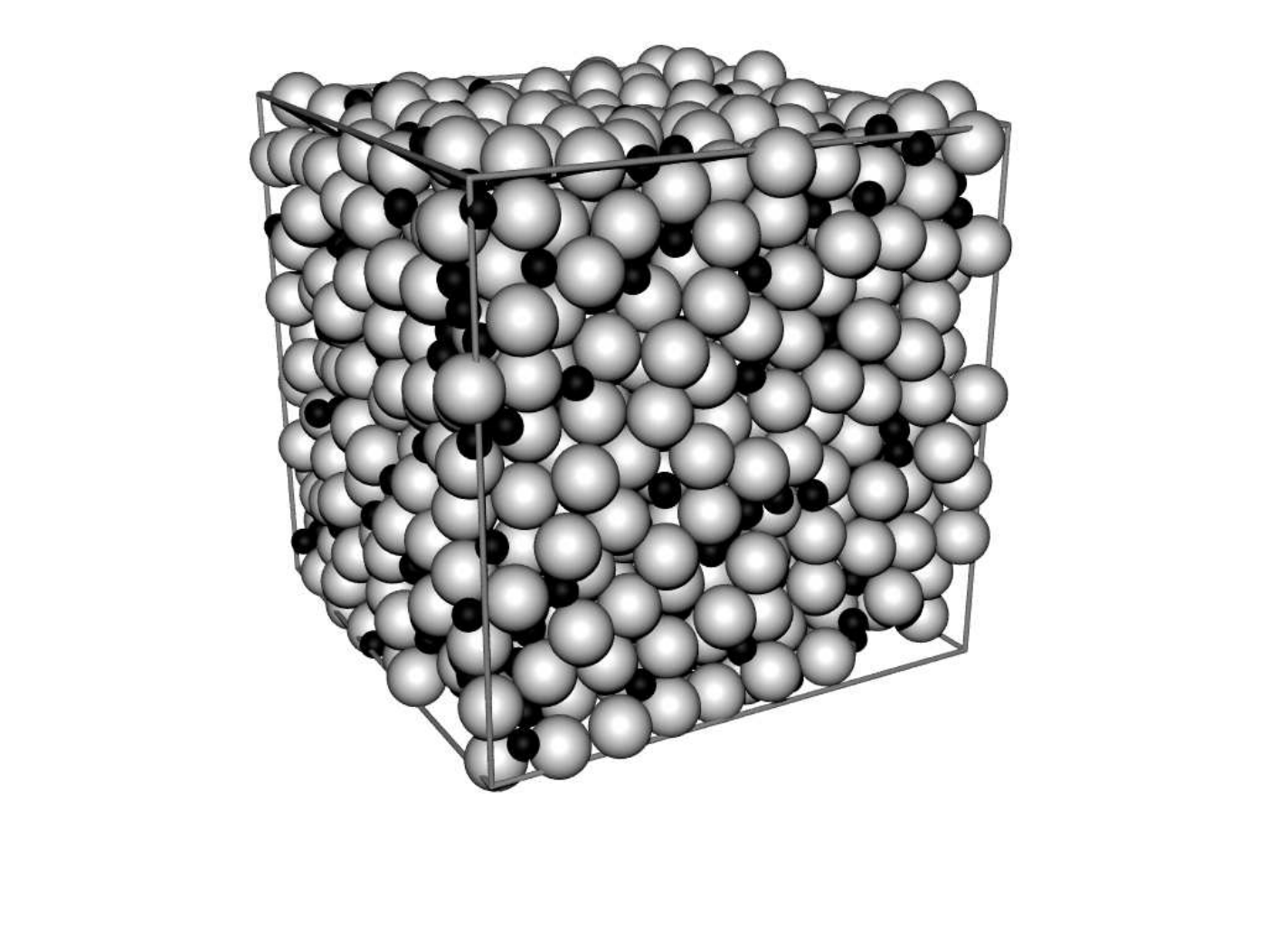}\label{c3d1038_ISO2}}
\subfigure[]{\includegraphics[width=0.40\textwidth, angle=0]{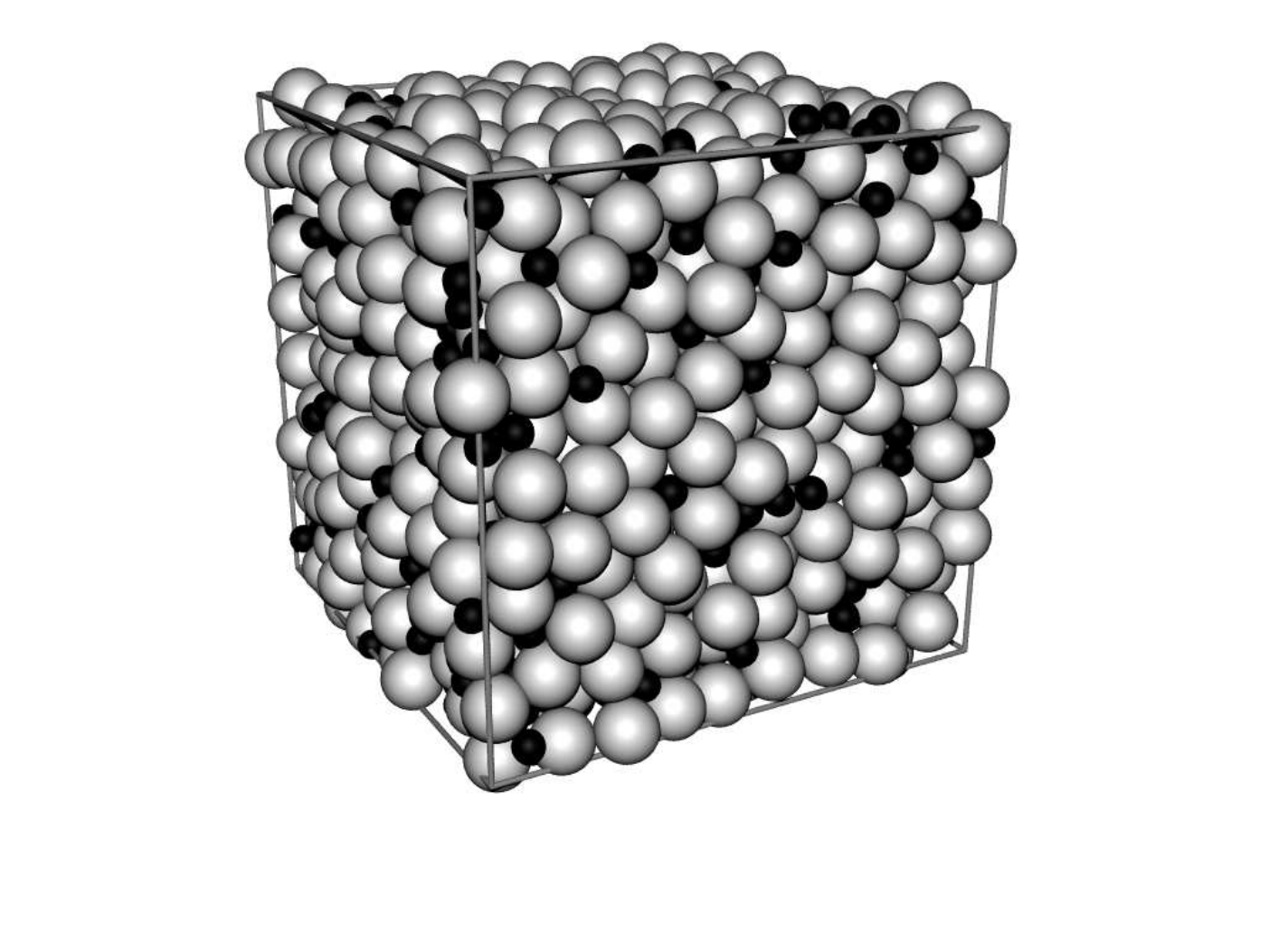}\label{c3d1025_ISO2}}\\
\subfigure[]{\includegraphics[width=0.45\textwidth, angle=0]{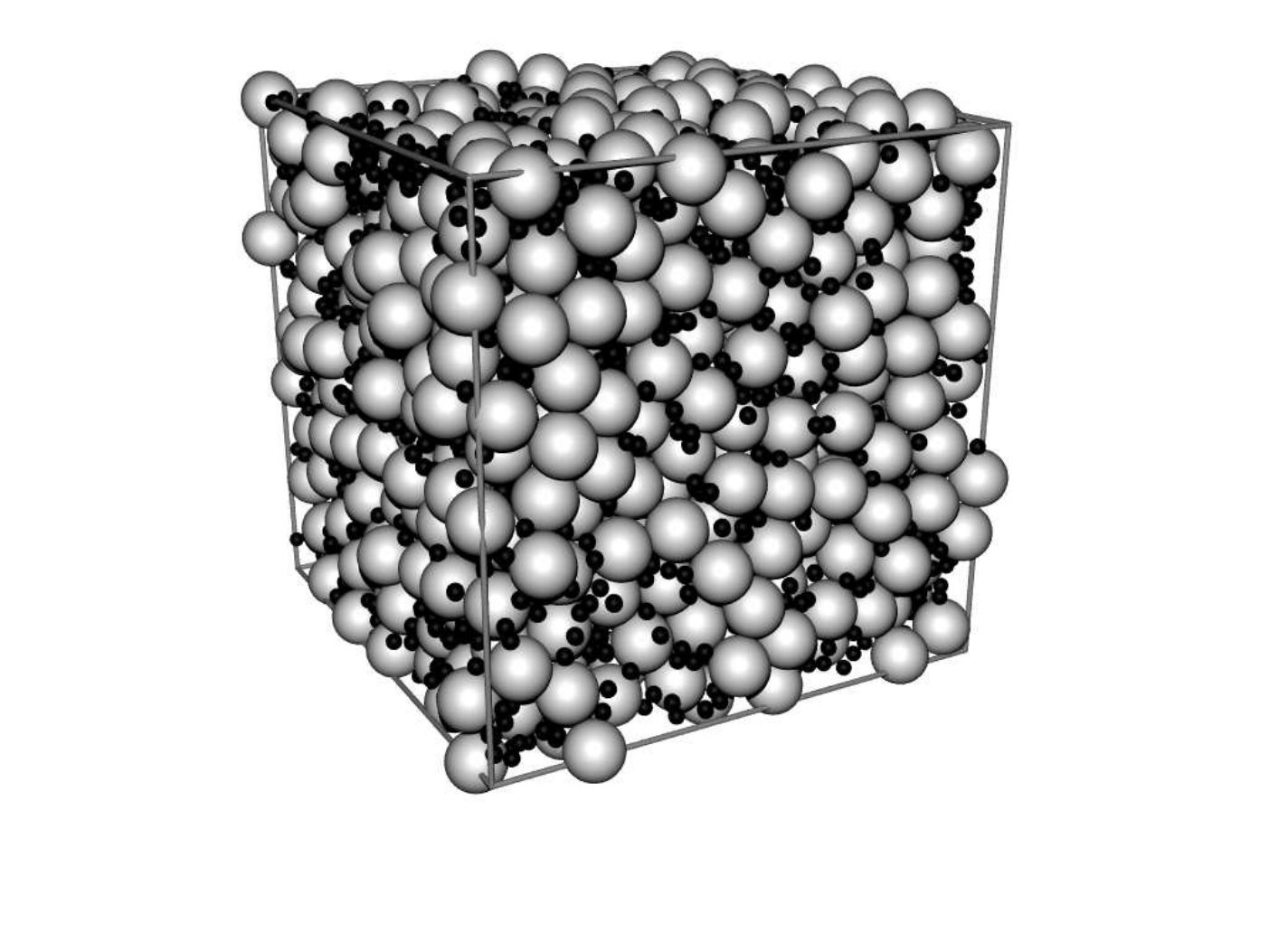}\label{c3d1038_ISO5}}
\subfigure[]{\includegraphics[width=0.40\textwidth, angle=0]{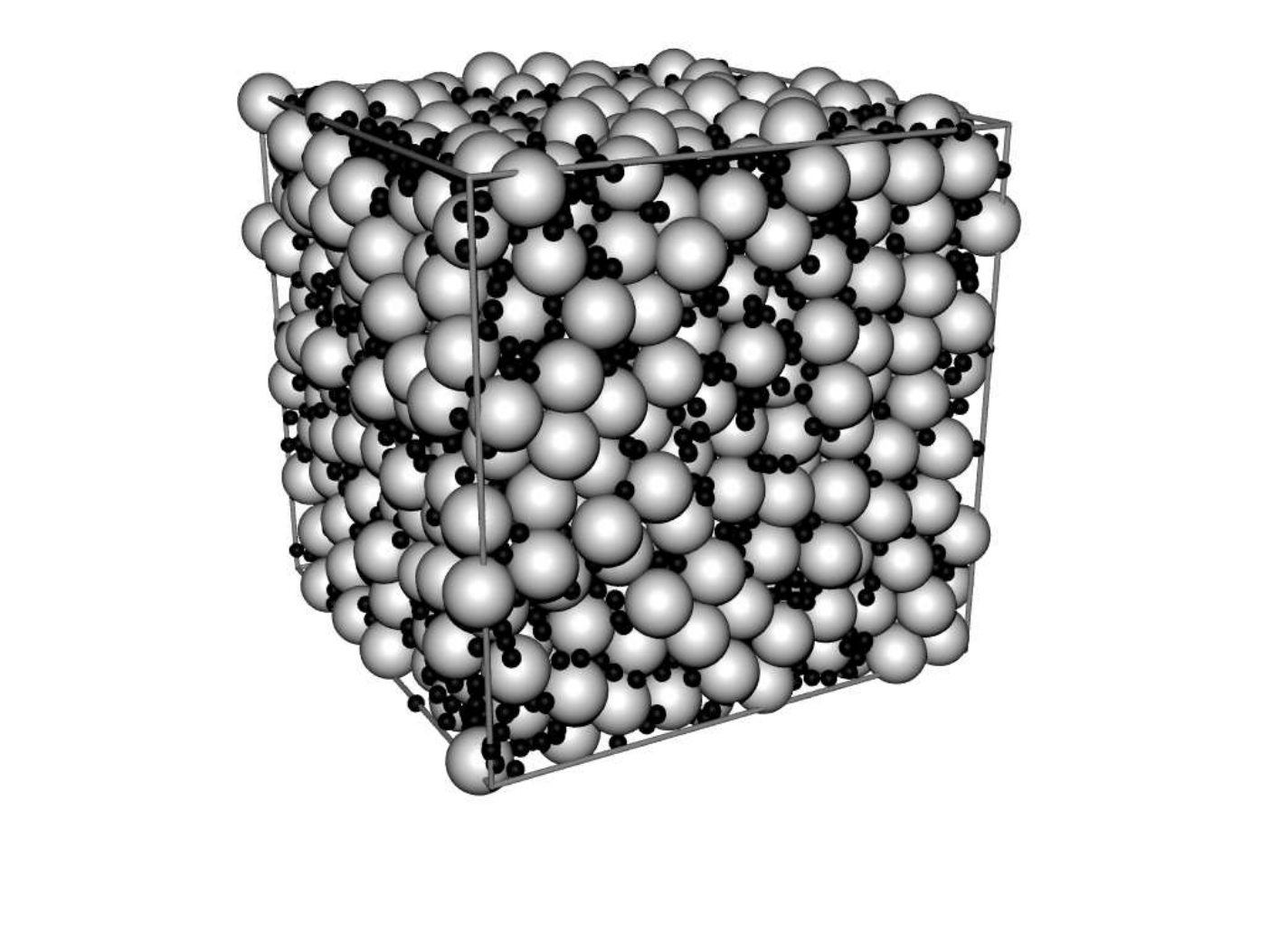}\label{c3d1025_ISO5}}\\
\subfigure[]{\includegraphics[width=0.45\textwidth, angle=0]{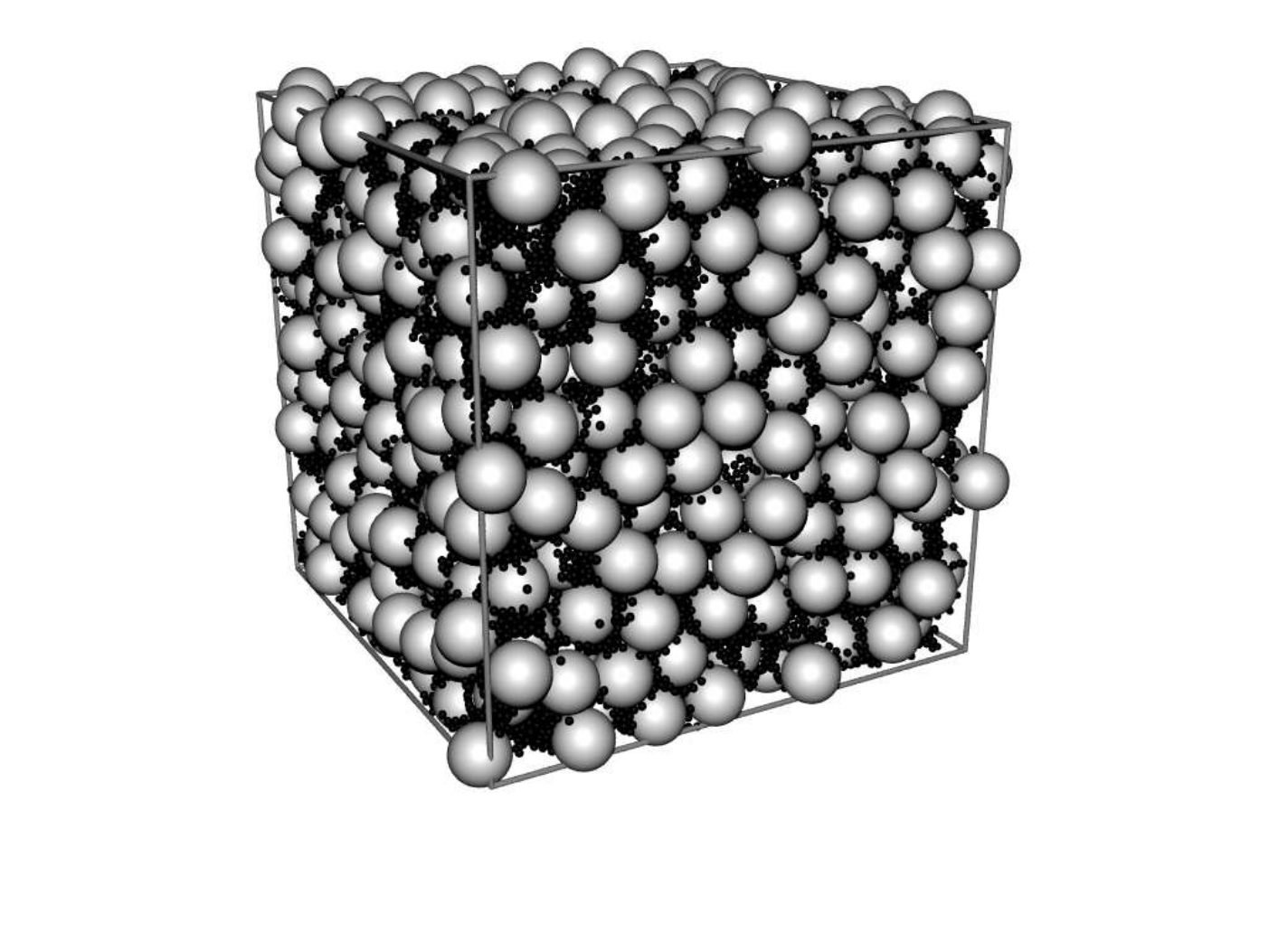}\label{c3d1038_ISO14}}
\subfigure[]{\includegraphics[width=0.40\textwidth, angle=0]{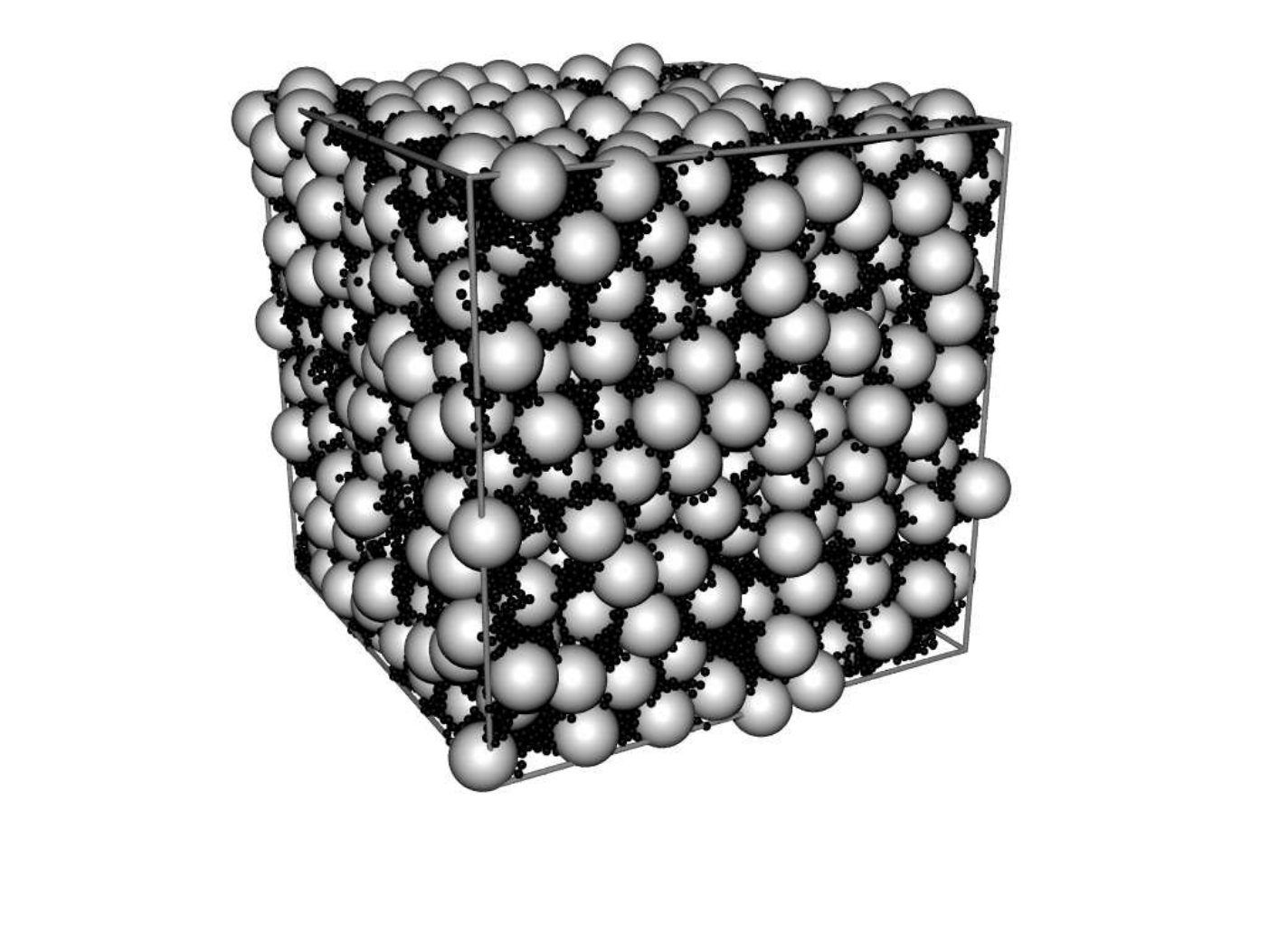}\label{c3d1025_ISO14}}
\caption{Snapshots of the composite material. 
White and black particles are particles of A (large) and B (small), respectively.
Different rows represent different $\beta= N_B^T/\left(N_A^T + N_B^T \right)=$ 0.075, 0.56, and 0.96, with size ratio $r_B/r_A=$ 0.5, 0.27 and 0.14.
Left and right columns correspond to total volume fractions $\nu=$ 0.69 (loose) and 0.82 (dense), respectively.}
\label{c3d}
\end{figure}

\begin{figure}[!ht]
\centering
\subfigure[]{\includegraphics[width=0.45\textwidth, angle=0]{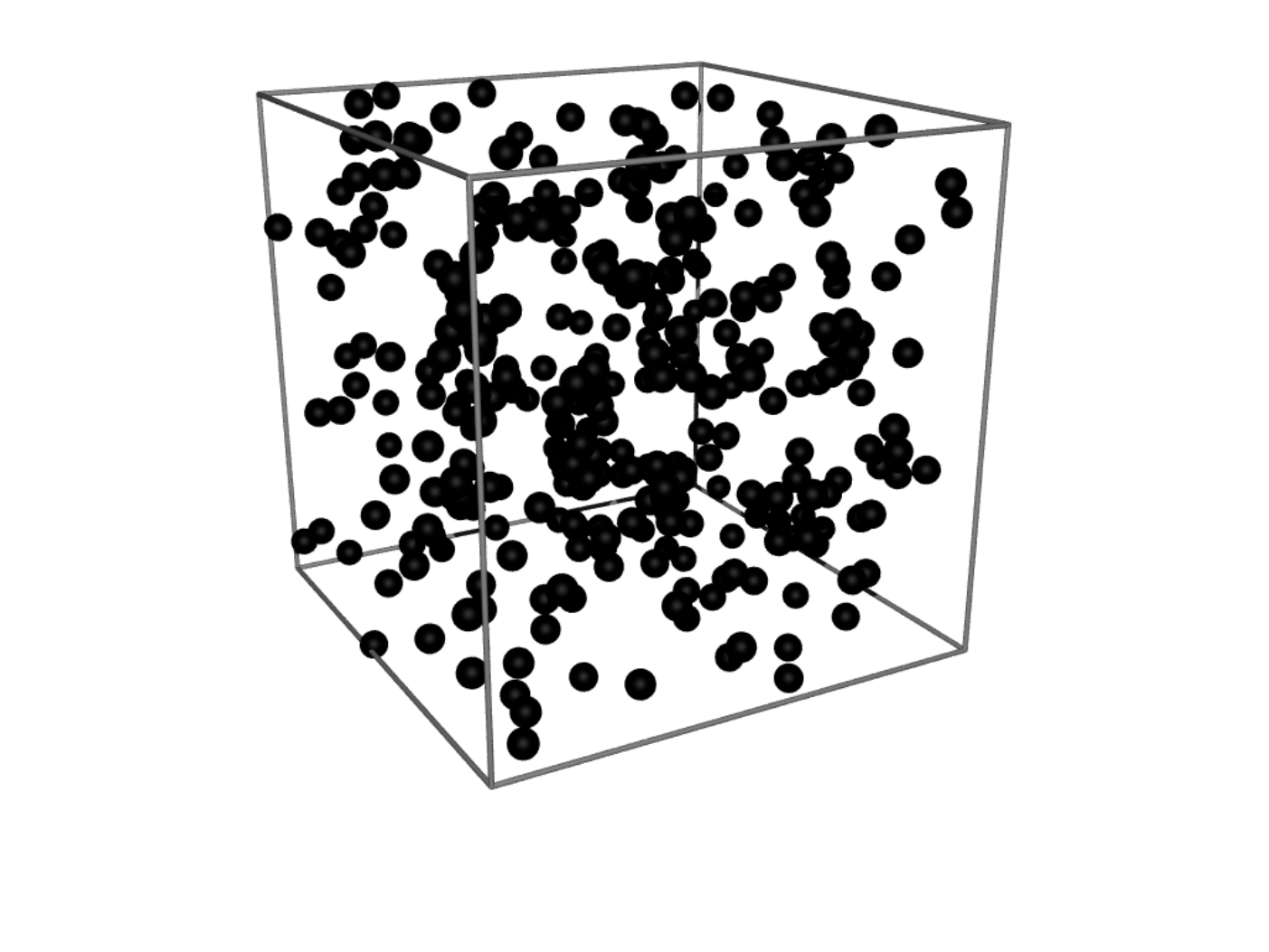}\label{new_c3d1038_ISO2}}
\subfigure[]{\includegraphics[width=0.40\textwidth, angle=0]{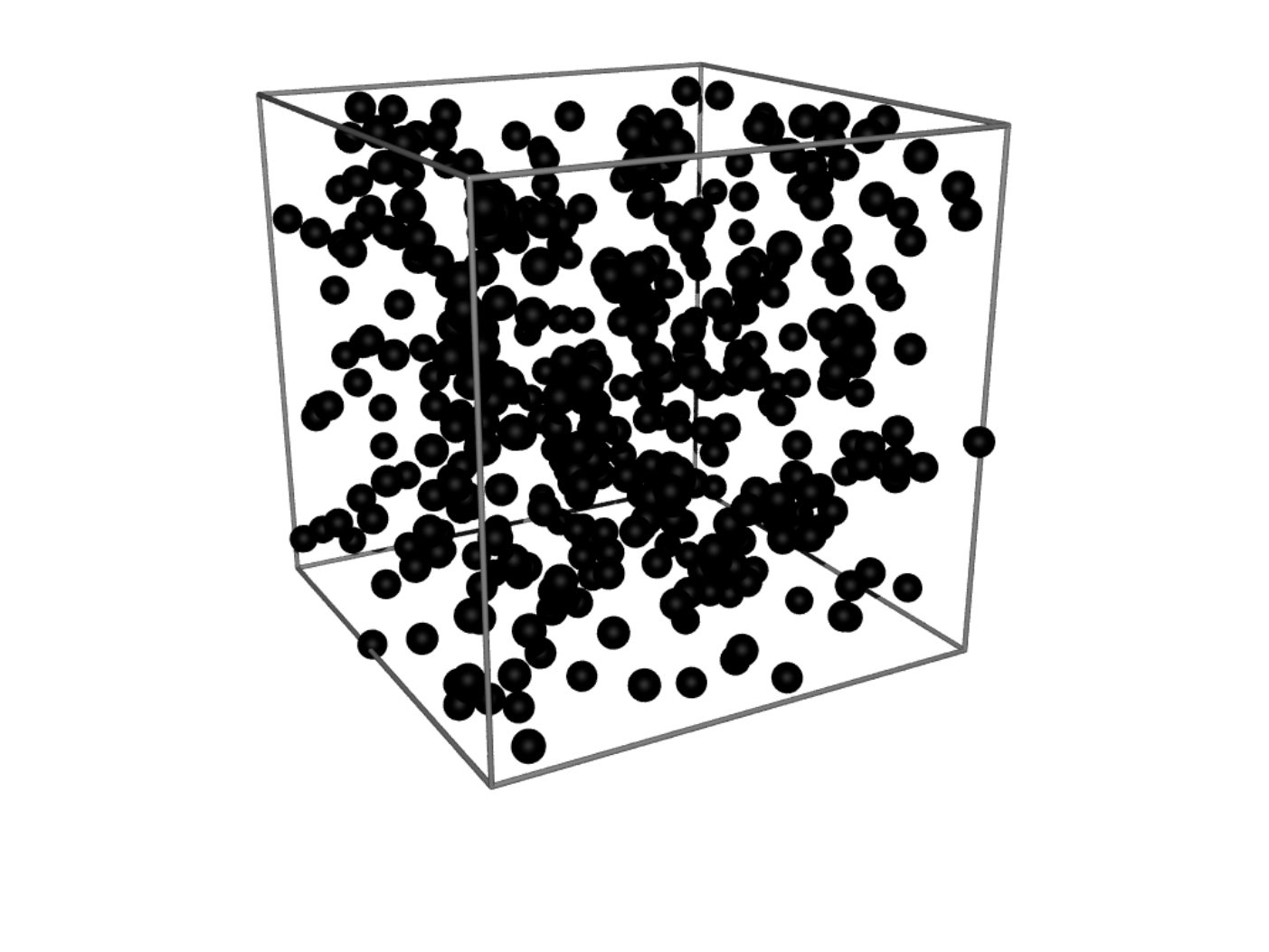}\label{new_c3d1025_ISO2}}\\
\subfigure[]{\includegraphics[width=0.45\textwidth, angle=0]{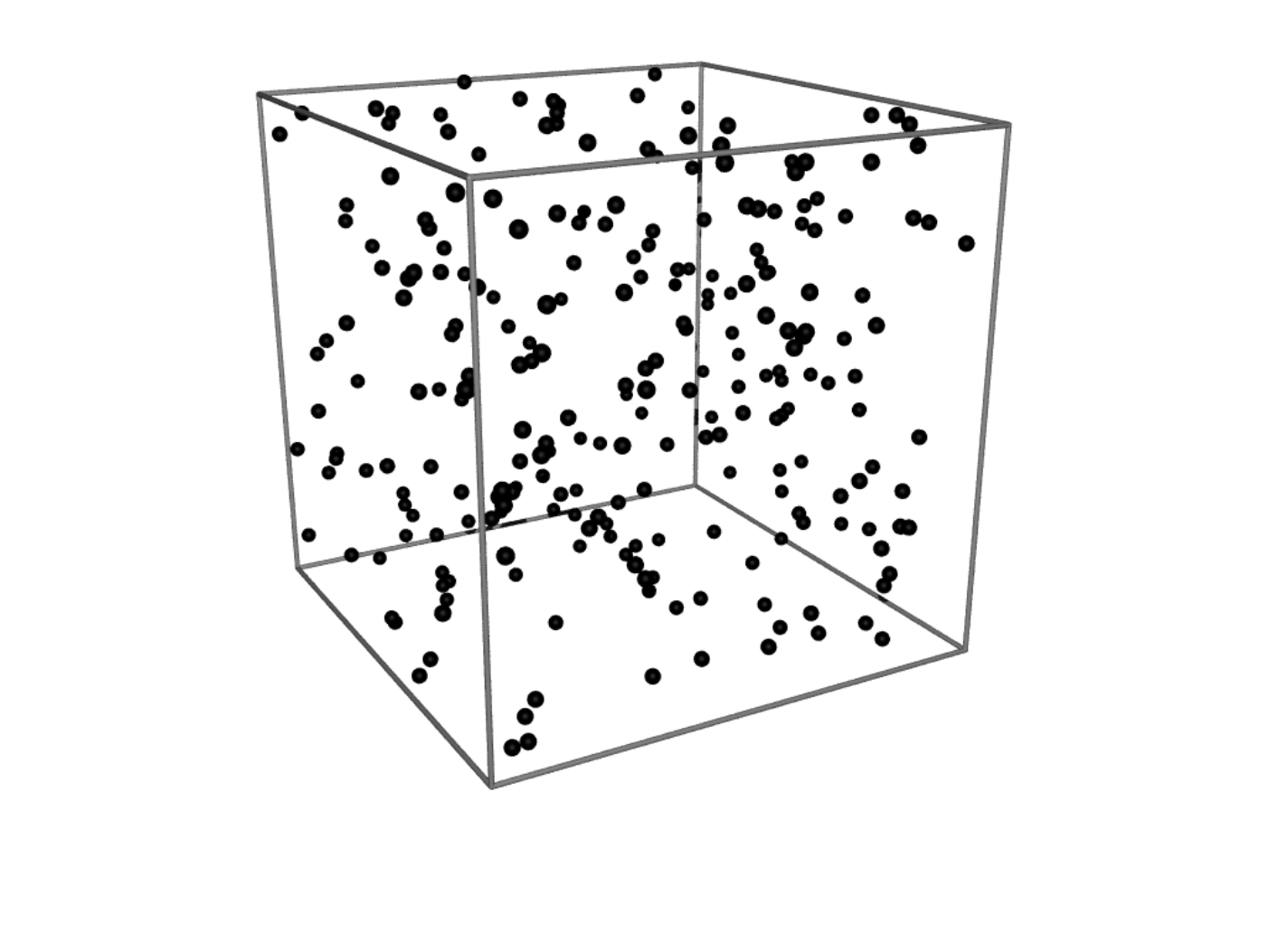}\label{new_c3d1038_ISO5}}
\subfigure[]{\includegraphics[width=0.40\textwidth, angle=0]{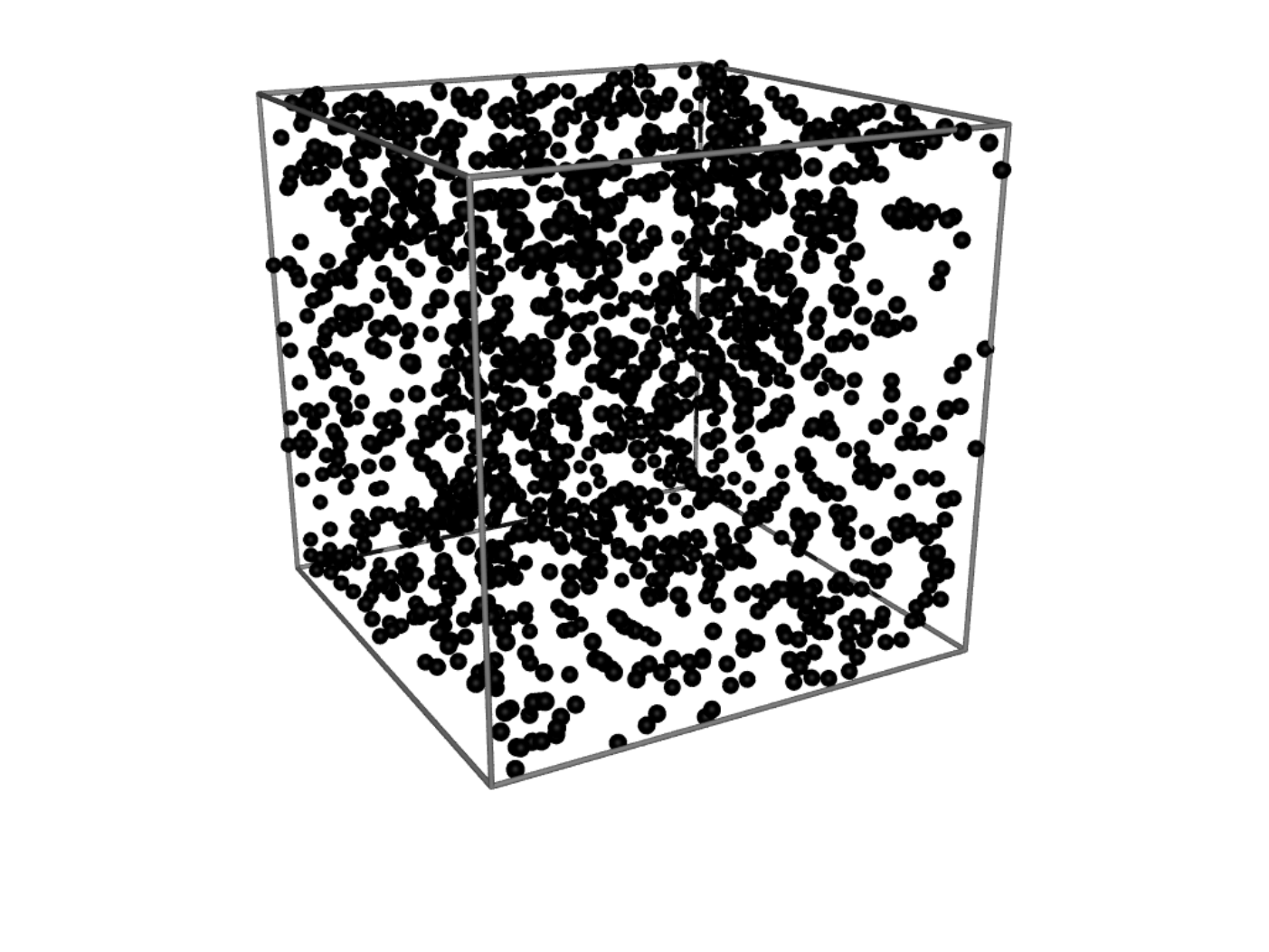}\label{new_c3d1025_ISO5}}\\
\subfigure[]{\includegraphics[width=0.45\textwidth, angle=0]{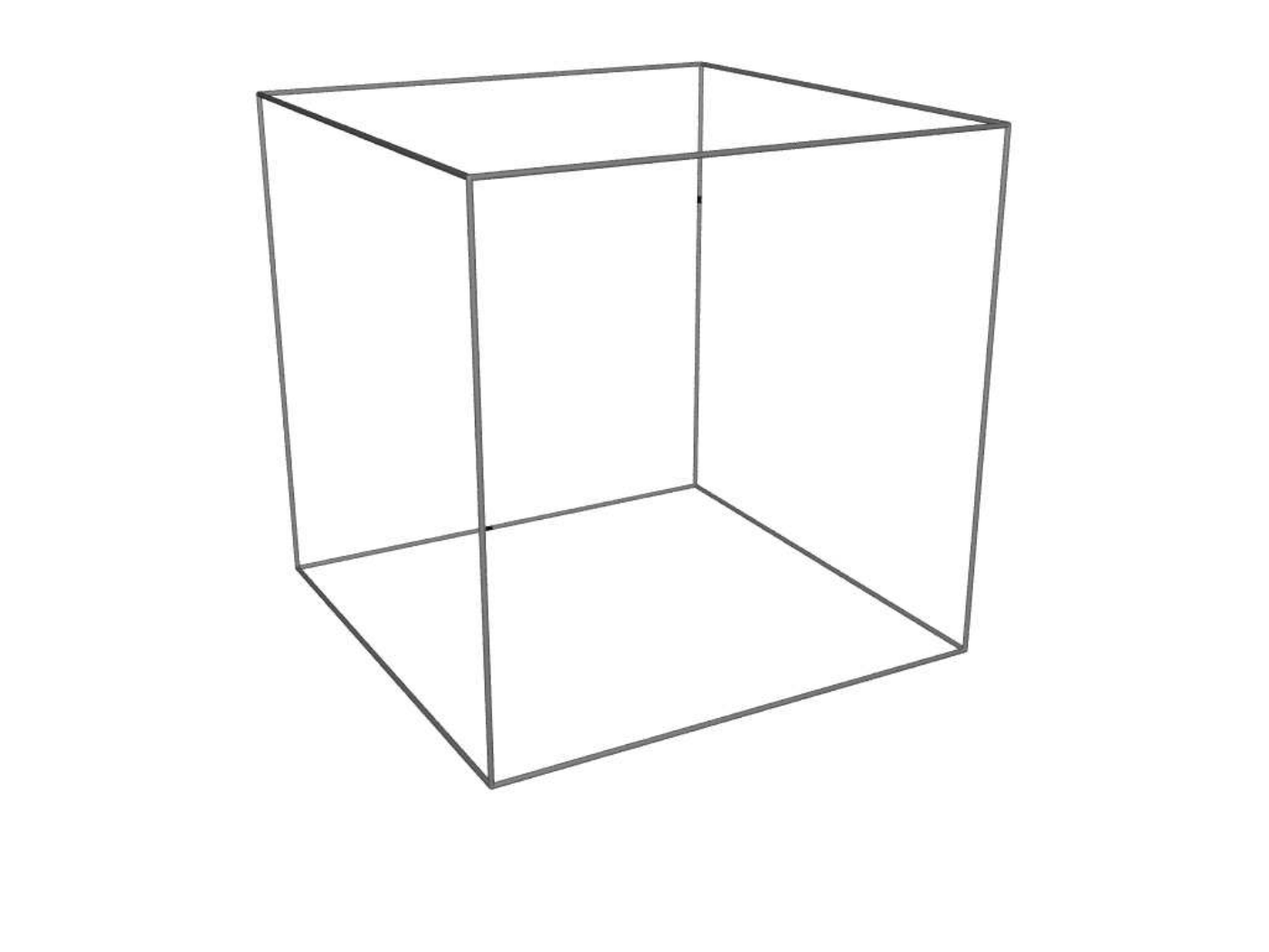}\label{new_c3d1038_ISO14}}
\subfigure[]{\includegraphics[width=0.40\textwidth, angle=0]{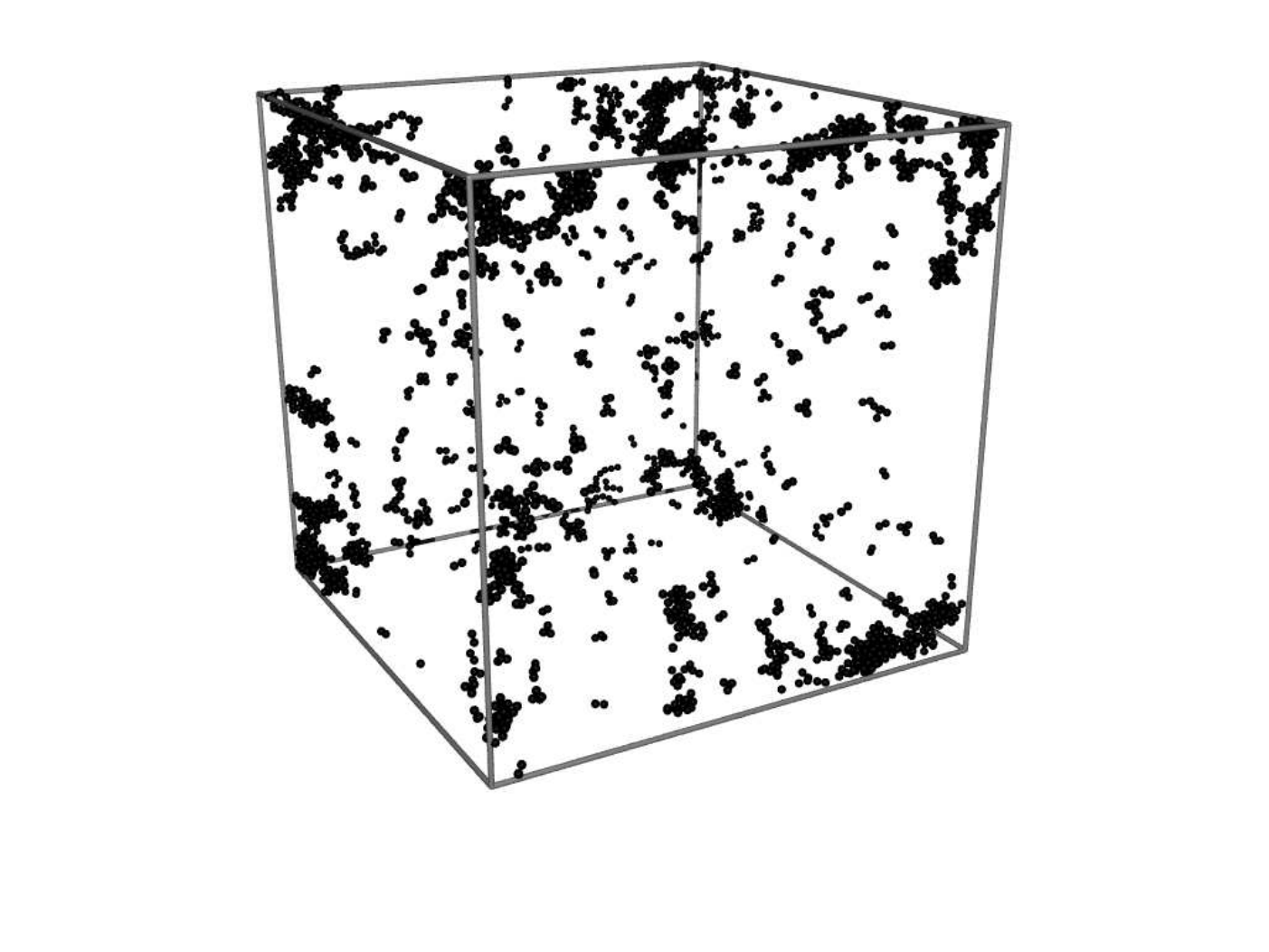}\label{new_c3d1025_ISO14}}
\caption{Snapshots of the fines (same as Fig.\ \ref{c3d}) without rattlers, i.e.,\ for clarity only particles B are shown. 
Different rows represent different $\beta= N_B^T/\left(N_A^T + N_B^T \right)=$ 0.075, 0.56, and 0.96, with size ratio $r_B/r_A=$ 0.5, 0.27 and 0.14.
Left and right columns correspond to total volume fractions $\nu=$ 0.69 and 0.82, respectively.}
\label{new_c3d}
\end{figure}

Each mixture, made of materials A and B as introduced in section\ \ref{sec:simmeth} is created and further compressed using a unique, well defined protocol. 
The preparation consists of three parts: 
(i) randomization, 
(ii) isotropic compression, and 
(iii) relaxation, 
all equally important to achieve the initial
mixtures for the following analysis.
The initial configuration is such that spherical particles of particles A and B, are randomly generated 
in a 3D box without gravity, with low volume fraction and rather large random velocities, such that they have
sufficient space and time to exchange places and to randomize themselves. 
This granular gas is then isotropically compressed, in order
to approach a direction independent initial configuration with target volume 
fraction $\nu_0 = 0.64$, sightly below the jamming volume fraction, i.e.\ the transition point from fluid-like behavior to 
solid-like behavior \cite{majmudar2007jamming,ohern2002random,makse2000packing,van2010jamming}. 
Isotropic compression is realized by a simultaneous inward movement of all the 
periodic boundaries of the system, with diagonal strain rate tensor
%$\strnrate = \epsisodot \left(-1,-1,-1\right),$ 
\[ \strnrate = \epsisodot\left( \begin{array}{ccc}
-1 & 0 & 0 \\
0 & -1 & 0 \\
0 & 0 & -1 \end{array} \right)\, , \] 
where $\epsisodot$ is the rate amplitude ($\epsisodot>0$ in our convention represents compression)  applied to 
the walls. 
This is followed by a relaxation period at constant volume fraction to allow 
the particles to fully dissipate their energy and to achieve a static configuration
in mechanical equilibrium, indicated by the drop in kinetic to potential energy ratio to almost zero. 
This relaxed state is further isotropically compressed until a target maximum volume fraction $\nu_{\rm {max}}=0.82$ is achieved. 
The simulations are continued with negative rate amplitude in the unloading mode, 
until the initial $\nu_0=0.64$ is reached. 

For each mixture, configurations at six different $\nu$ are picked from the unloading branch and relaxed, 
allowing to dissipate the kinetic energy and reach unjammed, non-overlapping stable packings.
\footnote[2]{Configurations from the unloading branch are more reliable since it is much less 
sensitive to the protocol and rate of deformation during 
preparation \cite{goncu2010constitutive,imole2013hydrostatic}.}
As an example, we show in Fig.\ \ref{c3d} isotropic samples 
with $\beta= N_B^T/\left(N_A^T + N_B^T \right)=$0.075, 0.56, and 0.96 for loose and dense samples with volume fraction $\nu=0.69$ and $0.82$ respectively.
% The relaxation of different snapshots is done for long time to ensure that mechanically stable particles have at least \textit{four} contacts. 
% Non stable particles have exactly zero contacts. 

% Unless not mentioned specifically, we will denote $N_A$ and $N_B$ as the number of mechanically stable particles of A and B respectively, and we will use them to compute micro-scopic and macroscopic quantities.
% This is due to the reason that, in the pores of a granular packing, one can imagine to put many non-interacting very small sized particles that will 
% not contribute to the mechanical strength of the packing and should not be counted in numbers. 

\section{Microscopic Quantities} 
\label{sec:micro}

In this section, we present the general definitions of averaged microscopic parameters including the coordination number and 
the fraction of rattlers.
% and geometrical heterogeneity. 

\subsection{Mechanically stable system} 
\label{sec:stable}

In order to properly link the macroscopic load carried by the sample with the active 
microscopic contact network, all particles that do not contribute 
to the force network are excluded from the analysis. Frictionless particles with less than 4 contacts are thus `rattlers', 
since they are not mechanically stable and hence do not participate to the force transmission \cite{imole2013hydrostatic, goncu2010constitutive, madadi2004fabric}.
From the snapshots in Fig.\ \ref{c3d}, where number of contacts of particle $p$ is $C_p\ge0$, all the particles with less than 4 contacts are removed.
The rattlers exclusion is an iterative process until all remaining particles have at least 4 contacts ($C_p^*\ge4$), 
that provides us a completely mechanically stable systems as shown in Fig.\ \ref{new_c3d} (only particles B are shown).

Unless mentioned explicitly, we will denote $N_A$ and $N_B$ as the number of mechanically stable particles of A and B, respectively, 
and use them to compute micro- and macroscopic quantities. 
Any superscript `\textit{T}' relates to the total number of particles of A and B, including the rattlers. 
% The same will be used for understanding the evolution of macroscopic properties with the micro-structure.

\subsection{Rattlers} 
\label{sec:rattlers}
\begin{figure}[!ht]
\centering
\includegraphics[width=0.4\textwidth, angle=0]{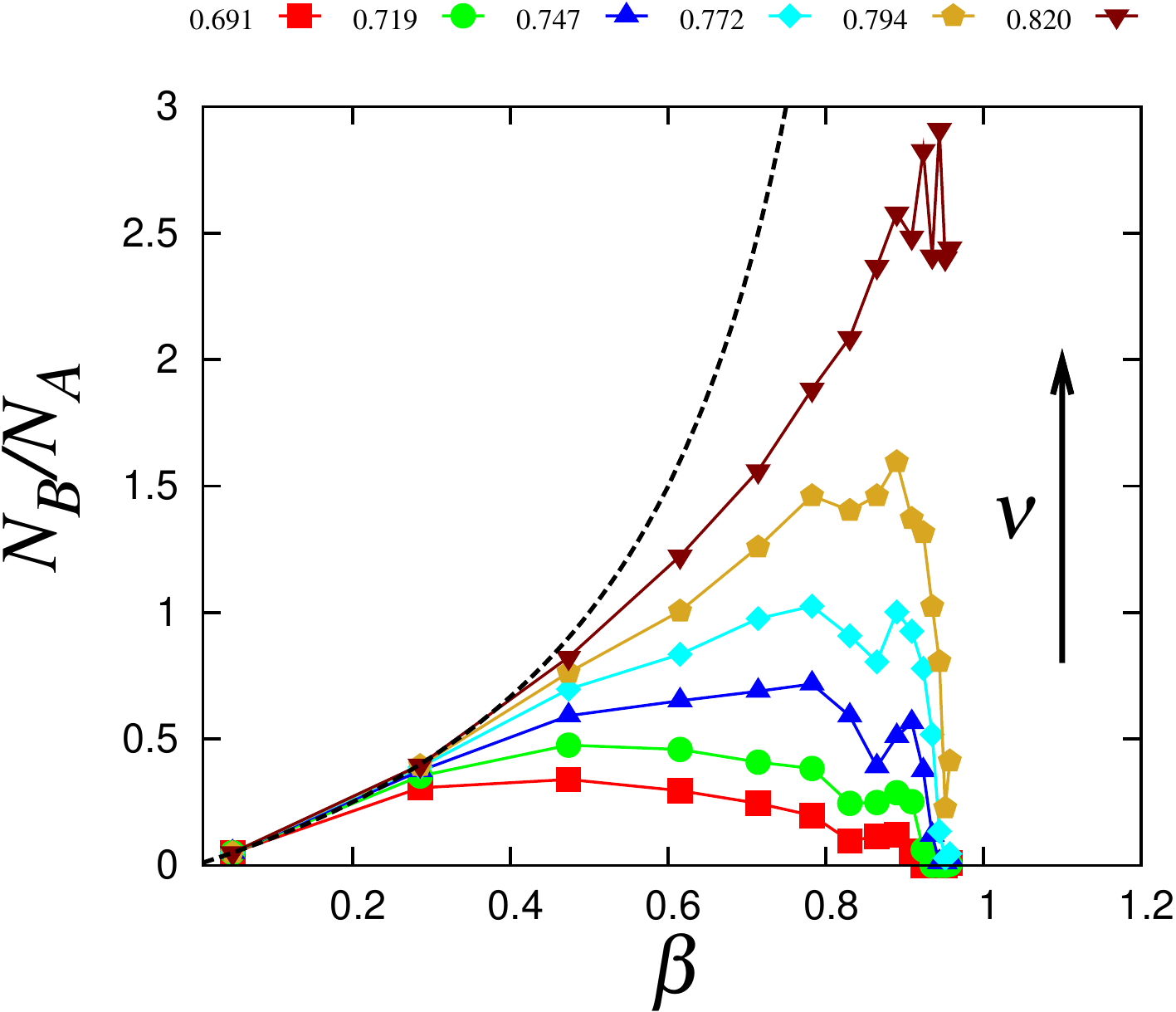}
\caption{Ratio of B particles with respect A: $N_B/N_A$, when the assembly contains no rattlers; the dashed line considers also rattlers (see Eq. (\ref{eq:NBTratio})), i.e.,\ $N_B^T/N_A^T$. 
All are plotted against the number fraction $\beta= N_B^T/\left(N_A^T + N_B^T \right)$. 
Different colors represent the volume fraction $\nu$ as shown in the legend and the arrow indicates increasing $\nu$. }
\label{NB_pvskappa_noscale_mu0p0}
\end{figure}

In Fig.\ \ref{NB_pvskappa_noscale_mu0p0}, we plot the number ratio of participating particles B with respect to A after removing the rattlers, i.e.,\ $N_B/N_A$.
For all cases, the assembly contains $95\%$ by volume of big particles of A. 
Thus, $N_A$ after removing rattlers is close to $N_A^T$, i.e.\ $N_A/N_A^T\approx1$. 
With decreasing size of B, i.e.,\ increasing $\beta$, an initial increase in the ratio $N_B/N_A$ is seen, followed by a maximum and a later decrease for all the volume fractions.
For small $\beta$, few B particles are present with size comparable to A. 
With increasing $\beta$ and for all $\nu$, the ratio $N_B/N_A$ increases as more B particles of smaller size are introduced in the system while the number of A stays constant.
For a fixed $\nu$, there is an average void size created by A that can be most efficiently filled by an optimal 
(just fitting) B\footnote[3]{Big particles A create voids filled by B. Different lattice arrangements of A provide the void size such that B touches A particles, giving the size ratio $r_B/r_A$ for the most compact packing. 
A simple approach to measure this ratio for triangular, tetrahedron, square and cubic lattices formed by A particle is shown in  appendix \ref{App:AppendixA}).}, 
the optimum size ratio $r_B/r_A$ corresponds to a maximum in $N_B/N_A$.
Indeed, when $\beta$ increases further, meaning more smaller particles B in the system, the number of active (non-rattlers) particles B decreases, 
as most of them become rattlers, `caged' in the voids of A \cite{kumar2014effects}. 
This can be seen comparing Fig.\ \ref{c3d1038_ISO5} with Fig.\ \ref{new_c3d1038_ISO5}. 
Therefore, the ratio $N_B/N_A$ decreases after the maximum. 

Another important observation is that with increasing $\nu$, the maximum in $N_B/N_A$ occurs at increasing $\beta$. 
This is because for increasing $\nu$, the base assembly A is more compressed with smaller void size. 
That is smaller B particles already fill efficiently the voids of A as non-rattlers. 
For the densest case $\nu=0.82$, the ratio $N_B/N_A$ seems to saturate for large $\beta$. However when $\beta \to1$, $N_B$ will decrease and hence the ratio $N_B/N_A$.
Due to the computational limitations, this observation can not be presented. 

Fig.\ \ref{NB_pvskappa_noscale_mu0p0} also shows the ratio $N_B/N_A$ including rattlers, same for all density represented by Eq. (\ref{eq:NBTratio}). 
$N_B^T/N_A^T$ is higher than $N_B/N_A$, since $N_B$ is smaller than $N_B^T$, while $N_A/N_A^T\approx1$. 
Note that the dashed line is closer to the dense systems, where the majority of B particles are active particles. 

\subsection{Coordination number} 
\label{sec:coord}

The classical definition of coordination number is $C = M/N^T$, 
where $M$ is the total number of contacts and $N^T$ is the total number of particles \cite{goncu2010constitutive, imole2013hydrostatic, kumar2014effects}. 
In order to quantify the active contact network (excluding rattlers), we use the corrected coordination number: $C^* = {M_4}/{N_4},$ where $M_4$ is the total number of contacts of 
the $N_4$ particles with at least 4 contacts, i.e.\ $C_p^*\ge4$ (see section\ \ref{sec:stable}). 
Note that, after excluding rattlers, the number of particles left in the system is $ (N_A+N_B) = N_4 < N^T=(N_A^T+N_B^T)$

\begin{figure}[!ht]
\centering
\includegraphics[width=0.4\textwidth, angle=0]{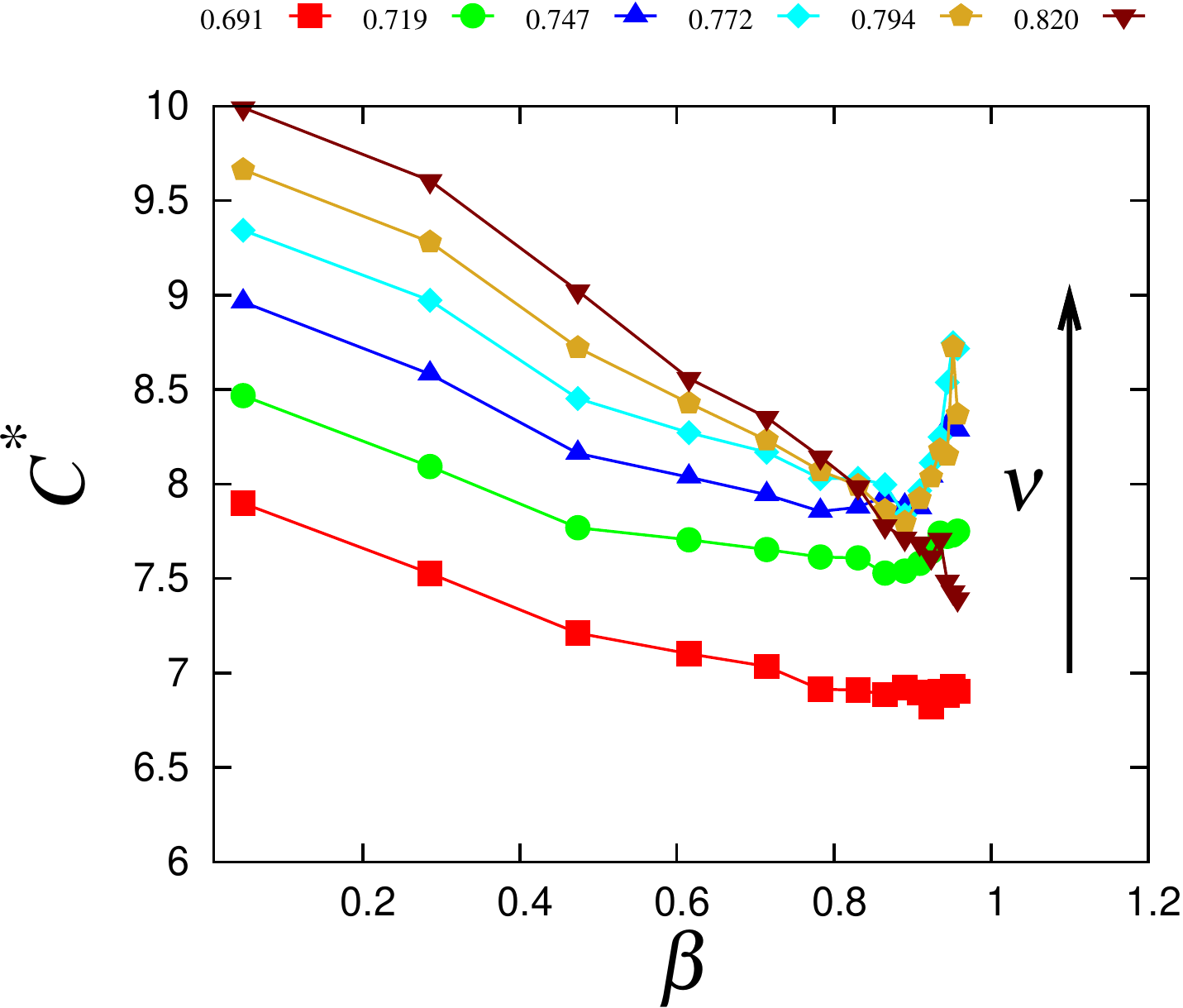}
\caption{Coordination number excluding rattlers plotted against the number fraction $\beta= N_B^T/\left(N_A^T + N_B^T \right)$. 
Different colors represent the volume fraction $\nu$ as shown in the legend and the arrow indicates increasing $\nu$.}
\label{cstar_pvskappa_noscale_mu0p0}
\end{figure}

Fig.\ \ref{cstar_pvskappa_noscale_mu0p0} shows the evolution of $C^*$ with $\beta$ for six different volume fractions.
As expected, for a given composition (fixed $\beta$), the total coordination number of the system increases with volume fraction $\nu$ 
as the system becomes more dense and particles are both closer and better coordinated.
For given density $\nu$, $C^*$ decreases continuously with $\beta$, since the number of non-rattler particles $N_4$ increases faster than the non-rattler contacts $M_4$; i.e.,\ $C^*$ decreases. 
At high $\beta$, an increase in $C^*$ is seen. 
This is associated with the drop in active particles B, $N_B$, in other words $N_B/N_A$, as shown in Fig.\ \ref{NB_pvskappa_noscale_mu0p0}.
%
% Considering the coordination number of individual species, $C^*_B$ decreases systematically with $\beta$ and reaches asymptotically the isostatic value 4, 
% while $C^*_A$ first increases and then decreases for high $\beta$.

\subsection{Dimensionless moments} 
\label{sec:moments}

\begin{figure}[!ht]
\centering
\subfigure[]{\includegraphics[width=0.26\textwidth, angle=0]{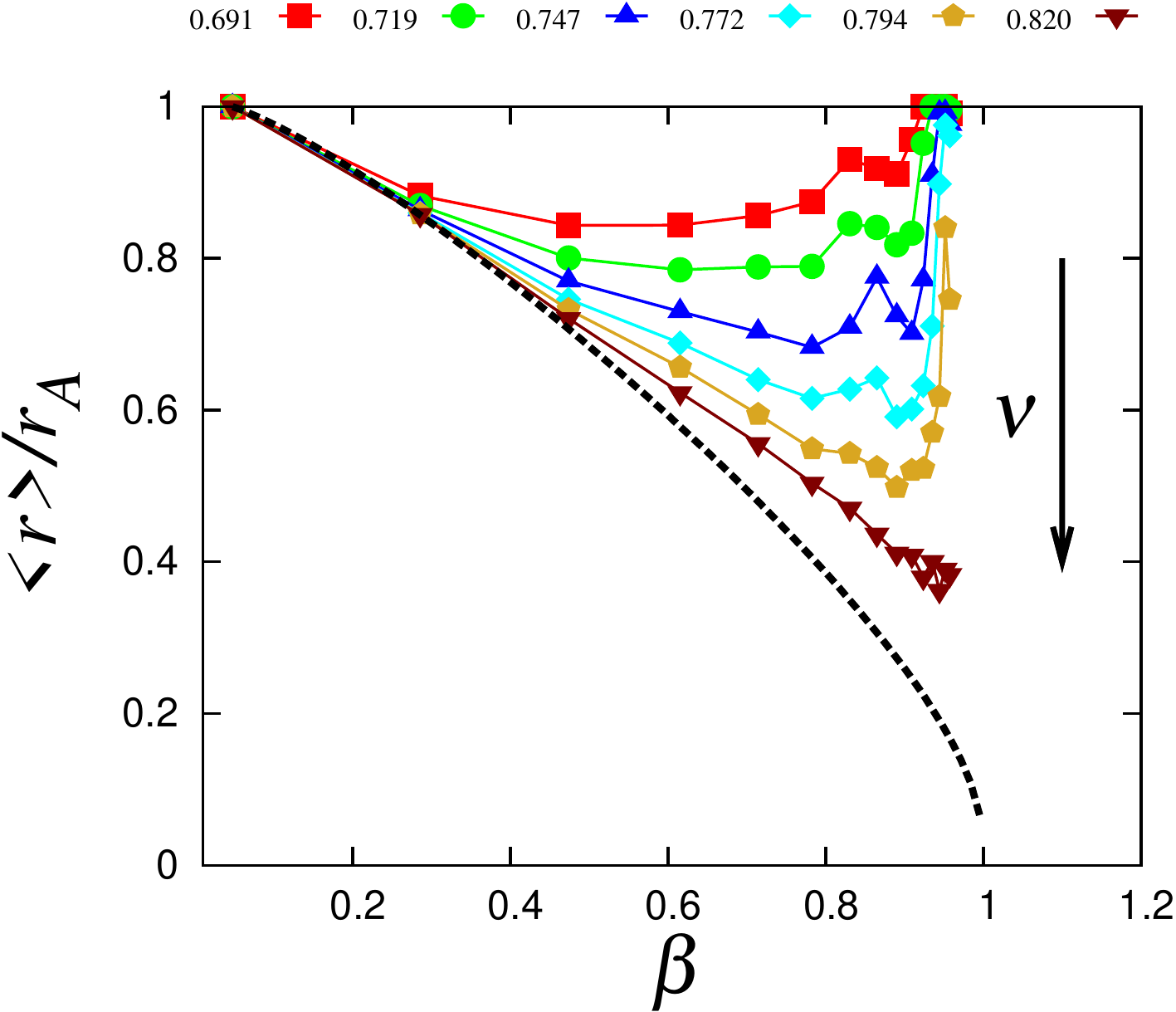}\label{meanR_pvskappa_noscale_mu0p0}}
\subfigure[]{\includegraphics[width=0.26\textwidth, angle=0]{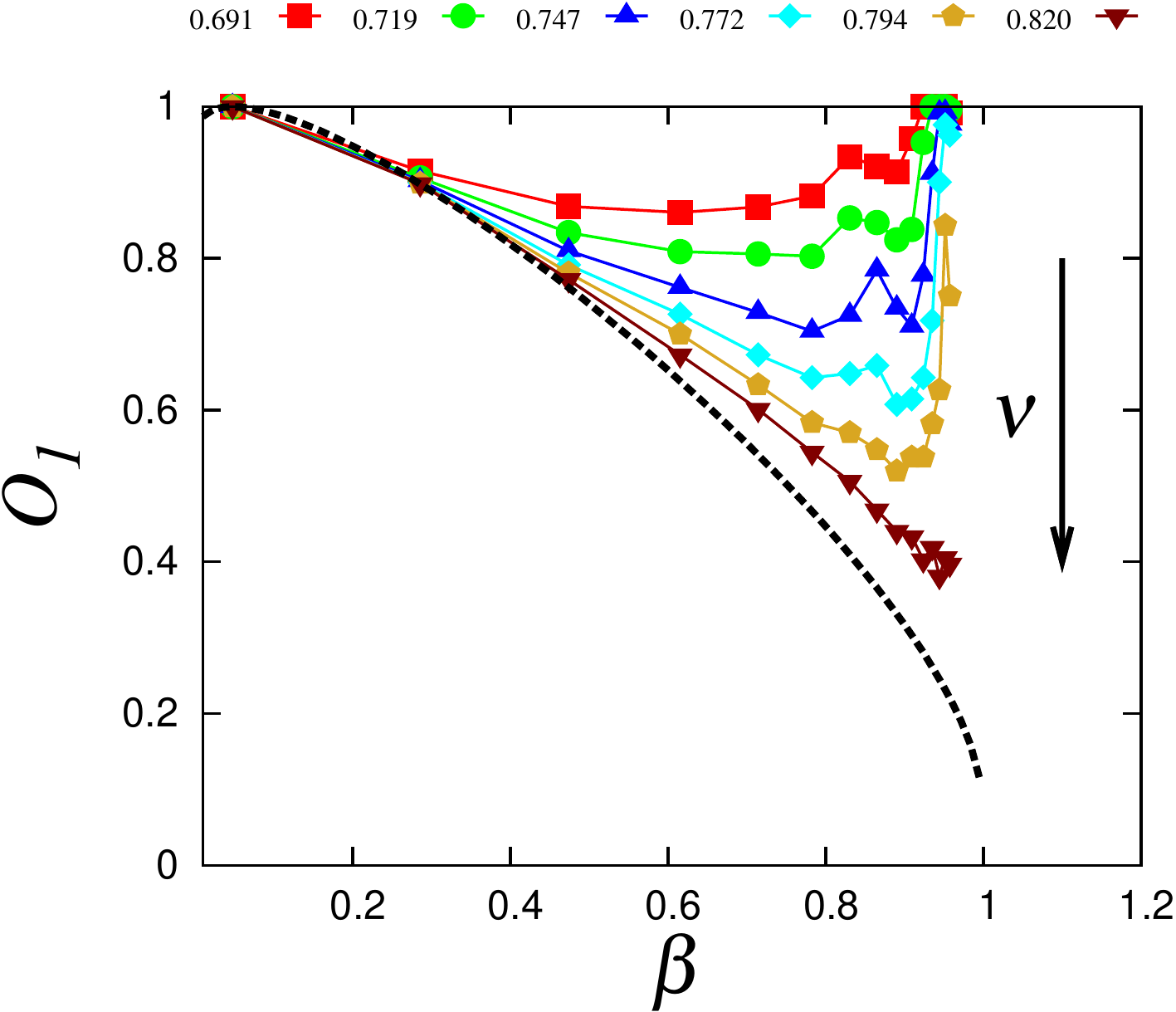}\label{O1_pvskappa_noscale_mu0p0}}
\subfigure[]{\includegraphics[width=0.26\textwidth, angle=0]{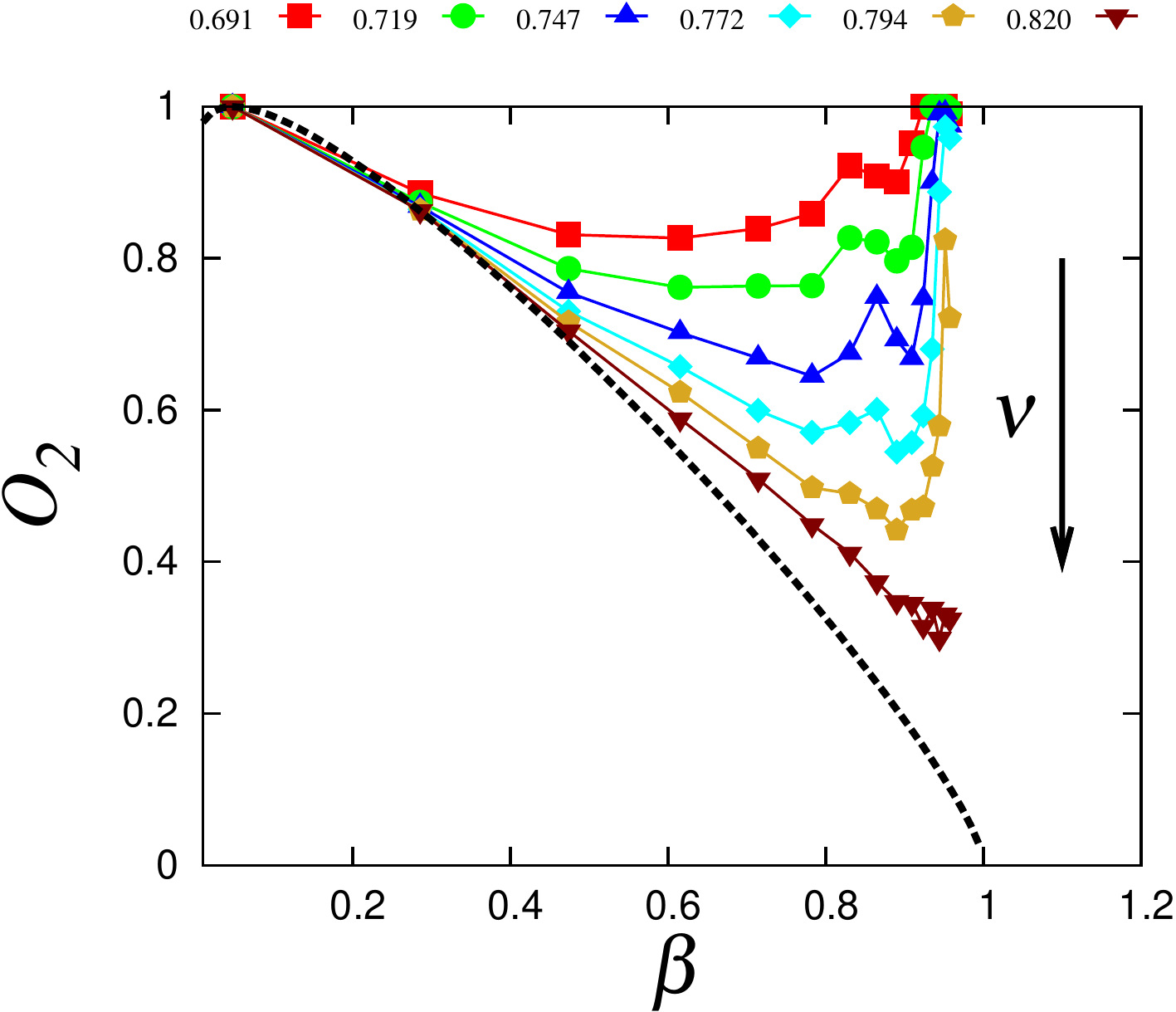}\label{O2_pvskappa_noscale_mu0p0}}
\caption{(a) Average radius $\langle r \rangle$ scaled $r_A$; 
dimensionless moments (b) $O_1$ and (c) $O_2$, measured using Eq.\ (\ref{eq:o1o2eqn}), excluding rattlers, plotted against the number fraction $\beta= N_B^T/\left(N_A^T + N_B^T \right)$. 
Different colors represent the volume fraction $\nu$ as shown in the legend. 
The dashed lines that consider also rattlers are Eqs.\ (\ref{eq:rmomentfull}) and (\ref{eq:o1o2eqnfull}). 
The arrows indicate increasing $\nu$.}
\label{O1O2}
\end{figure}

The average radius and moments are among the fundamental quantities needed to characterize the particle size distribution \cite{ogarko2012equation}. 
Given $f(r)$ as the particle radii (size) distribution, $f(r)dr$ is the probability
to find the radius between $r$ and $r+dr$, with a normalization condition $\int_0^\infty f(r)=1$ \cite{ogarko2012equation, goncu2010constitutive}. 
For a bidisperse distribution $f(r) = f_A \Delta\left(r - r_A\right) + f_B\Delta\left(r - r_B\right) $, 
where $f_A = N_A/\left(N_A + N_B \right)$ and $f_B = N_B/\left(N_A + N_B \right)$ are the number fractions of A and B and $\Delta(r)$ is the Dirac-delta function. 
While $N^T = N_A^T + N_B^T$, without superscript ($T$ denotes the total number of particles), 
$N_A + N_B = N_4$ is the total number of particles in the system with at least four contacts ($C_p^*\ge4$, see section\ \ref{sec:stable}). 
The $n^{\mathrm{th}}$ moment is $\langle r^n \rangle = \int_0^\infty r^nf(r)$. 
The mean particle radius for a bidisperse distribution is thus $\langle r \rangle = \int_0^\infty rf(r) = f_A r_A + f_B r_B$ 
and the $n^{\mathrm{th}}$ moment is $\langle r^n \rangle = f_A r_A^n + f_B r_B^n$ with $f_A + f_B=1$. 

Fig.\ \ref{meanR_pvskappa_noscale_mu0p0} shows the average radius of the system scaled with the radius of A; i.e.,\ $\langle r \rangle/r_A$ excluding rattlers. 
Starting from 1, $\langle r \rangle/r_A$ decreases with increasing $\beta$ due to the presence of smaller B particles
\footnote[4]{This is understood from the inequality: $\langle r \rangle/r_A =  f_A + f_B r_B/r_A = 1 -f_B +  f_B r_B/r_A = 1 - f_B(1-r_B/r_A) < 1$, 
since the second term is positive and smaller than unity.}. 
This decrease is faster for higher $\nu$ and shows an inverse trend with respect to $N_B/N_A$ in Fig.\ \ref{NB_pvskappa_noscale_mu0p0}. 
For $\beta\to1$, the size of B becomes very small compared to A so that they do not contribute, 
and the system excluding rattlers is mainly composed of A, so that $\langle r \rangle/r_A \to 1$.

The dimensionless moments of a polydisperse assembly $O_1$ and $O_2$ are defined as \cite{ogarko2012equation}:
\begin{equation}
\label{eq:o1o2eqn}
O_1=\frac{\langle r \rangle\langle r^2 \rangle}{\langle r^3 \rangle} \quad\mbox{and}\quad O_2=\frac{{\langle r^2 \rangle}^3}{{\langle r^3 \rangle}^2},
\end{equation}
where it was shown that $O_1$ and $O_2$ are needed to completely quantify the fluid-like behavior of a granular assembly well below jamming.
For our system, $N_A$ and $N_B$ change with $\beta$ and volume fraction $\nu$, hence 
$O_1$ and $O_2$ are different for different volume fractions. 
If the dimensionless moments $O_1$ and $O_2$ are known, the $2^{\mathrm{nd}}$ and $3^{\mathrm{rd}}$ dimensionless moments (moment scaled by $\langle r \rangle$) are:
\begin{equation}
\label{eq:r2r3eqn}
\frac{\langle r^2 \rangle}{{\langle r \rangle}^2} = \frac{O_2}{O_1^2}\quad\mbox{and}\quad 
\frac{\langle r^3 \rangle}{{\langle r \rangle}^3}=\frac{O_2}{O_1^3}.
\end{equation}
The $n^{\mathrm{th}}$ moment is always greater or equal to the $n^{\mathrm{th}}$ power of the mean radius, i.e.\ ${\langle r^n \rangle}/{{\langle r \rangle}^n} \ge 1$.
Therefore, $O_1^2 \le O_2$ and $O_1 \le 1$, as also shown in Ref.\ \cite{ogarko2012equation}.
For the full system including the rattlers, 
\begin{equation}
\label{eq:rmomentfull}
\langle r \rangle^T = (1-\beta) r_A  + \beta r_B  
\quad\mbox{and}\quad 
\langle r^2 \rangle^T=  (1-\beta) r_A^2 + \beta r_B^2
\quad\mbox{and}\quad 
\langle r^3 \rangle^T =  (1-\beta) r_A^2 + \beta r_B^2.
\end{equation}

% $\langle r \rangle^T = (1-\beta) r_A  + \beta r_B $, $\langle r^2 \rangle =  (1-\beta) r_A^2 + \beta r_B^2$ and $\langle r^3 \rangle =  (1-\beta) r_A^2 + \beta r_B^2$. 
Therefore, $O_1$ and $O_2$ for the full system become
\begin{equation}
\label{eq:o1o2eqnfull}
O_1^T=\frac{ \left[ (1-\beta) + \beta (r_B/r_A) \right] \left[ (1-\beta) + \beta (r_B/r_A)^2 \right]}{(1-\beta) + \beta (r_B/r_A)^3} 
\quad\mbox{and}\quad 
O_2^T=\frac{\left[ (1-\beta) + \beta (r_B/r_A)^2 \right]^3}{\left[ (1-\beta) + \beta (r_B/r_A)^3 \right]^2}.
\end{equation}
Inserting the radius ratio $r_B/r_A = \Phi^{1/3} \left(\frac{1-\beta}{\beta} \right)^{1/3}$ 
from Eq.\ (\ref{eq:radiusratio}), where $\Phi=0.05=\mathrm{const.}$ (see section\ \ref{sec:simmeth}), 
Eq.\ (\ref{eq:o1o2eqnfull}) can be re-written as
\begin{equation}
\label{eq:o1o2eqnfull1}
O_1^T=\frac{ \left[ (1-\beta)^{2/3} + \Phi^{1/3} \beta^{2/3}  \right]  \left[ (1-\beta)^{1/3} + \Phi^{2/3} \beta^{1/3}  \right] }{ (1 + \Phi) } 
\quad\mbox{and}\quad 
O_2^T=\frac{ \left[ (1-\beta)^{1/3} + \Phi^{2/3} \beta^{1/3} \right]^3 }{ (1 + \Phi)^{2} }.
\end{equation}
The asymptotic values for $O_1^T$ and $O_2^T$ when $\beta \to 1$ are $\Phi/(1+\Phi)$ and $\Phi^2/(1+\Phi)^2$, respectively. 
Fig.\ \ref{O1_pvskappa_noscale_mu0p0} and \ref{O2_pvskappa_noscale_mu0p0} show the evolution of $O_1$ and $O_2$ with $\beta$ and $\nu$. 
Both $O_1$ and $O_2$ are smaller than unity \cite{ogarko2012equation}, decreasing with $\beta$, and show a similar trend as $\langle r \rangle/r_A$. 
The dashed lines in Fig.\ \ref{O1O2} represent the granular assembly including the rattlers, i.e.\ only a single line for all volume fractions. 
%%%
%%%%
% \frac{r_B}{r_A} = \left( \Phi \frac{N_A^T}{N_B^T} \right)^{1/3} = \Phi^{1/3} \left(\frac{1-\beta}{\beta} \right)^{1/3}.
%%%%%
%%%%%

\subsection{Corrected volume fraction} 
\label{sec:volfrac}

\begin{figure}[!ht]
\centering
\subfigure[]{\includegraphics[width=0.4\textwidth, angle=0]{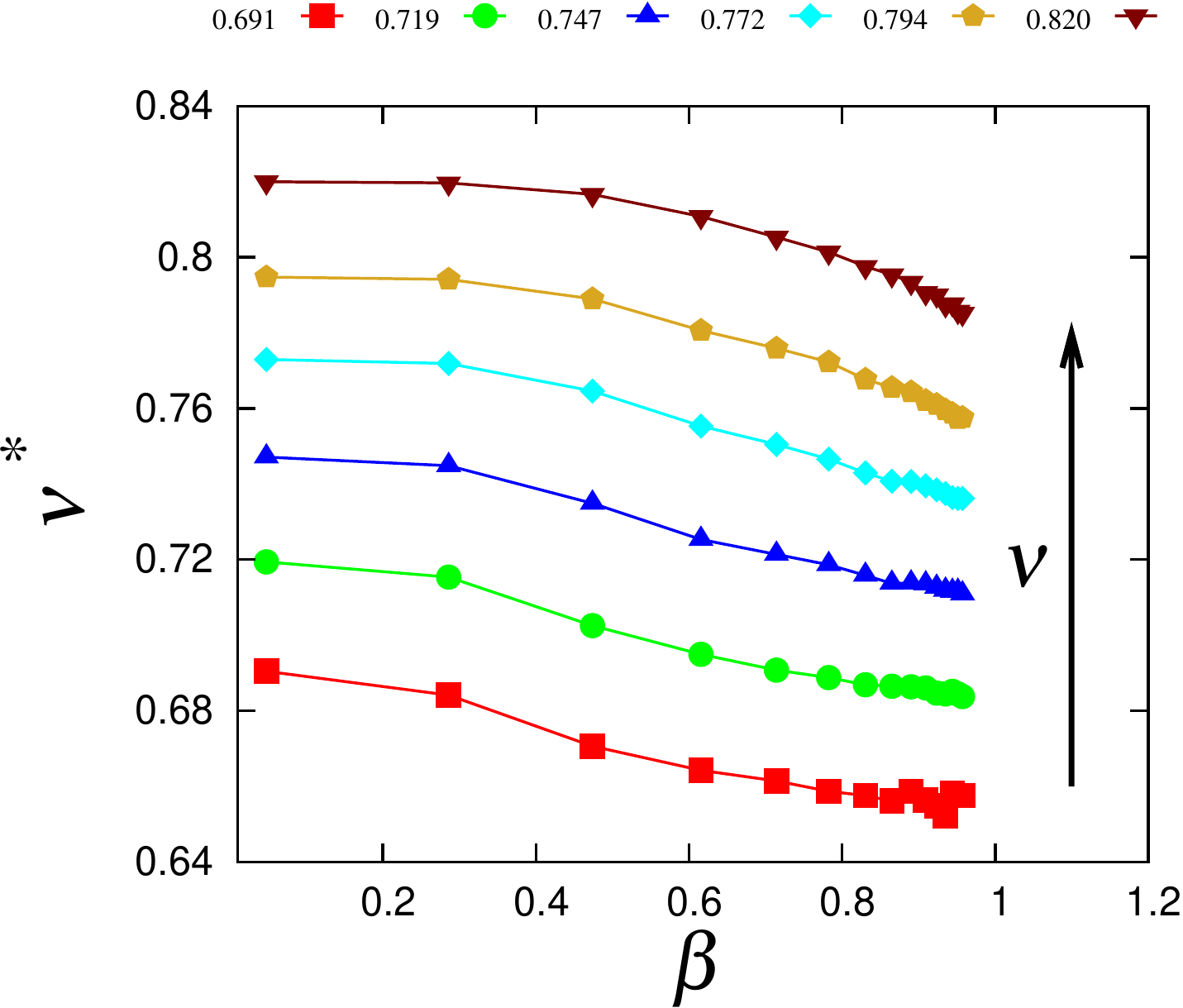}\label{nu_pvskappa_noscale_mu0p0}}
\subfigure[]{\includegraphics[width=0.4\textwidth, angle=0]{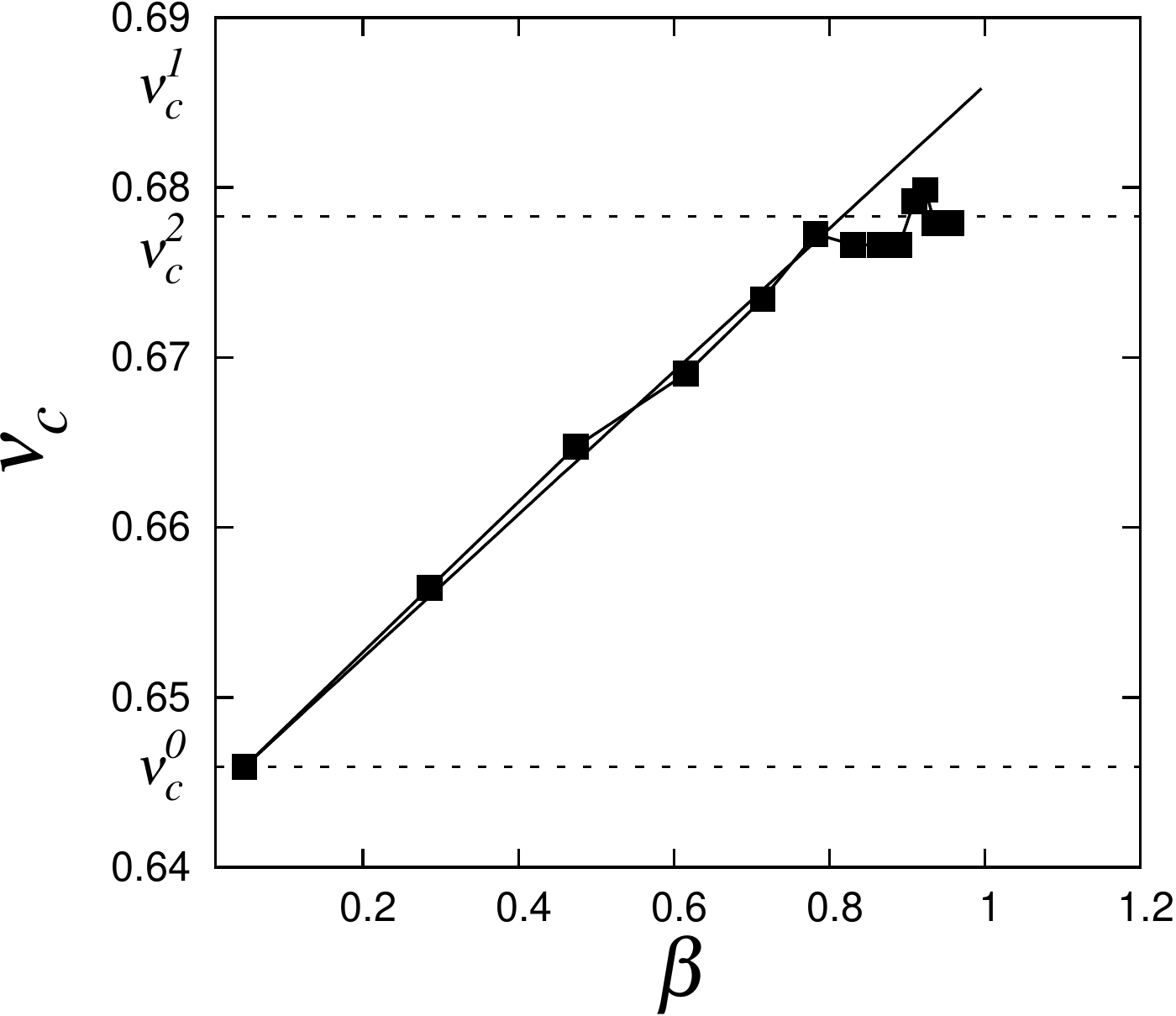}\label{nucvskappa_noscale_mu0p0}}
\caption{(a) Volume fraction $\nu^*$ of the system excluding rattlers plotted against the number fraction $\beta= N_B^T/\left(N_A^T + N_B^T \right)$. 
Different colors represent the volume fraction $\nu$ as shown in the legend and the arrow indicates increasing $\nu$. 
(b) Evolution of jamming point $\nu_c$ with $\beta$. The solid line is a linear fit to the simulation data using Eq.\ (\ref{eq:nuceqn}). 
The dashed horizontal lines are: 
$\nu_c^0$ for the smallest $\beta$, 
$\nu_c^1$ fit parameter obtained for $\beta\to1$ and
$\nu_c^2$ is the estimated value for $\beta\to1$ using Eq.\ (\ref{eq:nuceqn1}) when the system contains only particles A.}
\label{volnu}
\end{figure}

It is interesting to look closer at the behavior of the corrected volume fraction $\nu^*$, i.e.,\ 
the volume fraction excluding the non-active particles
\begin{equation}
\label{eq:volfracexcl}
\nu^* = N_4 \frac{4\pi}{3}\frac{ \langle r^3 \rangle}{V}. 
\end{equation}
Fig.\ \ref{nu_pvskappa_noscale_mu0p0} shows the corrected volume fraction $\nu^*$ versus $\beta$. 
For any volume fraction $\nu$, $\nu^*$ decreases continuously with $\beta$, 
since the volume fraction of rattlers (mainly B) increases with $\beta$. 
For decreasing size of B, more and more B particles are `caged' between the big particles A without having sufficient ($C_p^*\ge4$) 
contacts \cite{kumar2014effects}. 
For the reference case, when the radius of B is equal to that of A, (leftmost data points), $\nu^* \approx \nu$, 
and for the density close to the jamming point, $\nu^* \approx 0.98\nu$, as approximately $2\%$ of the particles are rattlers, 
in agreement with the values reported in \cite{kumar2014effects} for the monodisperse case.

For each mixture, we extract the jamming point $\nu_c$, i.e.,\ the volume fraction $\nu$ when the pressure $p$ of the mixture (defined in section\ \ref{sec:press}) approaches zero.
Fig.\ \ref{nucvskappa_noscale_mu0p0} shows $\nu_c$ increasing with $\beta$ and saturating for $\beta\to1$. 
This can be understood since the number of non-rattler particles decreases with $\beta$, as also seen in Fig.\ \ref{nu_pvskappa_noscale_mu0p0}, 
until they reach the number of $N_A^T$.
Therefore, with increasing $\beta$, one needs to compress the system further (or increase the volume fraction) 
to make sure particles achieve an overlapping, jammed configuration, leading to increase in $\nu_c$ with $\beta$. 
For $\beta\to1$ and volume fractions near $\nu_c$, all particles B are rattlers and therefore $\nu_c$ saturates for $\beta\to1$. 
Note that $\nu^*$ in Fig.\ \ref{nu_pvskappa_noscale_mu0p0} excludes the rattlers while $\nu_c$ in Fig.\ \ref{nucvskappa_noscale_mu0p0} includes the rattlers. 
The relation between $\nu_c$ and $\beta$ can be fitted by the linear relation:
\begin{equation}
\label{eq:nuceqn}
\nu_c = \nu_c^0 + (\nu_c^1 - \nu_c^0) \frac{\beta - \beta_\mathrm{min}}{1 - \beta_\mathrm{min}}, 
\end{equation}
where $\nu_c^0=0.646$ for the monodisperse case for $\beta_\mathrm{min} = 0.05/1.05=0.0476$. 
$\nu_c^1=0.682$ is the only fit parameter for $\beta=1$, obtained by fitting the simulation data in Fig.\ \ref{nucvskappa_noscale_mu0p0} up to $\beta=0.8$. 
Note that the $\nu_c^0=0.646$ value measured from the simulation for the monodisperse case is consistent with results $\nu_c^0$ in Ref.\ \cite{kumar2014effects} for different system size. 
Assuming that for $\beta\to1$ only A particles  contribute to te structure, we estimate the saturation volume fraction as 
\begin{equation}
\label{eq:nuceqn1}
\nu_c^2 = \nu_c^0 \left(1+ \Phi \right) =  0.678, 
\end{equation}
in consistent with the measured values for $\beta\to1$ as shown in Fig.\ \ref{nucvskappa_noscale_mu0p0}.

\section{Macroscopic Quantities} 
\label{sec:macro}

In the previous section, we focused on the averaged microscopic quantities; rattlers and coordination number. 
Next we focus on defining the averaged macroscopic quantities -- stress and fabric (structure), that 
reveal interesting bulk features and provide information
about the state of the packing via its response to applied deformation.

\subsection{Fabric} 
\label{sec:fabric}

 \begin{figure}[!ht]
\centering
\includegraphics[width=0.4\textwidth, angle=0]{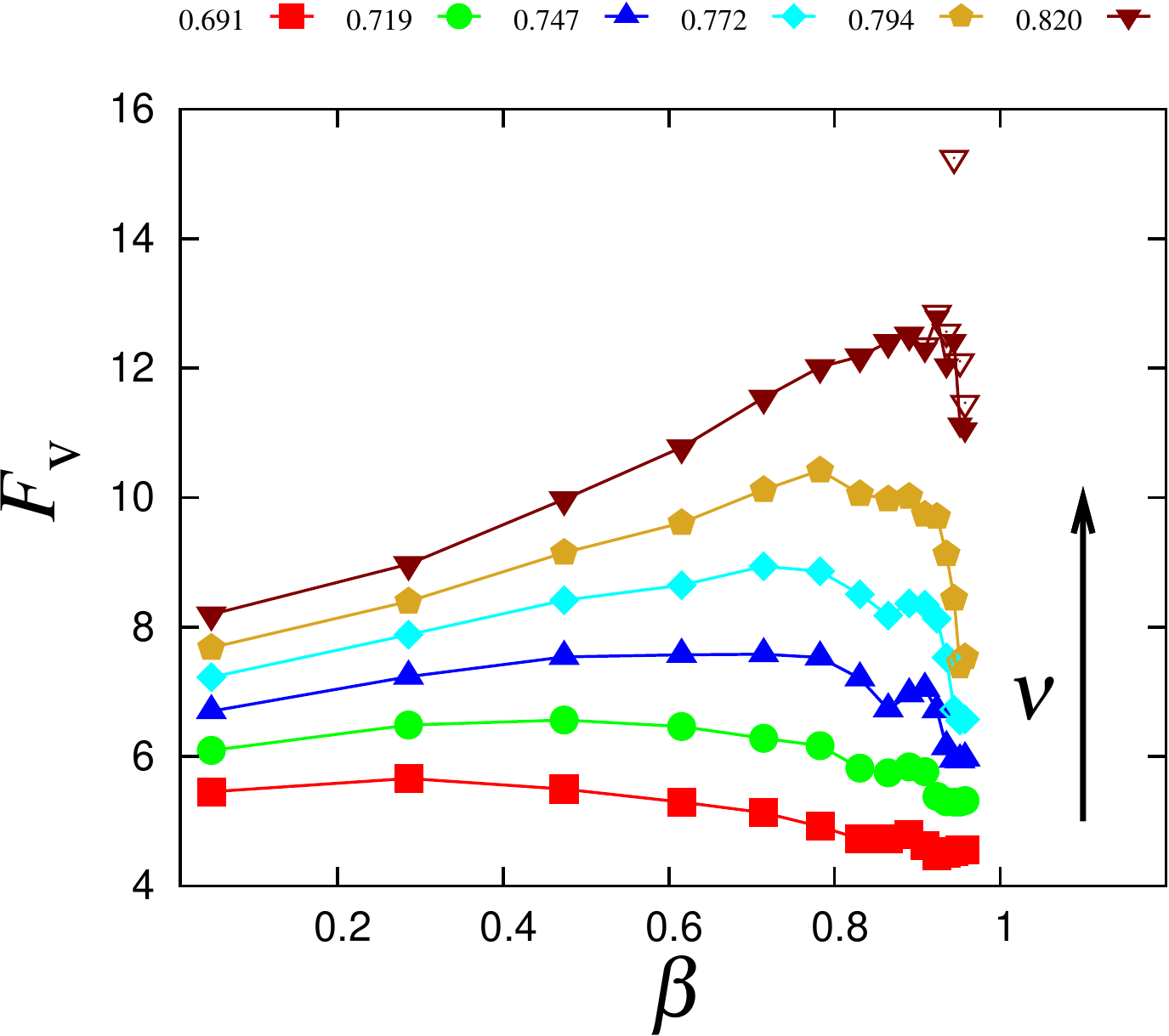}
\caption{Isotropic fabric $\Fv$ plotted against the number fraction $\beta= N_B^T/\left(N_A^T + N_B^T \right)$.
Different colors represent the volume fraction $\nu$ as shown in the legend and arrow indicates increasing $\nu$. 
Open symbols are corresponding $\Fv^T$ that includes the rattlers as well. }
\label{FvT_pvskappa_noscale_mu0p0}
\end{figure}

% \begin{figure}[!ht]
% \centering
% %
% \subfigure[]{\includegraphics[width=0.4\textwidth, angle=0]{Figures/Fv_pvskappa_noscale}\label{FvT_pvskappa_noscale_mu0p0}}
% \subfigure[]{\includegraphics[width=0.4\textwidth, angle=0]{Figures/FvT_pvskappa_noscale}\label{FvT_pvskappa_noscale_mu0p0}}
% %
% \caption{Isotropic fabric $\Fv$ plotted against the number fraction $\beta= N_B^T/\left(N_A^T + N_B^T \right)$.
% Different colors represent the volume fraction $\nu$ as shown in the legend and arrows indicate increasing $\nu$. 
% Open symbols are $\Fv^T$ that includes the rattlers as well. }
% \label{fabric_figures}
% \end{figure}

From the DEM simulations, one can determine the fabric tensor in order to characterize the geometry/structure of the static aggregate 
\cite{luding2005anisotropy, kumar2013evolution, goncu2010constitutive, shaebani2012influence}, defined as 
\begin{equation}
\label{eq:fabriceq}
\mathbf{F}^T=\frac{1}{V}\sum\limits_{p=1}^{N^T}V_p\sum_{c=1}^{C_p}\mathbf{n}^{c}\otimes\mathbf{n}^{c},
\end{equation}
where $V_p$ is the volume of particle $p$, which lies inside the
averaging system volume $V$, and $\mathbf{n}^{c}$ is the normal unit branch vector
pointing from center of particle $p$ to contact $c$. 
$C_p$ is the number of contacts of particle $p$ and $N^T$ represents the total number particles. 
In the case of isotropically compressed systems, the isotropic fabric $\Fv^T$ is 
the quantity of interest and is obtained by taking the trace of Eq.\ (\ref{eq:fabriceq}) as:
\begin{equation}
\label{eq:fabriceqtr1}
\Fv^T=\mathrm{tr}\left(\mathbf{F}^T\right) = \frac{1}{V}\sum\limits_{p=1}^{N^T}V_p\sum_{c=1}^{C_p}1 = \frac{1}{V}\sum\limits_{p=1}^{N^T}V_pC_p.
\end{equation}
Note that we exclude iteratively the rattlers from the system (see section\ \ref{sec:simmeth}), 
and observe that their contribution to the fabric is small (as shown in Fig.\ \ref{FvT_pvskappa_noscale_mu0p0}). 
Therefore, the isotropic fabric for non-rattler particles with stable non-rattler contacts ($C_p^*\ge4$) is:
\begin{equation}
\label{eq:fabriceqtr2}
\Fv= \frac{1}{V}\sum\limits_{p=1}^{N_4}V_pC_p^* \approx  \frac{1}{V}\sum\limits_{p=1}^{N^T}V_pC_p ,
\end{equation}
where $N_4 = \left(N_A + N_B\right) \le \left(N_A^T + N_B^T \right) = N^T$ is the total number of particles excluding the rattlers, as defined in section\ \ref{sec:simmeth}.

Fig.\ \ref{FvT_pvskappa_noscale_mu0p0} shows the evolution of $\Fv$ calculated using Eq.\ (\ref{eq:fabriceqtr2}) with $\beta$ for six volume fractions. 
The first important observation if that the contribution of rattlers to the fabric is small and $\Fv^T$ is very close to $\Fv$. 
For a given mixture (fixed $\beta$), $\Fv$ increases with volume fraction $\nu$, 
meaning that the system becomes more connected. On the other hand, for a given $\nu$, $\Fv$ first increases and then decreases, 
The maximum of $\Fv$ is correlated with the average voids created by particles A, that, 
for a given $\nu$ and size of B can be optimally filled (see appendix \ref{App:AppendixA}). 
The behavior of $\Fv$ is similar to that of number of non-rattlers particles of B, $N_B$, observed in Fig.\ \ref{NB_pvskappa_noscale_mu0p0}, since $\Fv \propto N_4$ (Eq.\ (\ref{eq:fabriceqtr2})). 

% This is due to the fact that for a given volume fraction, there is an average void created by the particles A, that can be optimally filled by fines B, leading to a maxima of $\Fv$ at that $\beta$. 
% This does not reflect in pressure directly as shown in Fig.\ \ref{pres_pvskappa_noscale_mu0p0}, where pressure decreases systematically with $\beta$ for all densities. 
% This can be also explained when looking into Fig.\ \ref{NB_pvskappa_noscale_mu0p0}, 
% where $N_A \approx N_A^T$  and $N_B$ increases with $\beta$ and later decreases. For particles B, due to their small size, $C_p^*$ is close to 4. 
% From Eq.\ (\ref{eq:fabriceqtr2}), $\Fv \propto N_4$; the behavior of $\Fv$ is similar to that of $N_B$ observed in Fig.\ \ref{NB_pvskappa_noscale_mu0p0}.
% The number of contacts increases with $\beta$ up to a certain limit, because of more small particles of B, but they contribute small to the stress and hence monotonic decrease pressure. 
% A direct relation between the two isotropic quantities, pressure and fabric shown in Fig.\ \ref{pvsFvT_pvskappa_noscale_mu0p0}, is not observed here, as presented in \cite{imole2013hydrostatic, kumar2014effects}. 

% % \subsubsection{Relation of fabric with packing properties} 
% % \label{sec:fabricreln1}

\begin{figure}[!ht]
\centering
\subfigure[]{\includegraphics[width=0.4\textwidth, angle=0]{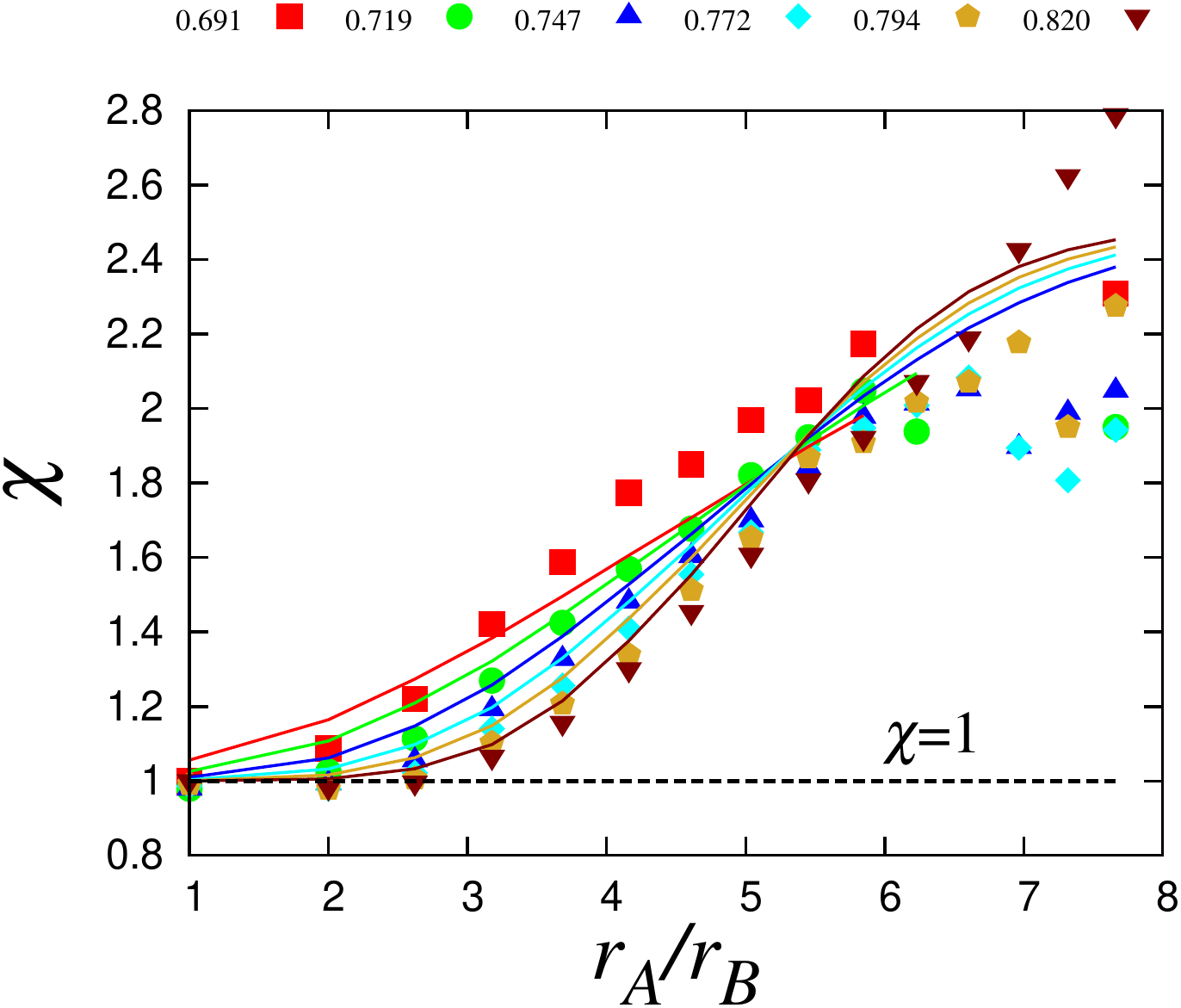}\label{gagbplot}}
\subfigure[]{\includegraphics[width=0.4\textwidth, angle=0]{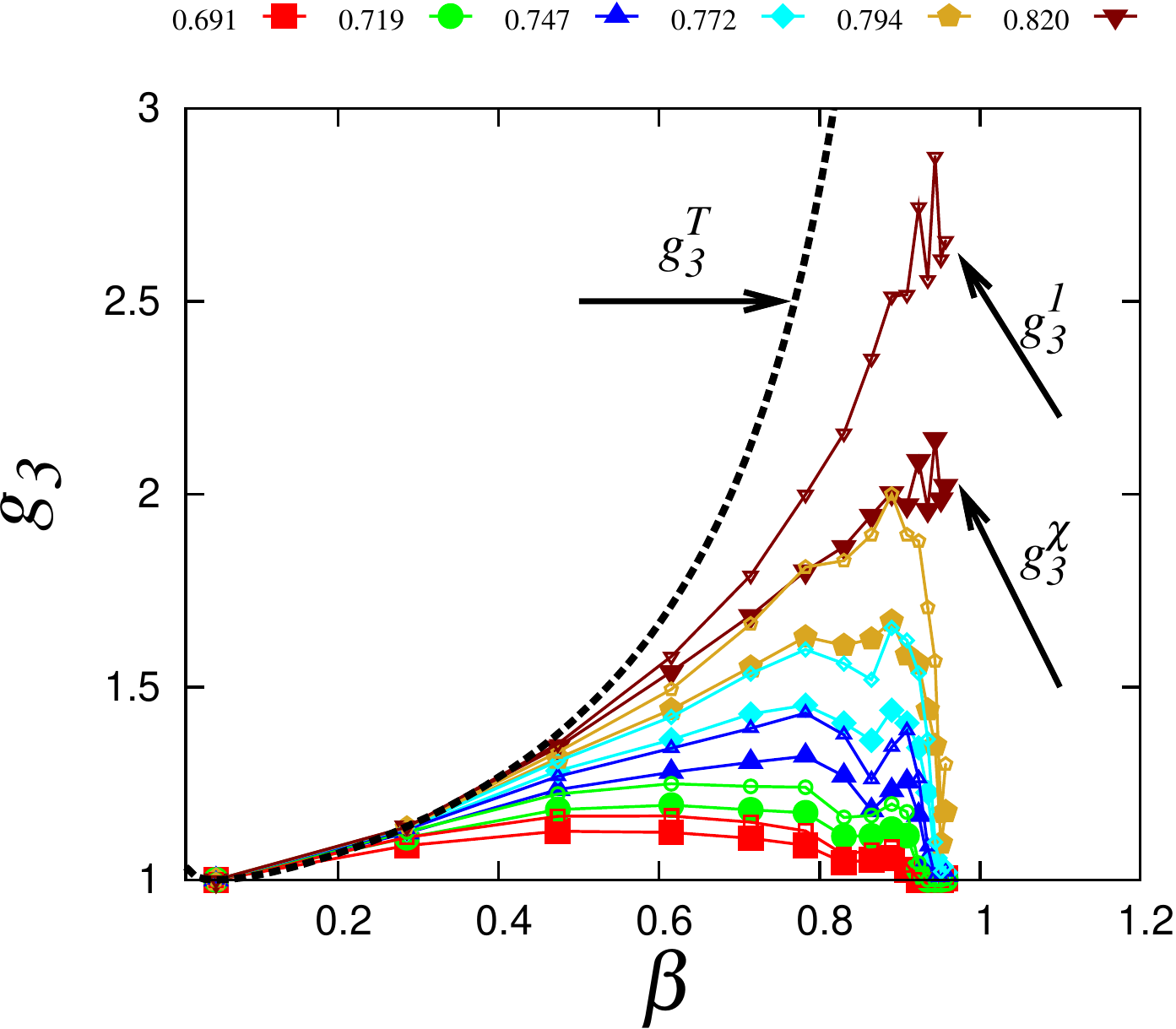}\label{g3F_pvskappa_noscale_mu0p0}}
\caption{(a) Ratio of linear compacities of particles B to A, $\chi $, versus the radius ratio $r_A/r_B$ measured from the 
DEM simulations (symbols) and the corresponding line is the analytical fit to the data using Eq.\ (\ref{eq:csAcsB1}). 
The dashed line is constant linear compacity, i.e., $\chi=1$.
(b) $g_3^\chi$ for fabric calculated using Eqs.\ (\ref{eq:g_3defnbi2}) with $\chi$ estimated using (\ref{eq:csAcsB1}) (solid symbols) 
and $g_3^1$ measured from constant linear compacity assumption $\chi=1$ (small open symbols), plotted 
against the number fraction $\beta= N_B^T/\left(N_A^T + N_B^T \right)$ excluding the rattlers. 
Different colors represent the volume fraction $\nu$ as shown in the legend. 
The dashed line is $g_3^T$ and considers also rattlers.  
}
\label{csinfoplot}
\end{figure}

We are interested in the relation of isotropic fabric with the system's mean packing properties 
e.g. volume fraction, average coordination number. 
An expression for isotropic fabric in terms on mean packing properties, similar to that given in Refs.\ \cite{shaebani2012influence,goncu2010constitutive, kumar2014effects}, as:
\begin{equation}
\label{eq:isofabriceq1}
\Fv  = g_3 \nu^* C^*,
\end{equation}
where $\nu^*$ and $C^*$ are the volume fraction and mean corrected coordination number of the system respectively, excluding the rattlers, as defined in section\ \ref{sec:stable}, 
and $g_3$ is related to the moments of size distribution
\footnote[3]{Refs.\ \cite{shaebani2012influence,goncu2010constitutive, kumar2014effects} used the corrected coordination 
number $C = M_4/N$ in Eq.\ (\ref{eq:isofabriceq1}), while is different than $C^* = M_4/N_4$ used in this study. 
It was checked that using $C$ in Eq.\ (\ref{eq:isofabriceq1}) instead of $C^*$ worsen the comparison with $\Fv$.}. 
For a bidisperse size distribution, $g_3$ excluding the rattlers is given as (see appendix \ref{App:AppendixB} for derivation): 
\begin{equation}
\label{eq:g_3defnbi2}
% g_3 = \frac{ r_A^3 c_s(r_A) \Omega_A^{-1}f_A  + r_B^3 c_s(r_B)\Omega_B^{-1}f_B }{\left( c_s(r_A)\Omega_A^{-1} f_A +  c_s(r_B) \Omega_B^{-1} f_B \right) \left( r_A^3f_A + r_B^3 f_B \right)  } 
g_3 := g_3^\chi = \frac{1}{ \langle r^3 \rangle  }\frac{ r_A^3 \Omega_A^{-1} f_A  + r_B^3  \Omega_B^{-1}f_B \chi}{\Omega_A^{-1} f_A +   \Omega_B^{-1} f_B \chi} 
= \frac{\langle r^3 \rangle_g}{\langle r^3 \rangle}  ,
\end{equation}
where $r_A$ and $r_B$ are the radii of A and B with number fraction $f_A$ and $f_B$ respectively. 
$\Omega(r) = 2\pi \left[1 - \sqrt{ 1 - {\langle r \rangle}^2/\left({r + \langle r \rangle}\right)^2} \right]$ 
is the space angle covered on a particle of radius $r$ by neighboring particles of radius $\langle r \rangle$.
$\chi$ is the ratio of the linear compacity 
(or the total fraction of shielded surface which is proportional to product of space angle and number of non-rattler contacts, defined in appendix \ref{App:AppendixB}) 
of B to A.
The unknown in the functional form of Eq.\ (\ref{eq:g_3defnbi2}) is $\chi$, the ratio of the linear compacities of B to A (see appendix \ref{App:AppendixB}). 
Fig.\ \ref{gagbplot} shows the evolution of the measured ratio $\chi $ with the size ratio $r_A/r_B$, as extracted from simulations for different volume fractions. 
$\chi $ increases with increasing size ratio and is dependent on the volume fraction $\nu$, in agreement with Ref.\ \cite{shaebani2012influence, goncu2010constitutive}. 
For fitting the data in Fig.\ \ref{gagbplot}, we propose
\begin{equation}
\label{eq:csAcsB1}
\chi = 1 + \frac{1}{2}\left(\frac{1}{\Lambda} -1 \right) \left[ 1 + \mathrm{erf}   \left( a(\nu) \left( \frac{r_A}{r_B} - b(\nu) \right) \right) \right],
\end{equation}
where $\mathrm{erf}(...)$ is the error function and $a(\nu) = 0.25 \left( \nu/\nu_c \right)^4$ and $b(\nu) = 4.5 + a(\nu)$ are empirical relations. 
$1/\Lambda\approx1/0.4$ is the maximum compacity ratio $\mathrm{max}(\chi) = 1/\Lambda$ and is reached near the jamming transition (see appendix \ref{App:AppendixB} for the bounds of linear compacity).

\begin{figure}[!ht]
\centering
\subfigure[]{\includegraphics[width=0.4\textwidth, angle=0]{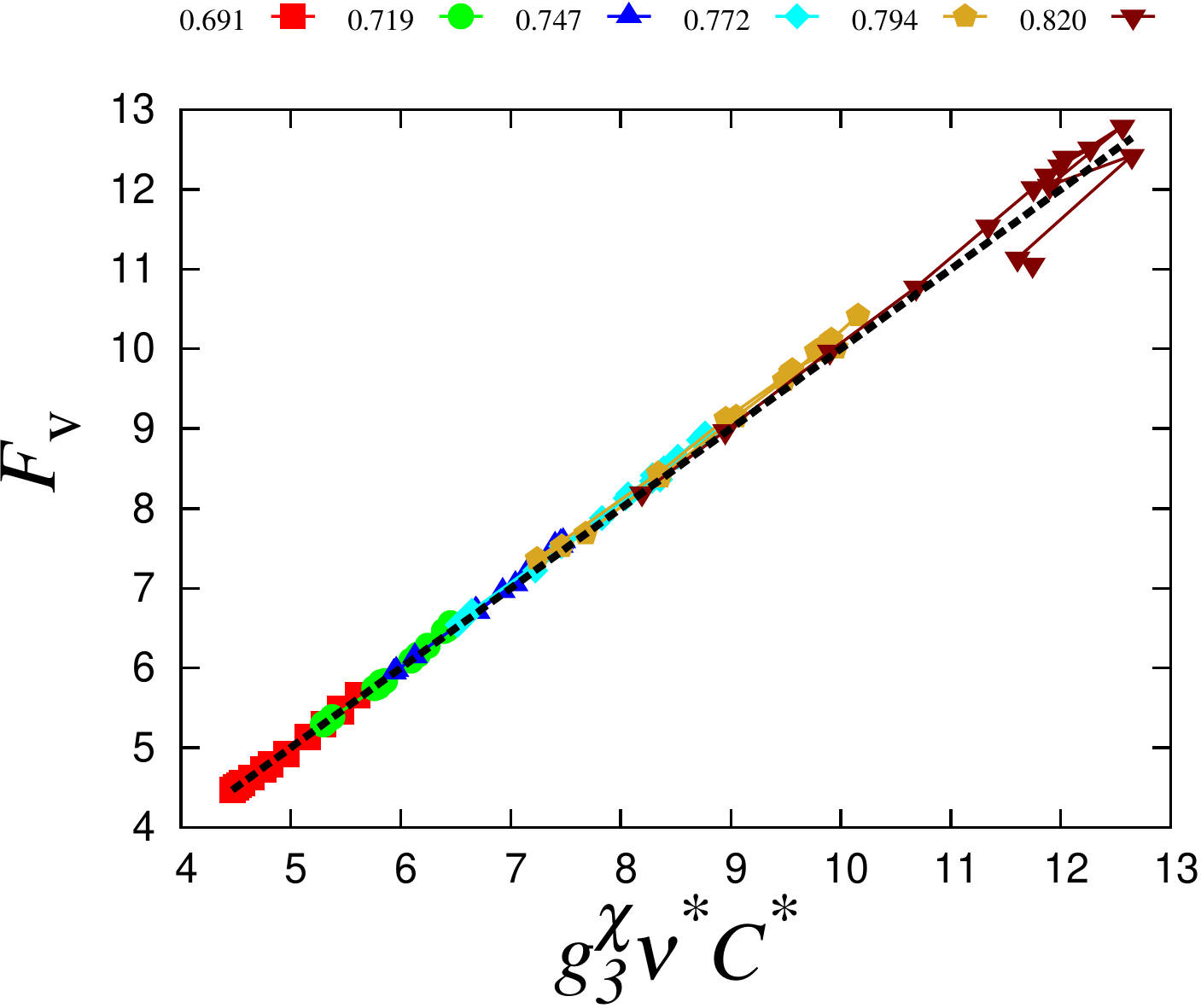}\label{fvg3nuCbetter_pvskappa_noscale_mu0p0}}
\subfigure[]{\includegraphics[width=0.4\textwidth, angle=0]{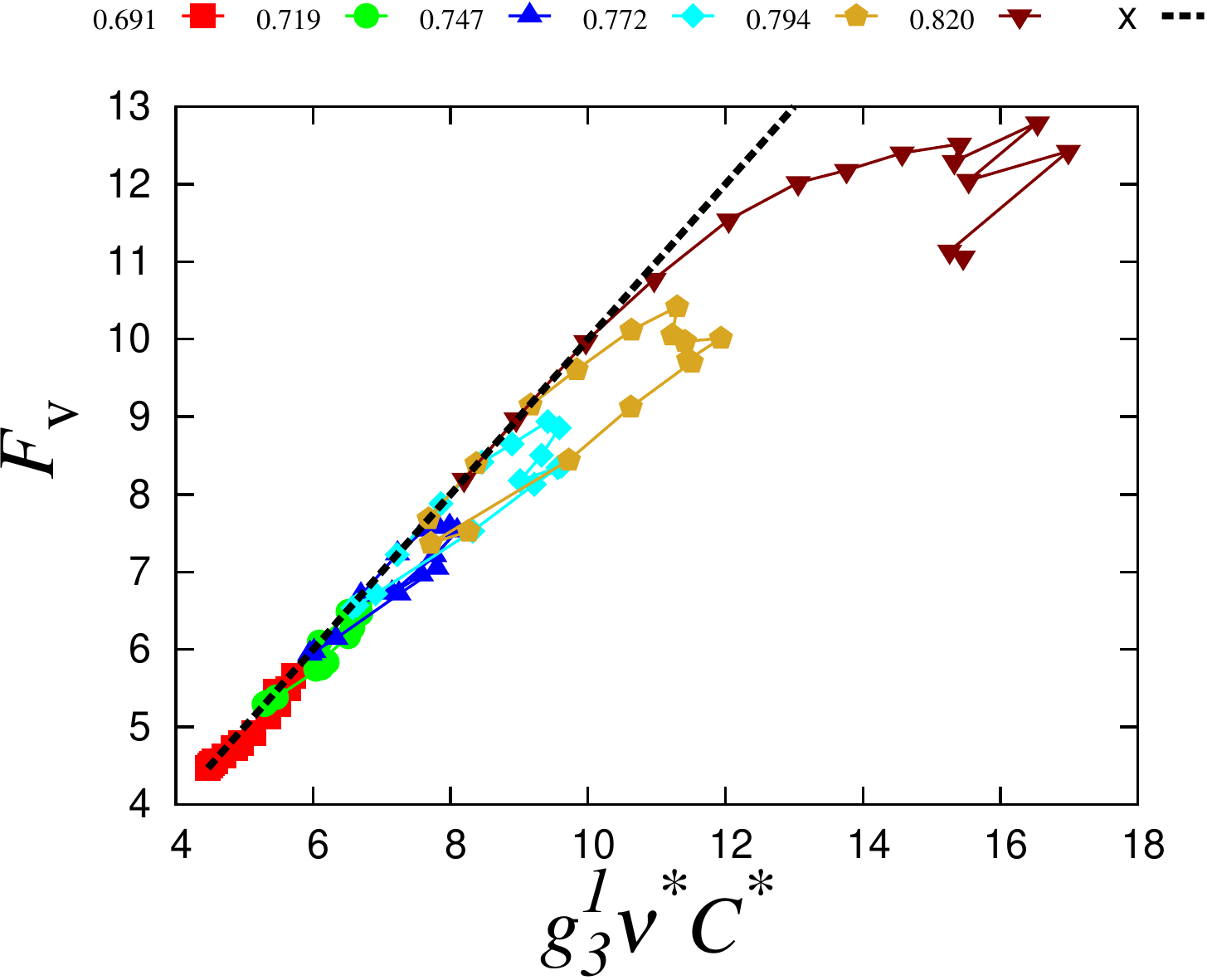}\label{fvg3nuC_pvskappa_noscale_mu0p0}}\\
\subfigure[]{\includegraphics[width=0.4\textwidth, angle=0]{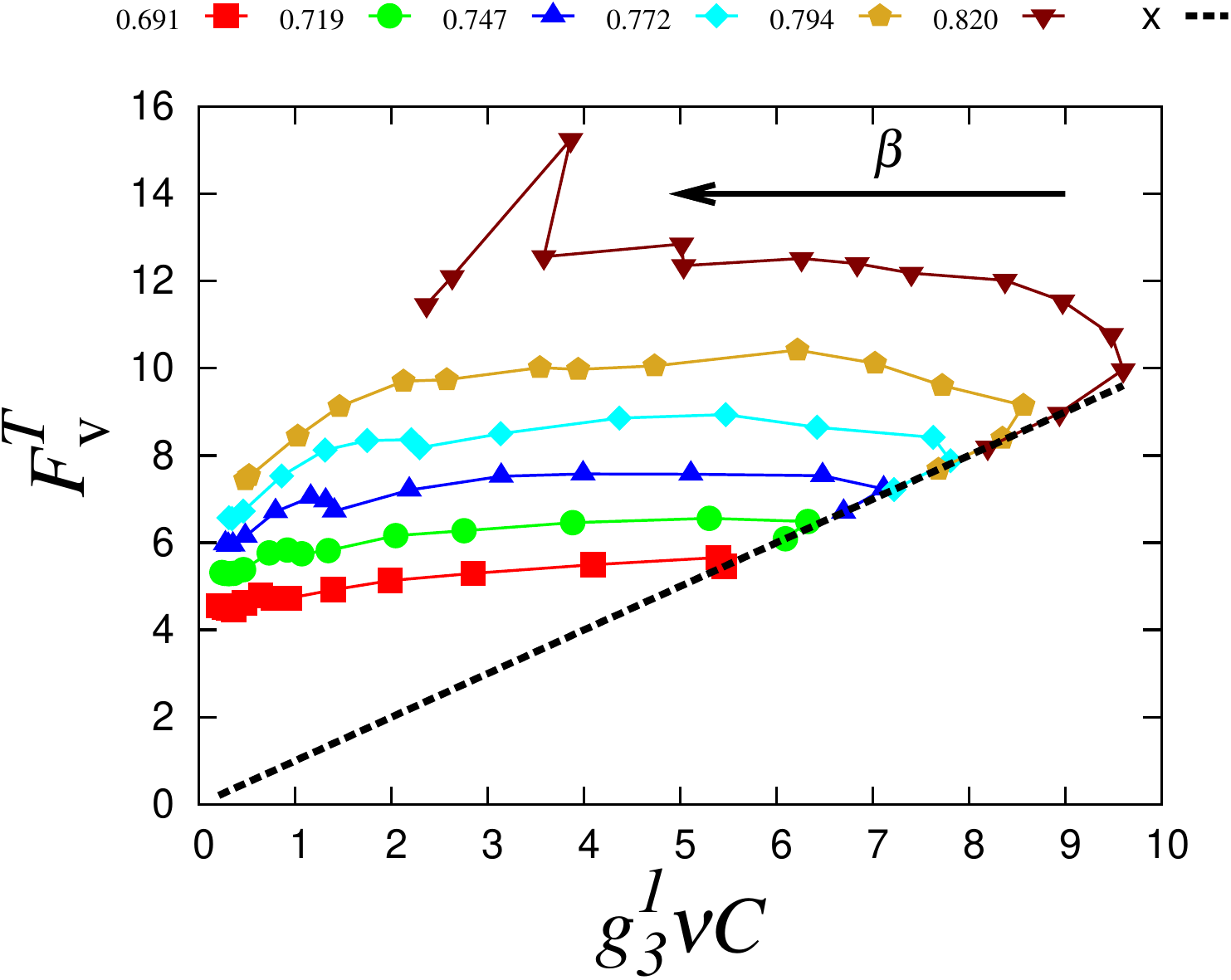}\label{fvTg3nuC_pvskappa_noscale_mu0p0}}
\caption{Isotropic fabric $\Fv$ plotted against Eq.\ (\ref{eq:isofabriceq1}) with $g_3^\chi$ calculated using Eq.\ (\ref{eq:g_3defnbi2}) with 
(a) non-constant linear compacity, with $\chi$ computed based on volume fraction and radius ratio using Eq.\ (\ref{eq:csAcsB1}) and (b) constant linear compacity, with $\chi=1$. 
The dashed black line has slope 1.
Different colors represent the volume fraction $\nu$ as shown in the legend. 
(c) The total isotropic fabric including the rattlers $\Fv^T$ compared to the relation presented in Ref.\ \cite{goncu2010constitutive}, and the arrow indicates increasing $\beta$. 
}
\label{fabric_plot}
\end{figure}

The measured $g_3^\chi$ using Eq.\ (\ref{eq:g_3defnbi2}) is plotted in Fig.\ \ref{g3F_pvskappa_noscale_mu0p0}. 
$g_3\chi$ is greater than 1 for all volume fraction, increasing with $\beta$ to a maximum followed by a decrease, similar as $N_B/N_A$ shown in Fig.\ \ref{NB_pvskappa_noscale_mu0p0}. 
$g_3$ measured assuming constant linear compacity \cite{goncu2010constitutive, shaebani2012influence, madadi2004fabric}, 
i.e., $g_3^1$ with $\chi=1$ is also plotted in Fig.\ \ref{g3F_pvskappa_noscale_mu0p0} shows similar trend as $g_3^\chi$ and is higher.
Asymptotic analysis for $\beta \to1$, considering all the particles present in the system and with constant linear compacity tells us that 
$g_3^T$ diverges as $(1-\beta)^{-2/3}$, in agreement with the behavior shown in Fig.\ \ref{g3F_pvskappa_noscale_mu0p0}.

Fig.\ \ref{fvg3nuCbetter_pvskappa_noscale_mu0p0} shows the relation of $\Fv$ with the mean 
packing properties via $g_3$ using Eq.\ (\ref{eq:isofabriceq1}), and a good agreement is observed with small errors up to 5\% for the highest densities and $\beta \to1$. 
Modification of the linear compacity helps to improve the relation of $\Fv$ with the mean packing properties, 
while a constant linear compacity assumption works only for low $\nu$ and up to intermediate $\beta$, as seen in Fig.\ \ref{fvg3nuC_pvskappa_noscale_mu0p0}. 
For dense states and high $\beta$, the constant linear compacity assumption can lead to up to 45\% error in the prediction of $\Fv$. 
This is due to the fact that very small B particles are present in large numbers, 
participating in the contact network, so that the assumption of the linear compacity independent with particle radii, is not valid anymore. 
The better understanding of the linear compacity that can account for large numbers of very small particles in highly polydisperse systems is subject of a future study. 
Finally, we attribute the poor agreement between $\Fv^T$ and $g_3^1 \nu C$ as used in Ref.\ \cite{goncu2010constitutive, imole2013hydrostatic, shaebani2012influence}, 
as shown in Fig.\ \ref{fvTg3nuC_pvskappa_noscale_mu0p0}, to the 
fact that homogeneous size distributions were used, not excluding one of the species strongly from the contact network.

\subsection{Pressure} 
\label{sec:press}

Besides the fabric, one can determine the static stress tensor as 
\begin{equation}
\label{eq:stresseq}
\pmb{\sigma}=\frac{1}{V}\sum_{c=1}^{M}\mathbf{l}^{c}\otimes\mathbf{f}^{c}, 
\end{equation}
which is the sum of the dyadic products between the branch vector $\mathbf{l}^{c} = l_c \mathbf{n}^{c}$ and 
the contact force $\mathbf{f}^{c} = k \delta_c \mathbf{n}^{c}$ over all the contacts (an exemplary two particle contact is shown in Fig.\ \ref{overlap}) 
in the system volume $V$, where the contribution of the kinetic energy has been neglected \cite{luding2005anisotropy, imole2013hydrostatic}. 
The isotropic component of the stress is the pressure $P = \mathrm{tr}(\pmb{\sigma})/3$. 
The non-dimensional pressure is defined as:
\begin{equation}
\label{eq:press}
p_n =  \frac{2 \langle r_A \rangle}{3 k} \mathrm{tr}(\pmb{\sigma}) = \frac{2 \langle r_A \rangle}{k} P , 
\end{equation}
scaled by constant ${2 \langle r_A \rangle}/{k}$. In bi-axial experiments, the pressure $ P$ can be measured, and hence $p_n$ can be estimated. 
Note that $p_n$ is used in the following to avoid dimensions of pressure. 
The size sensitive non-dimensional pressure is defined as \cite{goncu2010constitutive, imole2013hydrostatic, kumar2014effects}:
\begin{equation}
\label{eq:pressnon}
p = \frac{2 \langle r \rangle}{3 k} \mathrm{tr}(\pmb{\sigma}) = \frac{2 \langle r \rangle}{k} P.
\end{equation}
Note that in this work, the pressure calculated considering $M_4$ non-rattler contacts is very 
close to the one from $M$ contacts, as $M-M_4$ are temporary, very weak (rattler) contacts that barely contribute to the average stress.

\begin{figure}[!ht]
\centering
\subfigure[]{\includegraphics[width=0.4\textwidth, angle=0]{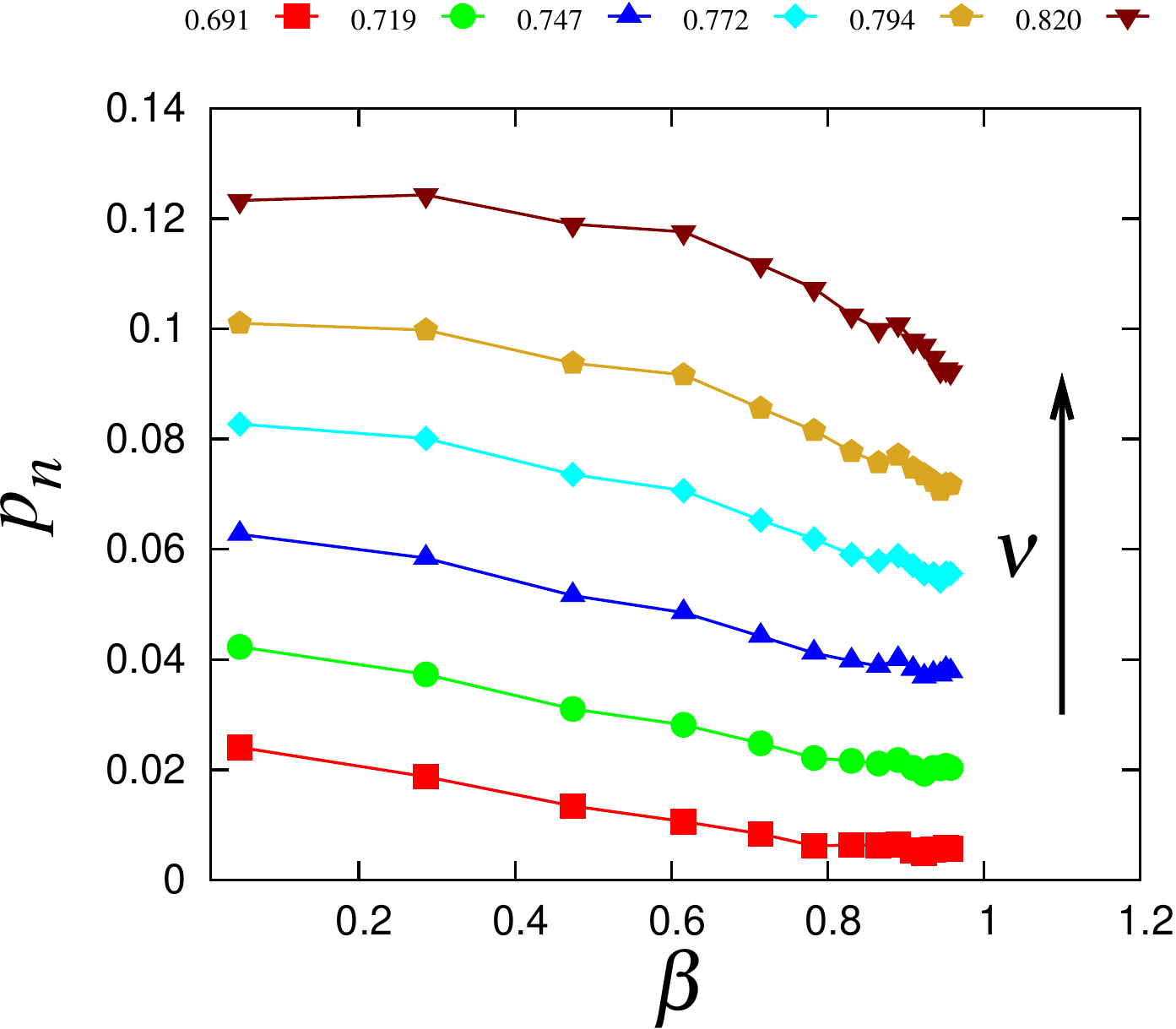}\label{pres_pvskappa_noscale_mu0p0}}
\subfigure[]{\includegraphics[width=0.4\textwidth, angle=0]{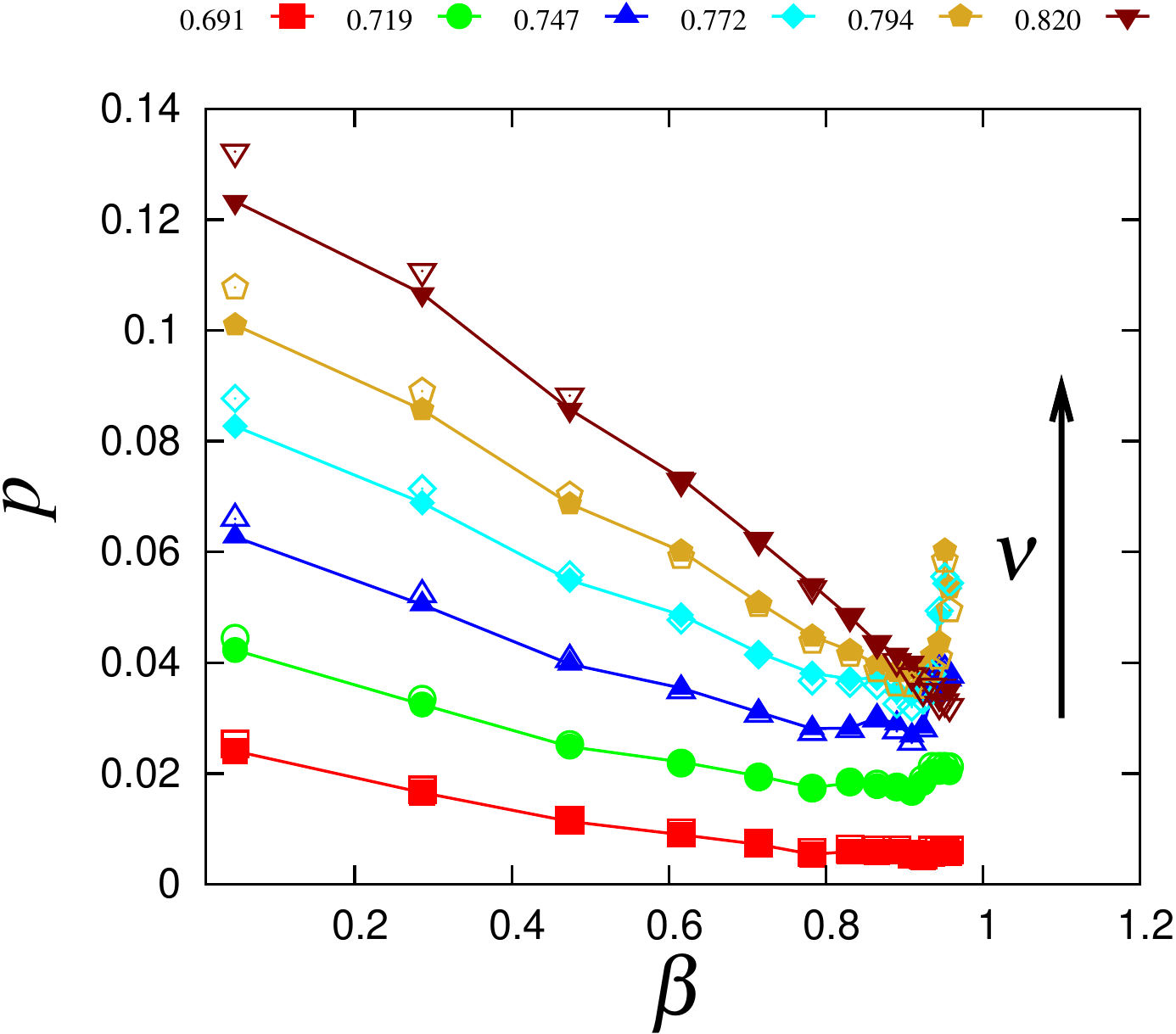}\label{nonpres_pvskappa_noscale_mu0p0}}\\
\subfigure[]{\includegraphics[width=0.4\textwidth, angle=0]{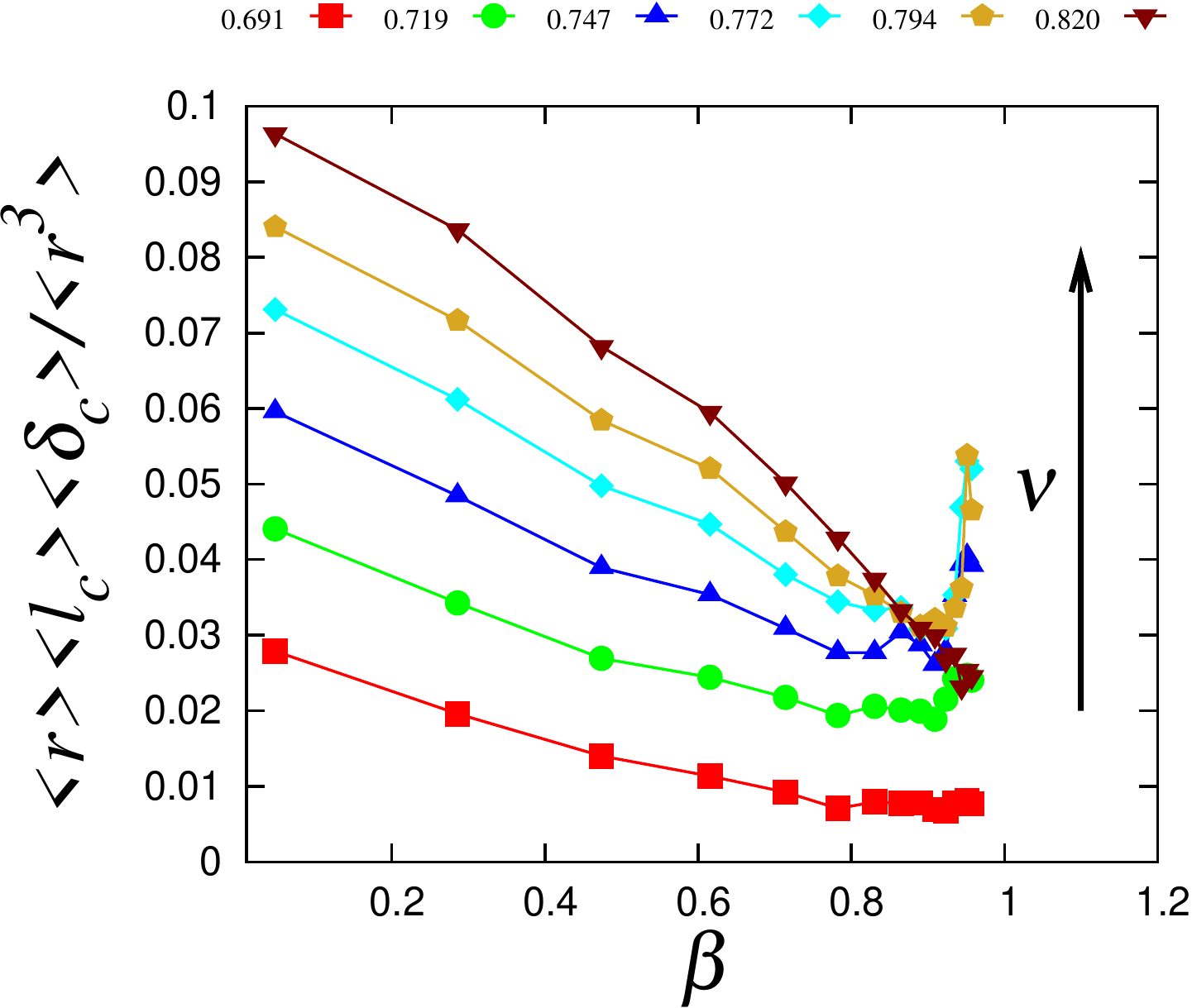}\label{meandelRvskappa_noscale_mu0p0}}
\subfigure[]{\includegraphics[width=0.4\textwidth, angle=0]{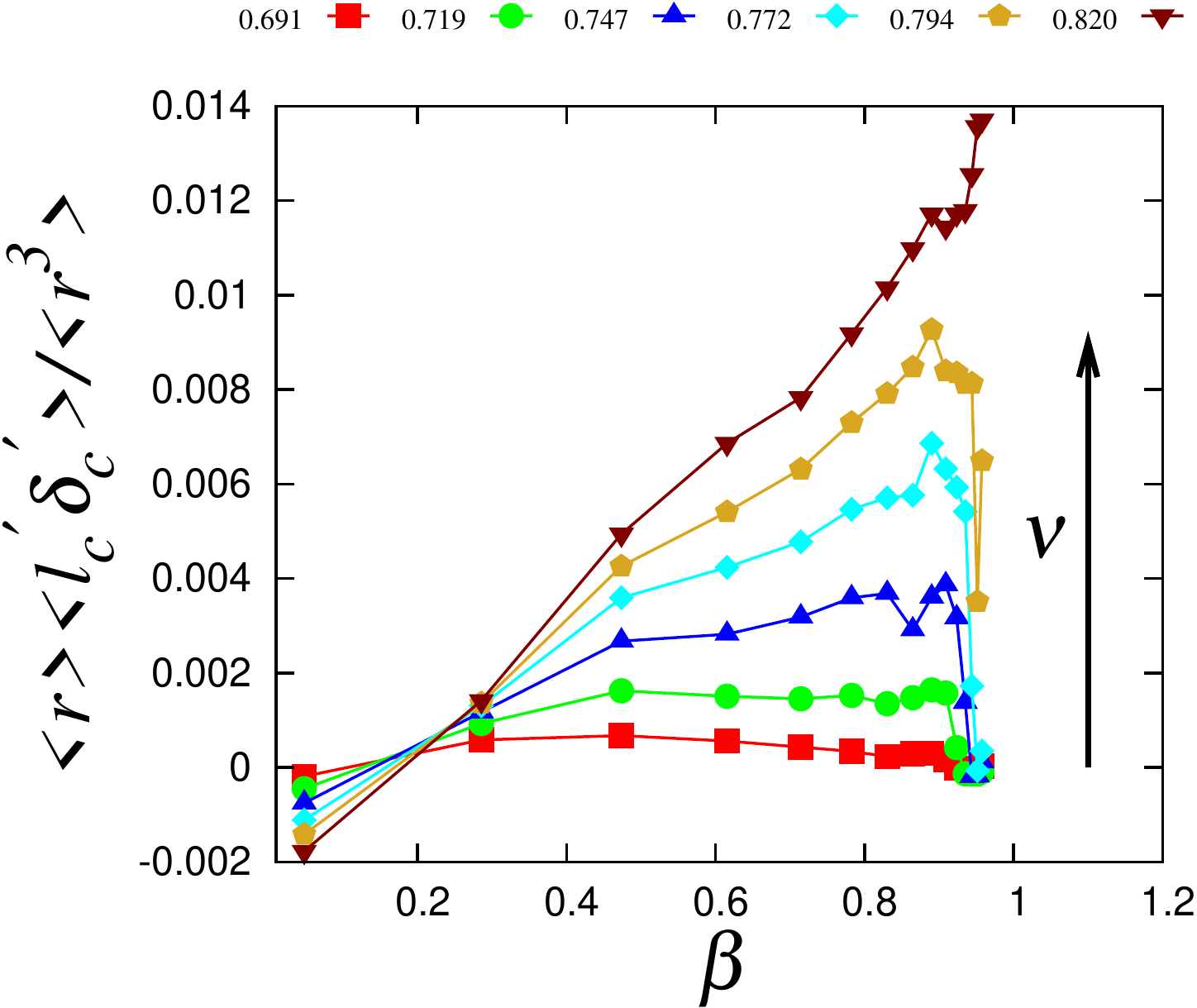}\label{meandelRflucvskappa_noscale_mu0p0}}
\caption{(a) Non-dimensional pressure (scaled by constant ${2 \langle r_A \rangle}/{k}$) $p_n$, 
(b) non-dimensional pressure $p$ scaled by ${2 \langle r \rangle}/{k}$ (solid symbols) and prediction using Eq.\ (\ref{eq:pstarmain}) (open symbols)
(c) product of mean radius ${\langle r \rangle}$, branch vector 
${\langle l_c \rangle}$ and particle overlap $\langle \delta_c \rangle$ scaled with the third moment ${\langle r^3 \rangle}$
and (d) product of mean radius ${\langle r \rangle}$ and the corresponding fluctuation term ${\langle l_c' \delta_c' \rangle}$ 
scaled with the third moment ${\langle r^3 \rangle}$, calculated using  Eq.\ (\ref{eq:stresseqmodify}), 
plotted against the number fraction $\beta= N_B^T/\left(N_A^T + N_B^T \right)$. 
Different colors represent the volume fraction $\nu$ as shown in the legend and arrows indicate increasing $\nu$.}
\label{pres2}
\end{figure}

Fig.\ \ref{pres_pvskappa_noscale_mu0p0} shows the evolution of the non-dimensional pressure $p_n$ with $\beta$ for six different volume fractions $\nu$ as shown in the legend.
For a given $\beta$, $p_n$ increases with $\nu$ as the particles are more compressed \cite{imole2013hydrostatic}. 
On the other hand, for a given density, $p_n$ systematically decreases with $\beta$. 
This observation is linked to the behavior of the corrected volume fractions $\nu^*$ (section\ \ref{sec:volfrac}), 
as seen in Fig.\ \ref{nu_pvskappa_noscale_mu0p0}, which also decrease systematically with $\beta$.
For moderate and large $\beta$, most of the contribution to pressure comes from particles A, while the contribution of B is negligible, as both overlap and radius are small 
and hence is their stress (proportional to both) 
becoming negligible for large $\beta$ (data not shown). 
Fig.\ \ref{nonpres_pvskappa_noscale_mu0p0} shows the evolution of the non-dimensional pressure $p$ using Eq.\ (\ref{eq:pressnon}) \cite{imole2013hydrostatic}. 
For the smallest $\beta$, where the system is composed of only A particles $p_n$ and $p$ are the same. 
For fixed $\nu$, $p$ decreases much faster than $p_n$ with $\beta$, since $p$ is a product of $p_n$ and $\langle r \rangle$, both decreasing with $\beta$. 
In this case, the behavior of $p$ differs from $\Fv$, see Fig.\ \ref{FvT_pvskappa_noscale_mu0p0}. 
Another important observation is that the behavior of $p$ has a similar trend as $C^*$ in Fig.\ \ref{cstar_pvskappa_noscale_mu0p0}. 
For a given mixture (fixed $\beta$), $\Fv$ increases with volume fraction $\nu$, as do both coordination number and pressure,

We try to better understand the evolution of the non-dimensional pressure by looking at the individual components that contribute to Eq.\ (\ref{eq:stresseq}). 
Due to the linear contact model used without any tangential component, the force and the branch vectors are parallel for all contacts (see Fig.\ \ref{overlap}). 
Hence, Eq.\ (\ref{eq:pressnon}) becomes \cite{goncu2010constitutive,shaebani2012influence}:
\begin{eqnarray}
\label{eq:stresseqmodify}
 p &=& \frac{2 \langle r \rangle}{3 k} \mathrm{tr}(\pmb{\sigma}) \approx \frac{2 \langle r \rangle}{3k}\frac{1}{V}\mathrm{tr}\left(\sum_{c=1}^{M_4}\mathbf{l}^{c}\otimes\mathbf{f}^{c}\right)   =  \frac{{2 \langle r \rangle}}{3V} \sum_{c=1}^{M_4} \mathrm{tr}\left(l_c\mathbf{n}^{c}\otimes\delta_c \mathbf{n}^{c}\right) \nonumber \\
   &=& \frac{{2 \langle r \rangle}}{3V} \sum_{c=1}^{M_4} \left[\langle l_c \rangle  + l_c' \right] \left[\langle \delta_c  \rangle + \delta_c' \right] \underbrace{\mathrm{tr}\left( \mathbf{n}^{c}\otimes \mathbf{n}^{c}\right)}_{\text{=1}}  = \frac{{2 \langle r \rangle}}{3V} M_4 \left[ {\langle l_c \rangle\langle \delta_c \rangle}   + {\langle l_c' \delta_c' \rangle}\right] \nonumber \\
   &=& \frac{{2 \langle r \rangle}}{3V} M_4 \left[ {\langle l_c \rangle\langle \delta_c \rangle}   + {\langle l_c' \delta_c' \rangle}\right]  \left[\frac{\nu^*}{(4 \pi/3) N_4 \langle r^3 \rangle/V}\right]  \nonumber \\ 
   &=&  \frac{C^*\nu^*} {2 \pi} \frac{\langle r \rangle}{\langle r^3 \rangle} \left[ {\langle l_c \rangle\langle \delta_c \rangle}   + {\langle l_c' \delta_c' \rangle} \right] ,
\end{eqnarray}
where the rattlers offer no contribution and the prime $'$ represents the fluctuations with respect to the average. 
The first term in Eq.\ (\ref{eq:stresseqmodify}) 
considers the average overlap $\langle \delta_c  \rangle$ and the 
average branch vector $\langle l_c  \rangle $. 
The second term is the contribution due to the correlated fluctuations in branch vector and overlap. 

Fig.\ \ref{meandelRvskappa_noscale_mu0p0} shows the evolution of the first term ${\langle r \rangle}{\langle l_c \rangle\langle \delta_c \rangle} /{\langle r^3 \rangle}$ in Eq.\ (\ref{eq:stresseqmodify}). 
For a given $\beta$ with increasing $\nu$, the average overlap $\langle \delta_c \rangle$ increases \cite{kumar2014effects}, 
$\langle l_c \rangle $ slightly decreases and ${\langle r \rangle}/{\langle r^3 \rangle}$ increases. 
Therefore ${\langle r \rangle}{\langle l_c \rangle\langle \delta_c \rangle} /{\langle r^3 \rangle}$ increases with $\nu$ 
(for fixed $\beta$) and decreases with $\beta$ (for fixed $\nu$), as seen in Fig.\ \ref{meandelRvskappa_noscale_mu0p0}, 
except for very high $\nu$'s and $\beta$'s. 
% up to $\beta\approx0.9$
% For a fixed $\nu$, both $\langle \delta \rangle$ and ${\langle l_c \rangle}$ decrease with $\beta$ while ${\langle r \rangle}/{\langle r^3 \rangle}$ slightly increases till $\beta\approx0.9$. 
% Therefore ${\langle r \rangle}{\langle l_c \rangle\langle \delta \rangle} /{\langle r^3 \rangle}$ decreases with $\beta$ up to $\beta\approx0.9$ value, as seen in Fig.\ \ref{meandelRvskappa_noscale_mu0p0}. 
%
% For the densest case $\nu = 0.82$ and $\beta \to1$, $N_B$ with $\beta$ will decrease further after maxima (section\ \ref{sec:rattlers}) 
% and hence ${\langle r \rangle}{\langle l_c \rangle\langle \delta \rangle} /{\langle r^3 \rangle}$ 
% will increase after minimum ($\beta>0.9$), similar like loose densities, as seen in Fig.\ \ref{meandelRvskappa_noscale_mu0p0}. 
% Due to the computational limitations, this can not be checked. 
%
% For a fixed $\nu$, both $\langle \delta \rangle$ and ${\langle l_c \rangle}$ decreases with $\beta$ while ${\langle r \rangle}/{\langle r^3 \rangle}$ slightly increases. 
% , since $r_B$ decreases with $\beta$, 
% leading to a decrease in ${\langle r \rangle}{\langle l_c \rangle\langle \delta \rangle} /{\langle r^3 \rangle}$ with $\beta$. 
The fluctuation factor ${\langle r \rangle}{\langle l_c' \delta' \rangle}/{\langle r^3 \rangle}$ increases with both $\nu$ and $\beta$, as seen in Fig.\ \ref{meandelRflucvskappa_noscale_mu0p0}. 
The common term $C^*$ has a similar trend as of ${\langle r \rangle}{\langle l_c \rangle\langle \delta_c \rangle} /{\langle r^3 \rangle}$, as seen in Fig.\ \ref{cstar_pvskappa_noscale_mu0p0}.
Comparing Fig.\ \ref{meandelRvskappa_noscale_mu0p0} and Fig.\ \ref{meandelRflucvskappa_noscale_mu0p0}, we conclude that the decrease in $p$ with $\beta$ observed in Fig.\ \ref{pres_pvskappa_noscale_mu0p0} 
is mainly associated with the decrease of both ${\langle r \rangle}{\langle l_c \rangle\langle \delta_c \rangle} /{\langle r^3 \rangle}$ and $C^*$, while the fluctuation term is very small.

\begin{figure}[!ht]
\centering
\subfigure[]{\includegraphics[width=0.4\textwidth, angle=0]{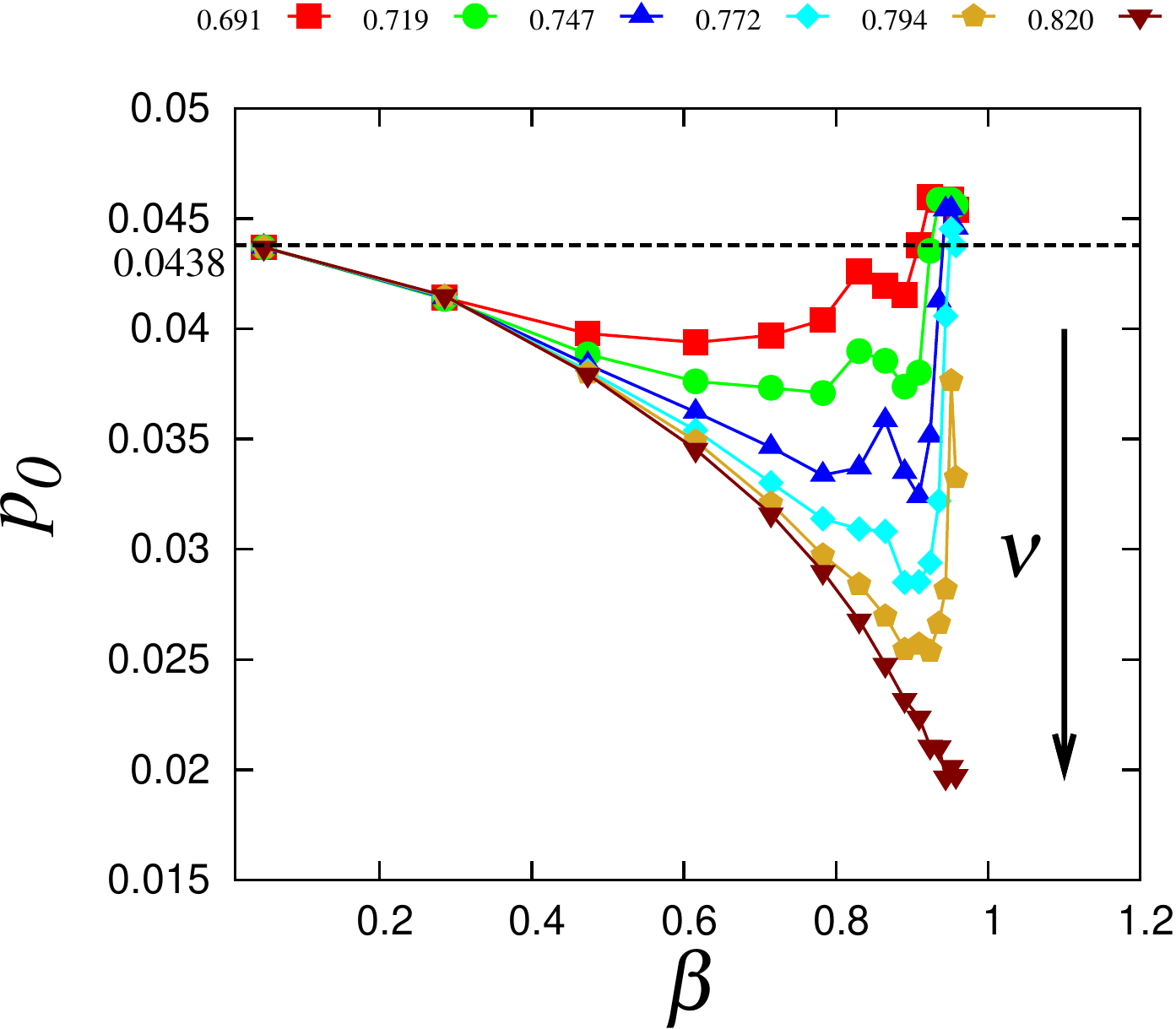}\label{p0meas_vskappa_noscale_mu0p0}}
\subfigure[]{\includegraphics[width=0.4\textwidth, angle=0]{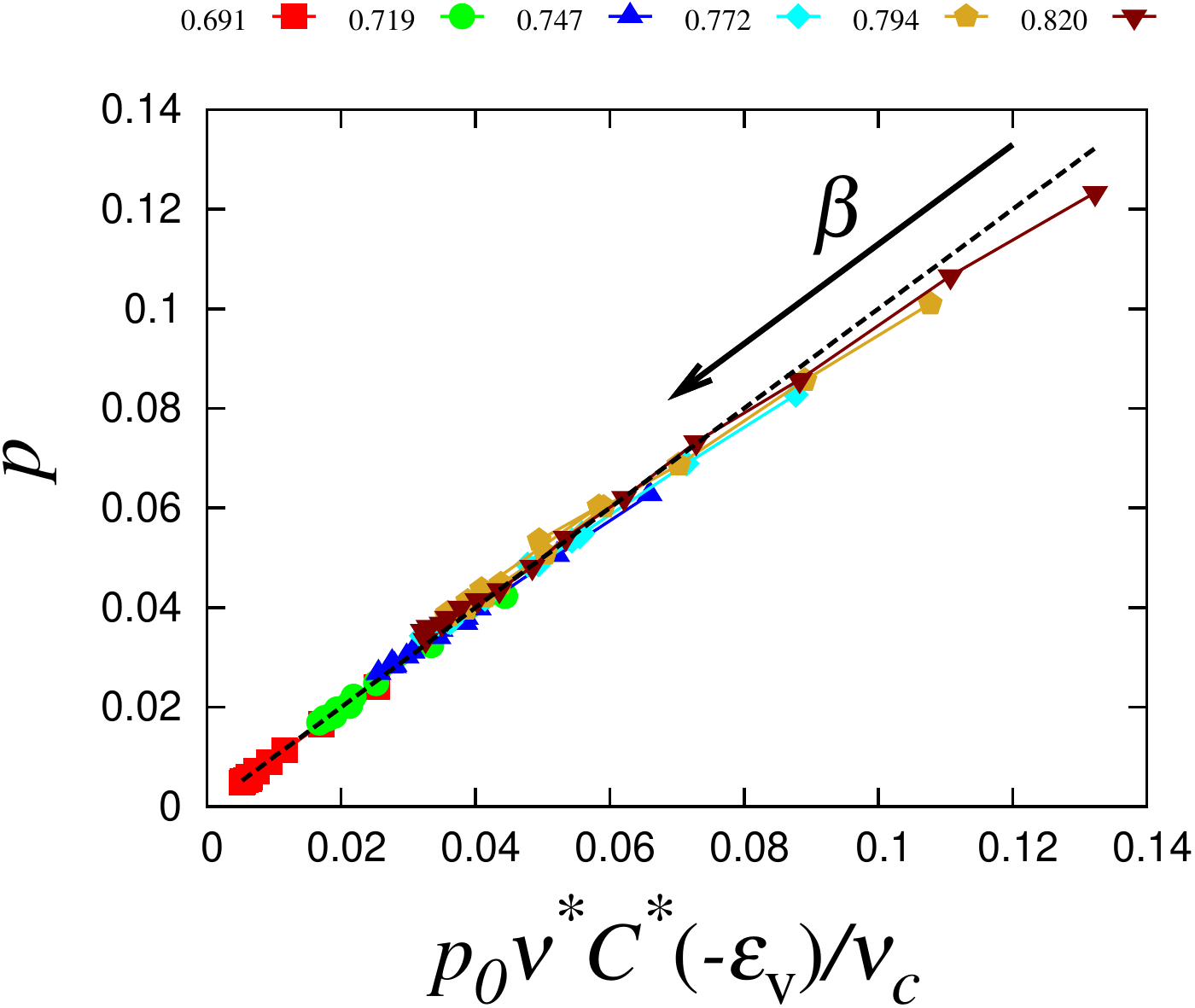}\label{nonpresest2_vskappa_noscale}}
\caption{(a) Measured $p_0$ plotted against the number fraction $\beta= N_B^T/\left(N_A^T + N_B^T \right)$. 
The dashed line represent the constant value of $p_0$ for the monodisperse case.
(b) Non-dimensional pressure $p$ plotted against Eq.\ (\ref{eq:pstarmain}) without the second term. 
The dashed line is the linear line with slope 1 and the arrow indicates increasing $\nu$. 
Different colors represent the volume fraction $\nu$ as shown in the legend. The arrow in (b) represents increasing $\beta$.}
\label{pres}
\end{figure}

The non-dimensional pressure $p$ can be written in the same form as given in Ref.\ \cite{goncu2010constitutive}:
\begin{equation}
p=p_0 \frac{ \nu^* C^*  }{\nu_c} (-\varepsilon_\mathrm{v})  \left [ 1-\gamma_p (-\varepsilon_\mathrm{v}) \right ]
\label{eq:pstarmain}
\end{equation}
where the quantity 
$(-\varepsilon_\mathrm{v})$ is the true or logarithmic volume change of the system proportional to the 
ratio of average overlap to the mean radius $\langle \delta_c  \rangle/\langle r  \rangle$ (see appendix \ref{App:AppendixC}).
$p_0\approx 0.043$ for uniform size distributions \cite{imole2013hydrostatic, kumar2014macroscopic, kumar2014effects, goncu2010constitutive}; 
however, $p_0$ is not constant for an arbitrary wide bidisperse size distributions, as the case in this work, as shown in Fig.\ \ref{p0meas_vskappa_noscale_mu0p0} 
(see appendix \ref{App:AppendixC} for more details about calculating $p_0$). 
Fig.\ \ref{nonpres_pvskappa_noscale_mu0p0} shows the prediction for the non-dimensional pressure $p$ using Eq.\ (\ref{eq:pstarmain}) without the second small term. 
Since, $\gamma_p$ is positive, the pressure is slightly over predicted, mainly in the dense regime and for the monodisperse case. 
Finally, Fig.\ \ref{nonpresest2_vskappa_noscale} shows a perfect prediction of the measured pressure $p$ when compared with Eq.\ (\ref{eq:pstarmain}), again without the non-linear term. 
Only for highly dense cases, a maximum error of few percent can be seen, which could be avoided by including the non-linearity with $\gamma_p$.

As final stage, we want to study how the stiffness of the granular assembly varies with the contribution of the fines.

\section{Bulk modulus} 
\label{sec:bulk}

For each granular mixture, examples are displayed in Fig.\ \ref{c3d}, we calculate the bulk modulus by first relaxing (see section\ \ref{sec:prepproc}) and then 
applying an incremental pure volumetric perturbation of small amplitude to the sample ($\mathrm{d}\nu \approx 0.00015$) \cite{kumar2014macroscopic}. 
The bulk modulus is then the ratio between the measured change in pressure and the applied strain $d\nu/\nu$, small enough to prevent irreversible contact rearrangements \cite{kumar2014macroscopic}: 
\begin{equation}
\label{eq:Bmeasure1}
K' = \nu \frac{\mathrm{d}P}{\mathrm{d}\nu} .
\end{equation}
The non-dimensional bulk modulus is thus \cite{imole2013hydrostatic, kumar2014macroscopic, goncu2010constitutive}:
\begin{equation}
\label{eq:Bmeasure}
K = \frac{{2 \langle r_A \rangle} }{k}  \nu \frac{\mathrm{d}P}{\mathrm{d}\nu} ,
\end{equation}
where ${2 \langle r_A \rangle}$ is the average particle radius of A which provides the backbone to the granular assembly and $k$ is the particle stiffness.
% Unless not mentioned specifically, we will denote non-dimensional bulk modulus as $K$ instead of $K_n$.
Just like the pressure $ P$, the bulk modulus $K'$ can be estimated in the bi-axial experiments, and hence $K$ can be calculated. 

\begin{figure}[!ht]
\centering
\subfigure[]{\includegraphics[width=0.4\textwidth, angle=0]{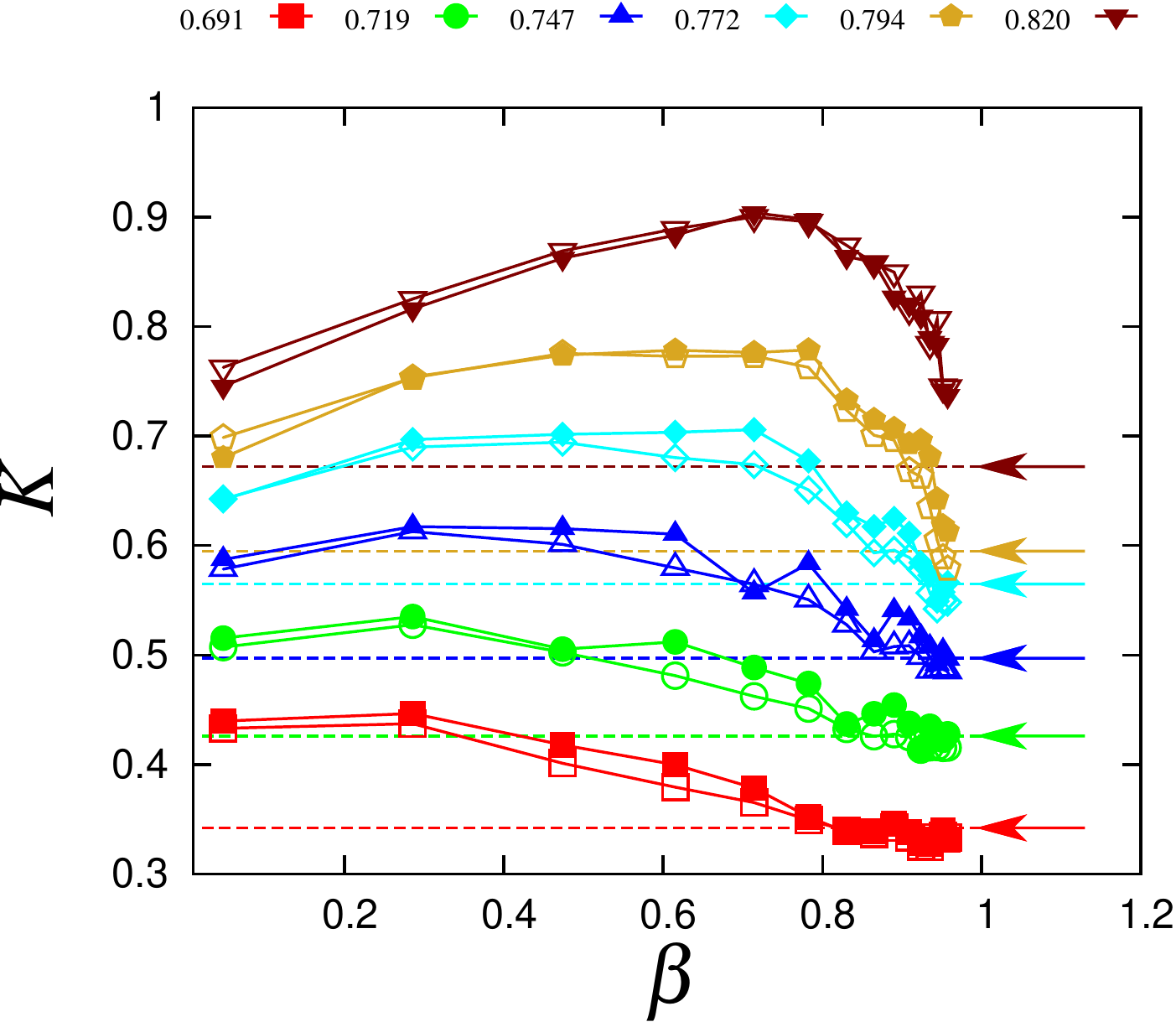}\label{bulkvskappa_noscale_mu0p0}}
\subfigure[]{\includegraphics[width=0.4\textwidth, angle=0]{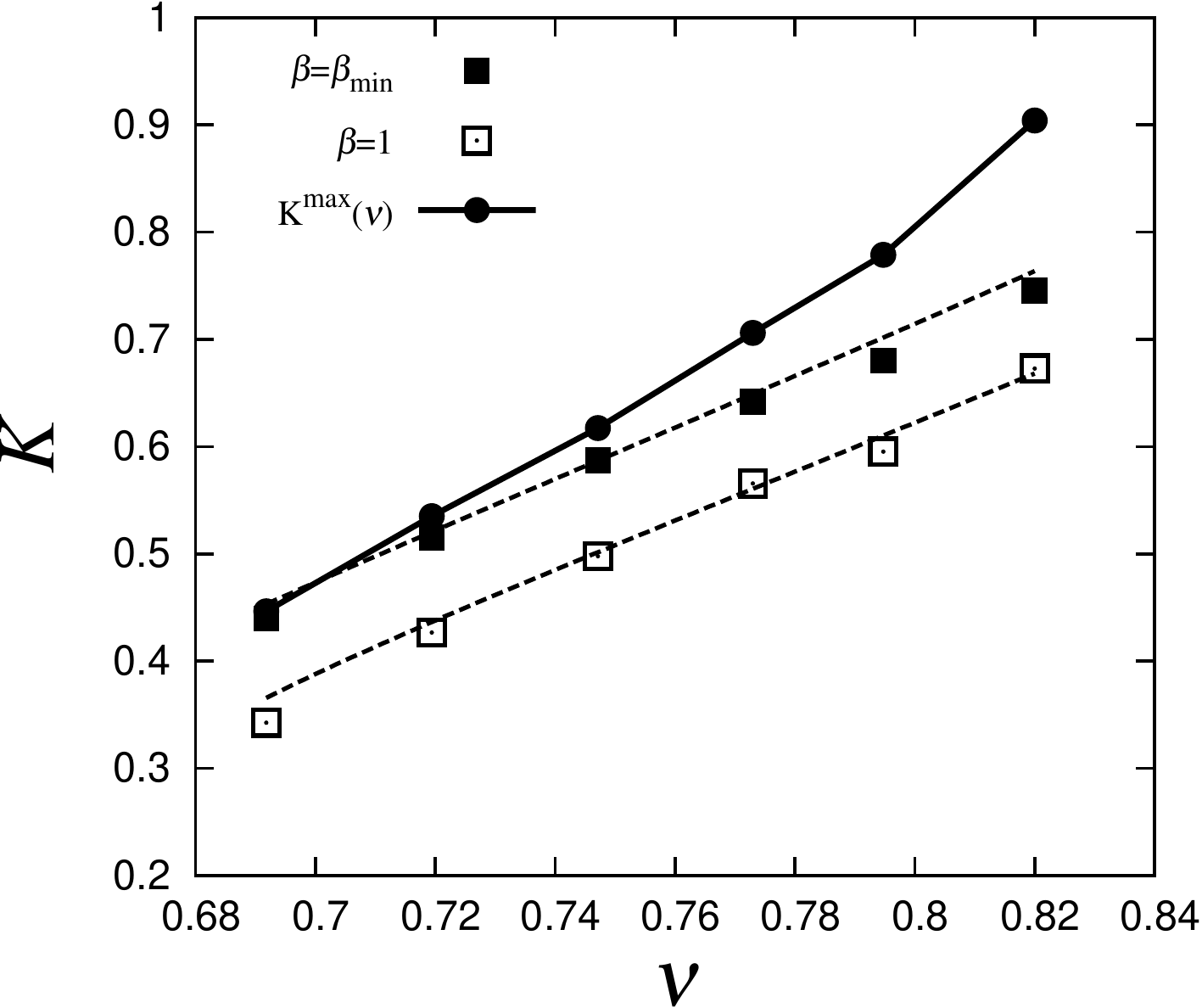}\label{rangebulkvskappa_noscale_mu0p0}} \\
% \subfigure[]{\includegraphics[width=0.4\textwidth, angle=0]{Figures/bulkscaledvskappa_noscale}\label{bulkscaledvskappa_noscale_mu0p0}}
\caption{(a) Bulk modulus (scaled by constant ${2 \langle r_A \rangle}/{k}$) $K$ measured using Eq.\ (\ref{eq:Bmeasure}) (solid symbols) and predicted using 
Eq.\ (\ref{eq:Bpredict}) for the whole system plotted against the number fraction $\beta= N_B^T/\left(N_A^T + N_B^T \right)$. 
Different colors represent the volume fraction $\nu$ as shown in the legend. 
The corresponding arrows show $K$ for $\beta=1$, i.e.\ the limit for infinitely small B particles, where the measurements are done after removing all the small particles and the static assembly consisted of only A. 
(b) $K$ for the two extreme cases: $\beta=\beta_\mathrm{min}$ (solid symbols) and $\beta=1$ (empty symbols), both are the monodisperse cases with the latter having 5\% fewer particles than the former. 
Lines passing through the data is Eq.\ (\ref{eq:Bmeasure}).
The dots represents the maximum $K$ obtained from (a) for a given density. 
}
\label{bulkinfo}
\end{figure}

Fig.\ \ref{bulkvskappa_noscale_mu0p0} shows the evolution of the bulk modulus $K$ plotted against $\beta$ for different volume fractions $\nu$. 
As expected, $K$ increases systematically with density $\nu$.
For loose states ($\nu$ = 0.69, 0.72), $K$ mostly decreases with increasing $\beta$ to the limit case $\beta\to1$, as discussed below. 
The behavior associated with denser states is much more interesting, as 
we observe an increase in $K$ to a maximum, followed by a decrease for larger $\beta$. 
Note that the value of $\beta$ where $K$ becomes maximum increases with increasing densities and the maximum also becomes stronger. 
From Fig.\ \ref{bulkvskappa_noscale_mu0p0}, we extract very important insights:
(i) The stiffness of a granular assembly can be manipulated by only substituting with a small amount 
of fines to the base material (in this case 5/105\%). 
(ii) We can control the ``direction'' of the change (stiffness enhanced or lowered) and the magnitude of change through the density and the size of the small particles. 
(iii) For loose material, there is no enhancement.
(iv) For dense material, for a given density, there is an ideal size of fines that leads to the maximum in the bulk stiffness. 

We associate the different trends observed for loose and dense systems with the ability of the fines (material B) to fill the voids formed by particles A. 
In the loose state, particles B are smaller than the average void size, so they act as rattlers, and do not contribute to the force network, leading to a decreasing bulk modulus with $\beta$.
With increasing density, the void size gets smaller and compatible with the size of particles B, 
and thus contribute to the active contact network (see appendix \ref{App:AppendixA}).

The maximum in $K$ observed in Fig.\ \ref{bulkvskappa_noscale_mu0p0} is different from that observed in the isotropic fabric $\Fv$ shown in Fig.\ \ref{FvT_pvskappa_noscale_mu0p0}. 
Thus in the case of a bidisperse granular mixture with wide size ratio, not only the contact network controls the bulk modulus, 
as the case for uniformly polydisperse systems, where the bulk modulus is directly associated with $\Fv$ \cite{kumar2014effects}.

Fig.\ \ref{rangebulkvskappa_noscale_mu0p0} shows the value of $K$ against $\nu$ for two extremes: 
smallest $\beta=\beta_\mathrm{min}$ and $\beta=1$ as shown in the legend, both represent monodisperse cases of only A particles. 
$K(\beta=\beta_\mathrm{min})$ is the left most data points in Fig.\ \ref{bulkvskappa_noscale_mu0p0}, increasing with $\nu$. 
For $\beta=1$, particles B are infinitely small and therefore do not participate in the contact network. 
Thus, the value $K(\beta=1)$ is obtained by removing B from the system. 
This leads to a slightly smaller volume fraction ($1/1.05$ times the original) and $K(\beta=1)$ is smaller than $K(\beta=\beta_\mathrm{min})$ and the two lines being parallel. 
The two data-sets for the limit cases in Fig.\ \ref{rangebulkvskappa_noscale_mu0p0} show good agreement when fitted by Eq.\ (\ref{eq:Bmeasure}). 
As already visible in Fig.\ \ref{bulkvskappa_noscale_mu0p0}, Fig.\ \ref{rangebulkvskappa_noscale_mu0p0} shows that 
the maximum stiffness measured for each volume fraction from Fig.\ \ref{bulkvskappa_noscale_mu0p0}, 
does not lie in between of the two extremes $K(\beta=1)$ is smaller than $K(\beta=\beta_\mathrm{min})$.

\begin{figure}[!ht]
\centering
\subfigure[]{\includegraphics[width=0.4\textwidth, angle=0]{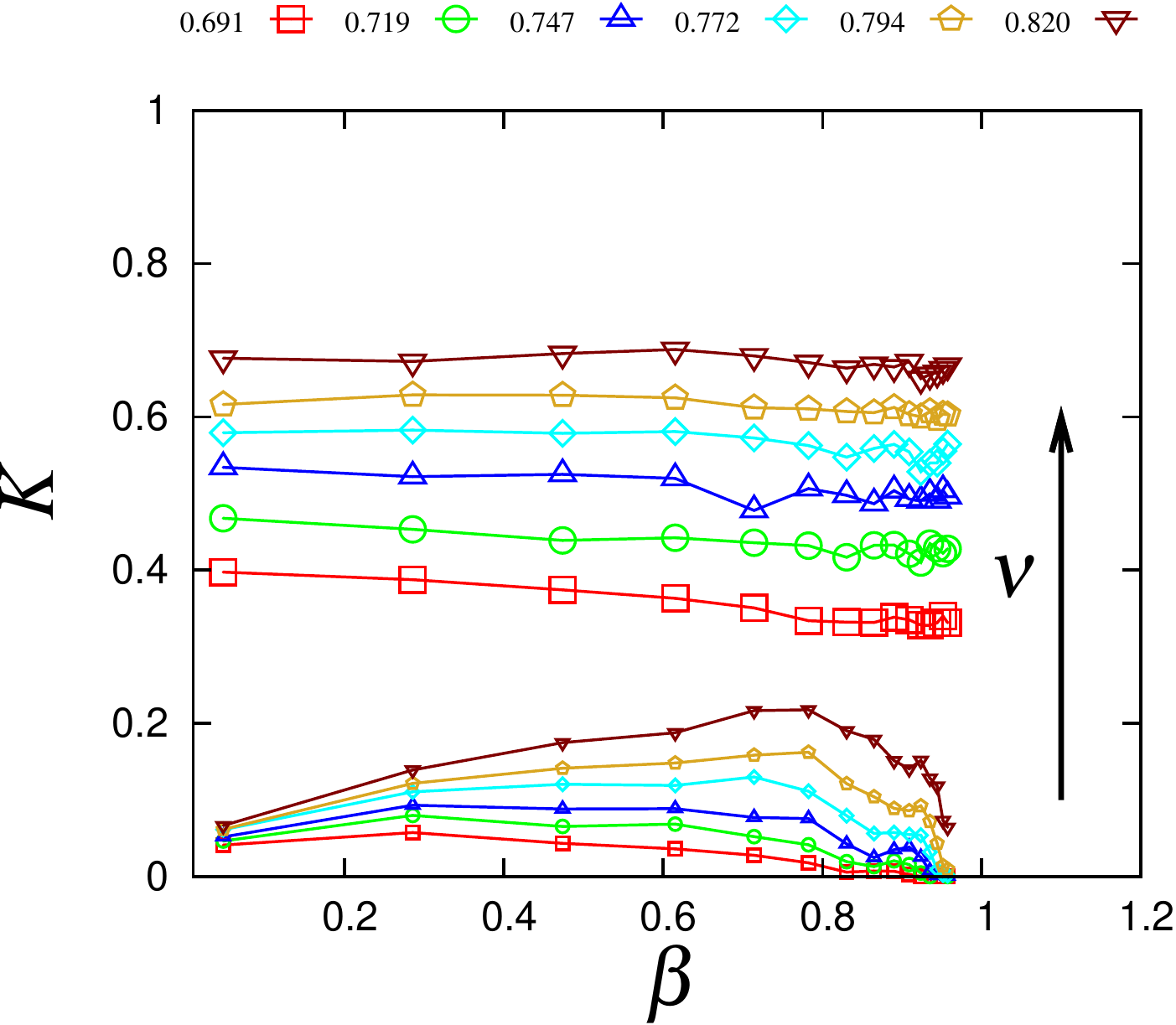}\label{bulkAABBAB_pvskappa_noscale_mu0p0}}
\subfigure[]{\includegraphics[width=0.4\textwidth, angle=0]{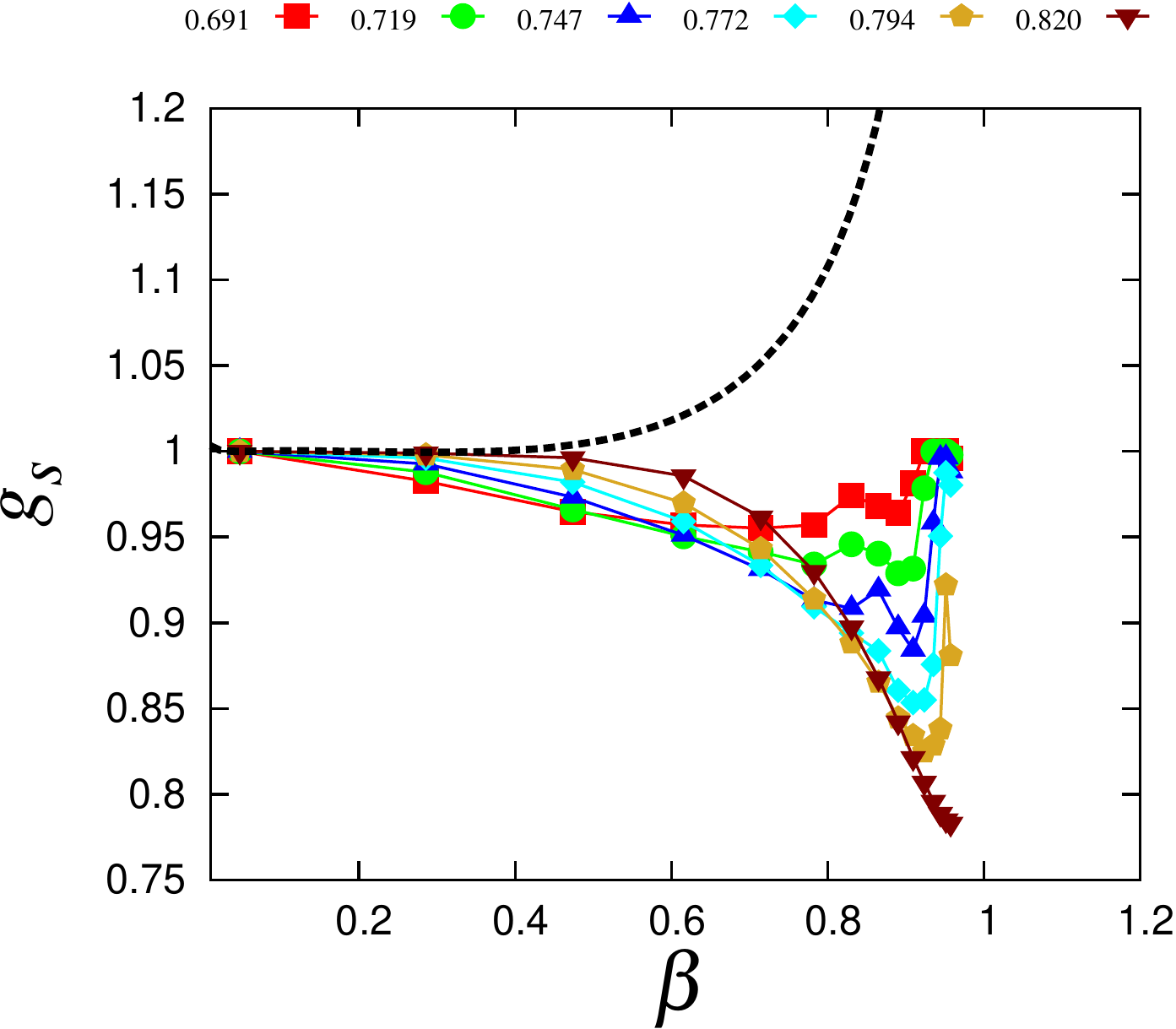}\label{g3s_pvskappa_noscale_mu0p0}}
\caption{(a) Partial bulk modulus (scaled by constant ${2 \langle r_A \rangle}/{k}$) for the AA (big symbols) and AB (small symbols) interactions, 
plotted against the number fraction $\beta= N_B^T/\left(N_A^T + N_B^T \right)$, from the same data as in Fig.\ \ref{bulkvskappa_noscale_mu0p0}. 
(b) $g_s$ calculated using Eq.\ (\ref{eq:g_sdefnbi2}), where the dashed line is Eq.\ (\ref{eq:g_sdefnbi2}) with constant linear compacity assumption, 
i.e., $\chi=1$ for the total mixture including the rattlers, see Eq. (\ref{eq:NBTratio}). 
}
\label{bulkmod}
\end{figure}

In order to understand the behavior of $K$ observed in Fig.\ \ref{bulkvskappa_noscale_mu0p0}, 
we look at the contributions to the bulk modulus of the three types of contacts present in the system, namely AA, AB and BB. 
It is straightforward to show that $K = K_{AA} + K_{AB} + K_{BB}$. 
Since the change in stress on B particles is very small in average, $K_{BB}$ is negligible (data not shown) 
when the granular assembly is subjected to small perturbation $\mathrm{d}\nu$, and hence $K \approx K_{AA} + K_{AB}$. 
Fig.\ \ref{bulkAABBAB_pvskappa_noscale_mu0p0} shows the bulk modulus for AA and AB interactions, $K_{AA}$ and $K_{AB}$.
$K_{AA}$ remains almost constant with $\beta$, except for the smallest $\nu$, where it slightly decreases. 
$K_{AB}$ remains small for loose system, as B particles mostly are rattlers. 
For high density, we observe an increase in $K_{AB}$ with $\beta$ followed by a decrease. 
% This means that the behavior in $K$ in Fig.\ \ref{bulkvskappa_noscale_mu0p0} is mainly related with the AB interactions. 
This signifies that the trend observed for $K$ in Fig.\ \ref{bulkvskappa_noscale_mu0p0} is mainly related 
to the behavior of AB interactions, while the actual value depends on the contributions of AA main network.

The radius of B, governed by $\beta$ at a particular volume fraction $\nu$, plays an important role not only in filling the voids of A, but also 
in contributing to the strong force network, leading to the maxima in bulk modulus $K$. 
% Relating $K$ with packing properties by introducing a correction $g_s$ for the stiffness matrix, similar to fabric, can significantly reduce the simulation effort 
% needed to extract the maximum bulk modulus at a given $\nu$. Different size distributions are focus of future research. 
Now we want to relate the bulk stiffness $K$ with the packing properties, in a similar fashion of Eqs.\ (\ref{eq:isofabriceq1}) and (\ref{eq:pstarmain}) as adopted for fabric and pressure respectively.
We use the relation proposed in \cite{shaebani2012influence} to link $K$ 
to the polydispersity and the mean packing properties of the sample \cite{kumar2014macroscopic, goncu2010constitutive}: 
\begin{equation}
\label{eq:Bpredict}
K =  \frac{{2 \langle r_A \rangle} p_0^*  g_s \nu^* C^*}{ \nu_c \sqrt{\langle r^2 \rangle}} \left[ 1 - 2\gamma_p \left(-\epsiso\right) + \left(-\epsiso\right)\left( 1 - \gamma_p \left(-\epsiso\right) \right)\frac{\partial \mathrm{ln}\Fv}{\partial\left(-\epsiso\right)} \right], 
\end{equation}
% where $p_0=0.0415$, $\gamma_p \approx 0.0320$ and $\nu_c=0.658$ for the deviatoric mode are taken from Ref.\ \cite{imole2013hydrostatic, kumar2014effects}. 
where $p_0^* =0.043$, $\gamma_p=0.2$ are constant fit parameters taken from Refs.\ \cite{imole2013hydrostatic, goncu2010constitutive, kumar2014effects} 
and the $\nu_c$ is the jamming volume fraction. 
$g_s$ is the size distribution factor and for our bidisperse distribution is given by (see appendix \ref{App:AppendixB}) \cite{shaebani2012influence}: 
\begin{equation}
\label{eq:g_sdefnbi2}
g_s =  \frac{ \langle r \rangle}{\langle r^3 \rangle } \frac{r_A^2 \Omega_A^{-1} f_A +  r_B^2 \chi \Omega_B^{-1} f_B}{  \Omega_A^{-1}f_A + \chi \Omega_B^{-1}f_B }
= \frac{\langle r \rangle \langle r^2 \rangle_g}{\langle r^3 \rangle}.
\end{equation}
where the same modification for $\chi$ as given in Eq.\ (\ref{eq:csAcsB1}) has been adopted 
\footnote[5]{For a uniform distribution, $g_s$ is very close to unity and is independent of the width of the distribution. 
That may be the reason it did not appear in Refs.\ \cite{kumar2014macroscopic,goncu2010constitutive}.}. 
Fig.\ \ref{g3s_pvskappa_noscale_mu0p0} shows the variation in $g_s$ with $\beta$  for different volume fractions $\nu$. 
$g_s$ starts from 1 and decreases with $\beta$, with $g_s=1$ recovered only for small densities at larger $\beta$, 
when the system behaves like an assembly of only A particles and B do not contribute to the contact network (rattlers), see also in Fig.\ \ref{new_c3d}. 
For dense states, $g_s$ decreases continuously and reaches 0.75. 
The asymptotic value of $g_s$ where also rattlers are considered diverges with $1-\beta$ with power law -1/3, dotted lines in Fig.\ \ref{g3s_pvskappa_noscale_mu0p0}. 
% This is an interesting difference between the system with and without rattlers, where one increases and diverges while the other one shows a decreasing trend. 
%
It is worthwhile to notice that the $g_s$ in Eq.\ (\ref{eq:g_sdefnbi2}) 
is different from $g_3$ as used in Eq.\ (\ref{eq:g_3defnbi2}) for isotropic fabric $\Fv$, meaning that bulk modulus and fabric depend on the polydispersity in a different fashion. 
In Fig.\ \ref{bulkvskappa_noscale_mu0p0}, the prediction of Eq.\ (\ref{eq:Bpredict}) is reported and an excellent agreement is found. 
It is noteworthy that neglecting the particle polydispersity via $g_s$ over-predicts the bulk modulus as much as $25\%$. 
Note that in Eq.\ (\ref{eq:Bpredict}) a constant $p_0^*$ is used, while this is not the case for extreme polydispersity in our system (see Fig.\ \ref{p0meas_vskappa_noscale_mu0p0}). 
Future research will focus developing analytical relation for $K$ from Eq.\ (\ref{eq:pstarmain}), where the dependence of $p_0$ on volume fraction $\nu$ is also considered.

% Fig.\ \ref{bulkvskappa_noscale_mu0p0} shows an excellent prediction of bulk modulus $K$ with $\beta$ (small symbols) using Eq.\ (\ref{eq:Bmeasure}).
% Slight modification with $\sqrt{\langle r^2 \rangle}$ is used for the scaling of $K$ instead of ${\langle r \rangle}$ as used in Refs.\ \cite{kumar2014macroscopic}. 
% Eq.\ (\ref{eq:Bmeasure}) also involves a term $g_s$ that accounts for the polydispersity in the system. 
% It is noteworthy that neglecting the particle polydispersity via $g_s$ over-predicts the bulk modulus as much as $25\%$. 
% %
% Note that in Eq.\ (\ref{eq:Bpredict}) a constant $p_0^*$ is used, while the future research focus is to develop a analytical relation directly from 
% Eq.\ (\ref{eq:pstarmain}), where $p_0$ is no more a constant and is highly dependent on the volume fraction $\nu$. 

\section{Qualitative Results}
This paragraph is devoted to the qualitative discussion of the 
findings of this study.  

Starting from a base assembly consisting of particles A, a certain volume
is substituted with smaller size particles B, while the total volume is kept constant. 
We restrict ourselves
to a small amount of additives (5/105=4.76\%), i.e.,\ much less
material than would be necessary to fill the pore-space in the base
material and focus on the effect of particle size-ratio. 
We study the two limits of either similar sizes ($r_B \sim r_A$)
and of very small sized B ($r_B \ll r_A$), as well as the interesting regime in between the limits. 

Substituting A with similar sized particles B is unlikely to change the system properties significantly, since the new particles fully participate 
in the mixture (besides a few rattlers). Thus, we observe 
a small effect of the polydispersity due to the new species on the bulk properties. 
On the contrary, when substituting with very small particles B, the 
mechanical properties of the mixture are practically the same as 
of the base material A alone, since the B particles 
are so small that they can move in between particles A,
freely passing through the pore-throats and thus escaping 
the pores whenever necessary to reduce their stress.
In the intermediate size-regime (roughly $1/2 > r_B/r_A > 1/5$)
a little volume of particles B can change the mixture 
properties considerably, providing systems with higher mechanical
bulk modulus as compared to the mere interpolation between the two
limit cases.

Assume a pore-size distribution with most pores (formed by A) 
between a cubic and a hexagonal local structure, such that they can 
accommodate particles of sizes between $r_8=(\sqrt{3} -1) r_A \approx 0.732 r_A$ and $r_6=(\sqrt{3/2} -1) r_A \approx 0.225 r_A$, respectively 
(see appendix \ref{App:AppendixA} for discussion on the pore sizes corresponding to different packing arrangements). 
Thus, the typical pore-size is $r_p/r_A \approx (r_8+r_6)/2r_A \approx 0.48$,
while there are no pores outside of the range.

There are two possibilities to scan this range of pores for a given 
volume of substitution. One can either change $(i)$ the size of particle B, $r_B$ (or $\beta$) 
or $(ii)$ the size of the pores (by changing the volume of the sample).

$(i)$ Assuming fixed volume of the system,
for $r_B>r_8$, any particle B sitting in a pore between particles A will mechanically 
contribute to the packing, and we are in the large size of particle B limit.
When decreasing $r_B$ below $r_8$, more and more
pores will lose the mechanical contact with their
(caged or trapped) particles B, until, for $r_B \sim r_6$, practically
all pores filled with single particles B have lost contact with their cages.
However, pores filled with more than one particle B still could
contribute to the force network, so that the number of particles B becomes
important (which is considerably increasing with decreasing size $r_B$).
When $r_B<r_4=(\sqrt{2} -1) r_A \approx 0.414 r_A$ also multiple B particles lose their efficiency, since they
can escape through square pores and for $r_B<r_3=(2/\sqrt{3} -1) r_A \approx 0.155 r_A$ even through
the smallest triangle pores, i.e.\ we are in the small $\beta$ limit.
 
$(ii)$ For a fixed size of particles B, increasing the volume fraction (decreasing
the volume) will reduce the available pore sizes and thus shift the whole
phenomenology towards smaller $r_B$. All pores become smaller and,
for the largest densities used, the smallest pore-throats $r_3$ are almost closed
so that the escape mechanism is hindered, but not blocked since there are
still other (e.g.\ square) shaped throats in the (disordered, non-crystalline) 
packings. The reduction in pore size is proportional to the typical AA-contact
deformation, which in turn is proportional to the pressure (first order) and
thus to the density relative to the jamming density.
%SL is this true? 

Combining the two cases, the maximal bulk stiffness $K$ of the packings, as function of volume fraction, 
occurs at $r_B \propto r_p (1-\delta_{AA}/r_A) \approx r_p (1-\mathrm{ln}\left(\nu/\nu_c\right))$.
Furthermore, we attribute the increase of the maximal $K$ with increasing compression to the reduced mobility of 
the small particles due to smaller pores and pore throats.

\begin{figure}[!ht]
\centering
\includegraphics[width=0.5\textwidth, angle=0]{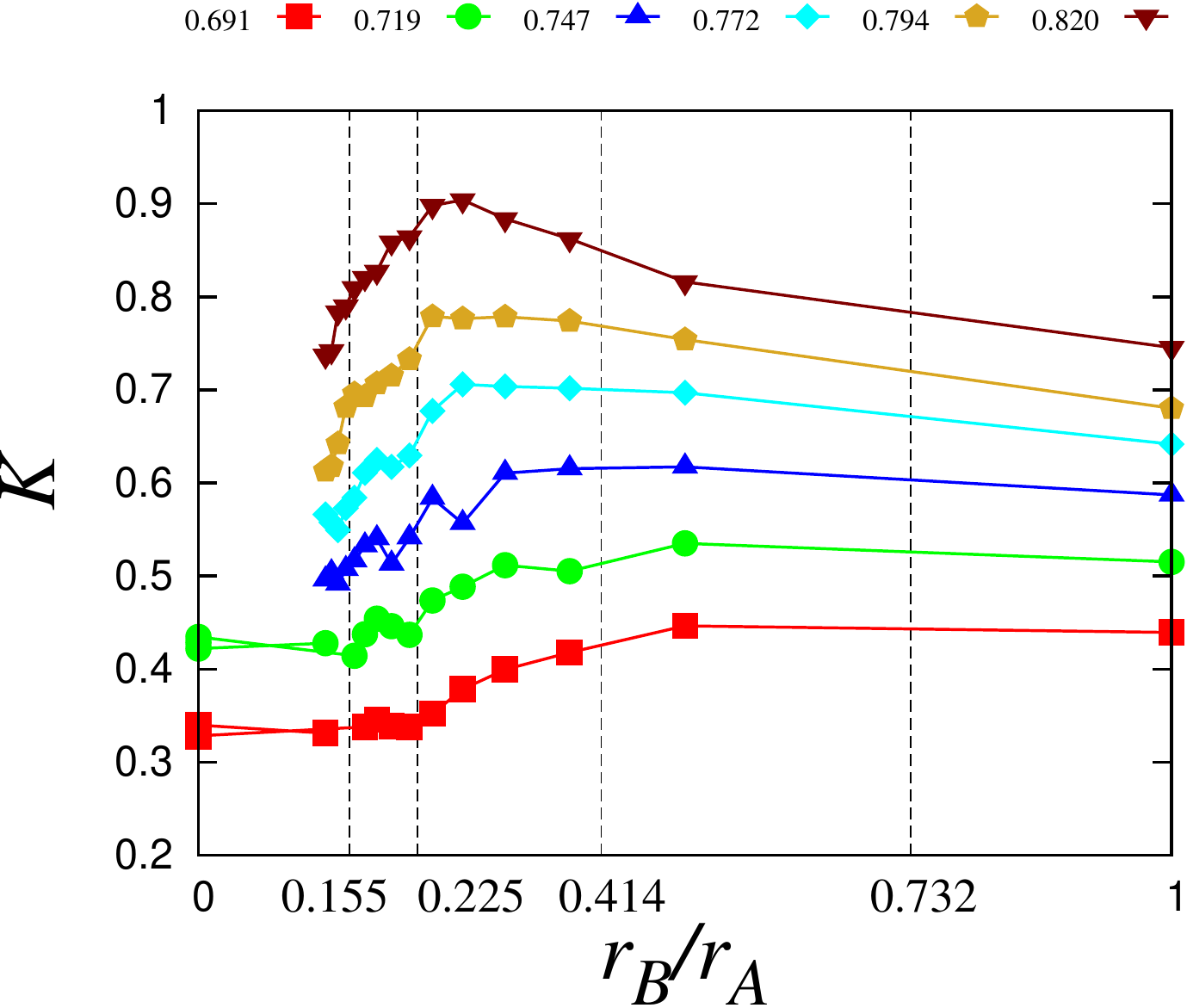}
\caption{Bulk modulus (scaled by constant ${2 \langle r_A \rangle}/{k}$) measured using Eq.\ (\ref{eq:Bmeasure}) 
plotted against size ratio $r_B/r_A$. %Note that the $x$-axis is in log scale.
The dashed vertical lines represent the radio ratio when particle B fills the void by A formed in triangular, tetrahedron, square and cubic lattices respectively, as shown in Fig.\ \ref{lattice}.
Different colors represent the volume fraction $\nu$ as shown in the legend. 
For the two lowest loose states, all B particles are rattlers and hence $r_B=$ in the system. }
\label{bulkvssizeratio_noscale}
\end{figure}

Fig.\ \ref{bulkvssizeratio_noscale} shows the variation in the bulk modulus $K$, 
against $r_B/r_A$ for different volume fractions. 
For dense assemblies, $K$ attains a maxima near the size ratio corresponding to tetrahedron lattice, 
meaning that the dense state is more likely to create such a configuration and this is the efficient arrangement. 
For the system becoming looser, the maximum of $K$ moves towards high size ratios $r_B/r_A$ corresponding to a cubic-like configuration.

\section{Conclusion and Outlook}
In this study, we use DEM simulations to study the bulk properties of a granular assembly, initially composed of monodisperse particles. 
A fixed volume of 5\% of these monodisperse particles are substituted with 5/105=4.76\% of fines, in order to create a bidisperse mixture. 
%
% We maintain a constant volume of the fines and systematically vary the radius (and hence number) of fines. 
%
The focus is on the manipulation of the properties of industrial mixtures with minimal costs/alterations by introducing a minimal amount of fines in the assembly. 
The system is characterized by the number ratio $\beta = N_B^T/\left(N_A^T + N_B^T \right)$ of fines to the total number of particles and 
we study the combined effects of $\beta$ and volume fraction on the micro- and macroscopic properties of the mixture. 
% Bigger particles present in large volume creates voids at a given volume fraction. 
% The effective filling of these voids by smaller sized fines enhance the mechanical strength of the mixture. 
% On the other hand, if the pore size is bigger than the size of fines, most fines does not contribute to the contact network and the bulk stiffness get reduced. 

Important highlights are extracted regarding microscopic and macroscopic (bulk) information of granular mixtures. 
% Firstly, all the particles with less than \textit{four} contacts are iteratively removed to generate a mechanically stable structure. 
%
% For a given volume fraction, the structure of big particles create an voids. 
% These voids, available to the fines can enhance the mechanical strength to the aggregate once they achieve more than four contacts with the neighbors. 
% For small $\beta$, the size of fines is comparable to the big particles and contribute to the contact network for all volume fractions. 
% For larger $\beta$, leading to smaller size of fines, most fines are rattlers. 
%
% The volume fraction excluding the rattlers $\nu^*$ decreases monotonically with $\beta$ for all volume fractions. 
The static pressure due to particle interactions and the coordination number (excluding rattlers) decrease monotonically with $\beta$, with small variations for $\beta \to 1$. 
% For $\beta\to1$, the coordination number increases a little since all fines become rattler and the system approaches a monodisperse case with higher coordination number. 
% The non-dimensional pressure, a representation of average overlap in system shows a similar trend like coordination number. 
%
% The jamming volume fraction, extracted when pressure approaches zero, increases with $\beta$ and reaches an asymptote for $\beta\to1$.
%
The isotropic fabric $\Fv$, a measure of the contact network density, 
decreases with $\beta$ for loose systems (in agreement with pressure behavior), since large pores created by big particles provide space 
for fines to be `caged'. 
The behavior for higher densities is different, as $\Fv$ first increases with $\beta$ and then decreases for $\beta \to1$.
In the first stage, the system is more coordinated and fines efficiently pack the voids, while when $\beta\to1$, most fines are rattlers, and thus $\Fv$ decreases.
The fabric is well described by the relation $\Fv = g_3 \nu^*C^*$, as introduced in Refs.\ \cite{imole2013hydrostatic, goncu2010constitutive, shaebani2012influence, kumar2014effects}, 
when $g_3$ is properly modified with a non constant compacity accounting for large polydispersity. 

Finally, we focus on the effective bulk modulus $K$, measured by applying small volumetric perturbations to the system. 
The behavior of $K$ is different from both pressure and fabric. 
For loose systems, a monotonous decrease is observed, while for denser systems, 
$K$ first increases, reaching a maximum, whose value depend on the density of the sample, and later decreases.
$\beta=1$ can be thought as the case of infinitely small fines and thus resembles the monodisperse case with volume fraction $1/1.05$ with respect to the original case. 
% Finally, the bulk modulus is linked with the mean packing properties using Eq.\ (\ref{eq:Bmeasure}) via $g_s$, accounting for high size dispersity in the system.
%

% \textbf{Outlook}\\
In this study, we focus on the bulk properties of a granular assembly when only a small volume fines is included. 
%
% % The operating volume fraction affects the size of voids and optimum size of fines can fill these voids efficiently. 
% % The presence of fines in granular packing, which are often associated with degradation of strength in the material, are here turned into an asset for functionality. 
% % %
% % Estimating the number of non-rattlers particles of the mixture that provides the mechanical strength to the mixture remains an open challenge. 
% % % Using probabilistic approach and possible particle packings can help to predict the non-rattlers of different components in the system. 
% % %
% % A better understanding of splitting the components of macroscopic properties of AB interaction into its individual components is also necessary. 
%
Next natural step is studying the influence of the volume of fines on the material behavior, 
that is to explore different regimes of coarse-fine mixtures. 
Finally, the focus can be moved towards other loading paths (uniaxial compression or shear) to study the effect fine content on the evolution of the full elastic tensor. 

%XXX SL relation to fatih + Nishant and relation to VO work
\section*{Acknowledgement}
We would like to thank Marco Ramaioli, University of Surrey for suggesting the initial idea and scientific discussions. 
We acknowledge the financial support of 
European Union funded Marie Curie Initial Training Network, FP7 (ITN-238577), 
PARDEM (www.pardem.eu), and by the NWO-STW VICI grant 10828.

% \clearpage

% \pagebreak

\appendix

% \setcounter{equation}{0}
% \setcounter{figure}{0}
% 
% 
% \renewcommand{\theequation}{A.\arabic{equation}}  
% \renewcommand{\thefigure}{A.\arabic{figure}}
% \renewcommand{\thesubfigure}{\thefigure.(\alph{subfigure})}
% \makeatletter
% \renewcommand{\p@subfigure}{}
% \renewcommand{\@thesubfigure}{\thesubfigure\hskip\subfiglabelskip}
% \makeatother

\setcounter{equation}{0}
\setcounter{figure}{0}
\renewcommand{\theequation}{A.\arabic{equation}}  
\renewcommand{\thefigure}{A.\arabic{figure}}

\renewcommand{\thesubfigure}{\thefigure.(\alph{subfigure})}
\makeatletter
\renewcommand{\p@subfigure}{}
\makeatother

\section{Radius ratio in different lattice configurations} \label{App:AppendixA}

\begin{figure}[!ht]
\centering
\subfigure[]{\includegraphics[width=0.3\textwidth, angle=00]{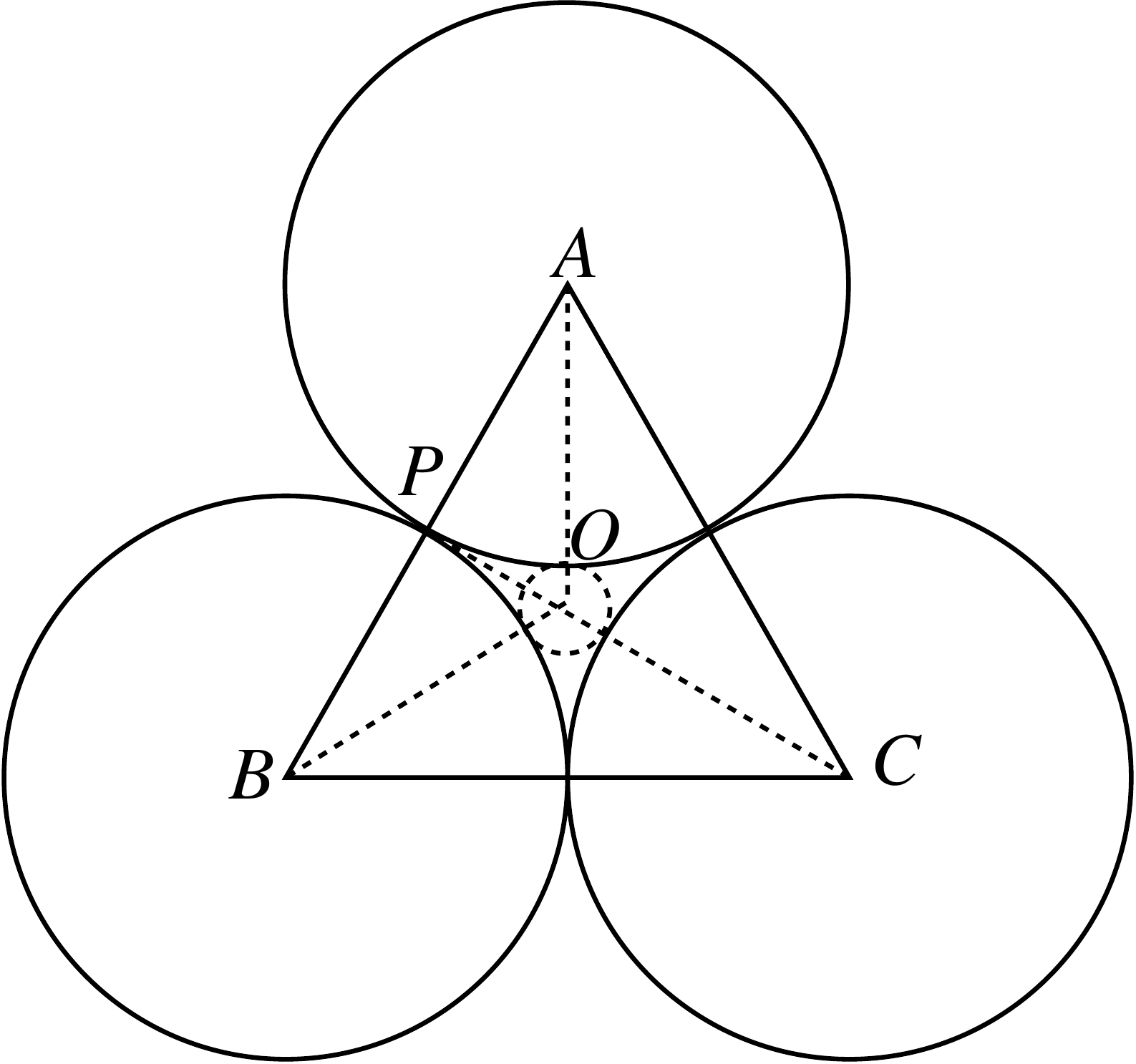}\label{tri}}\hspace{12mm}
\subfigure[]{\includegraphics[width=0.3\textwidth, angle=00]{Final_Figs/Lattice_tri_tri}\label{tetra}}\\
\subfigure[]{\includegraphics[width=0.3\textwidth, angle=00]{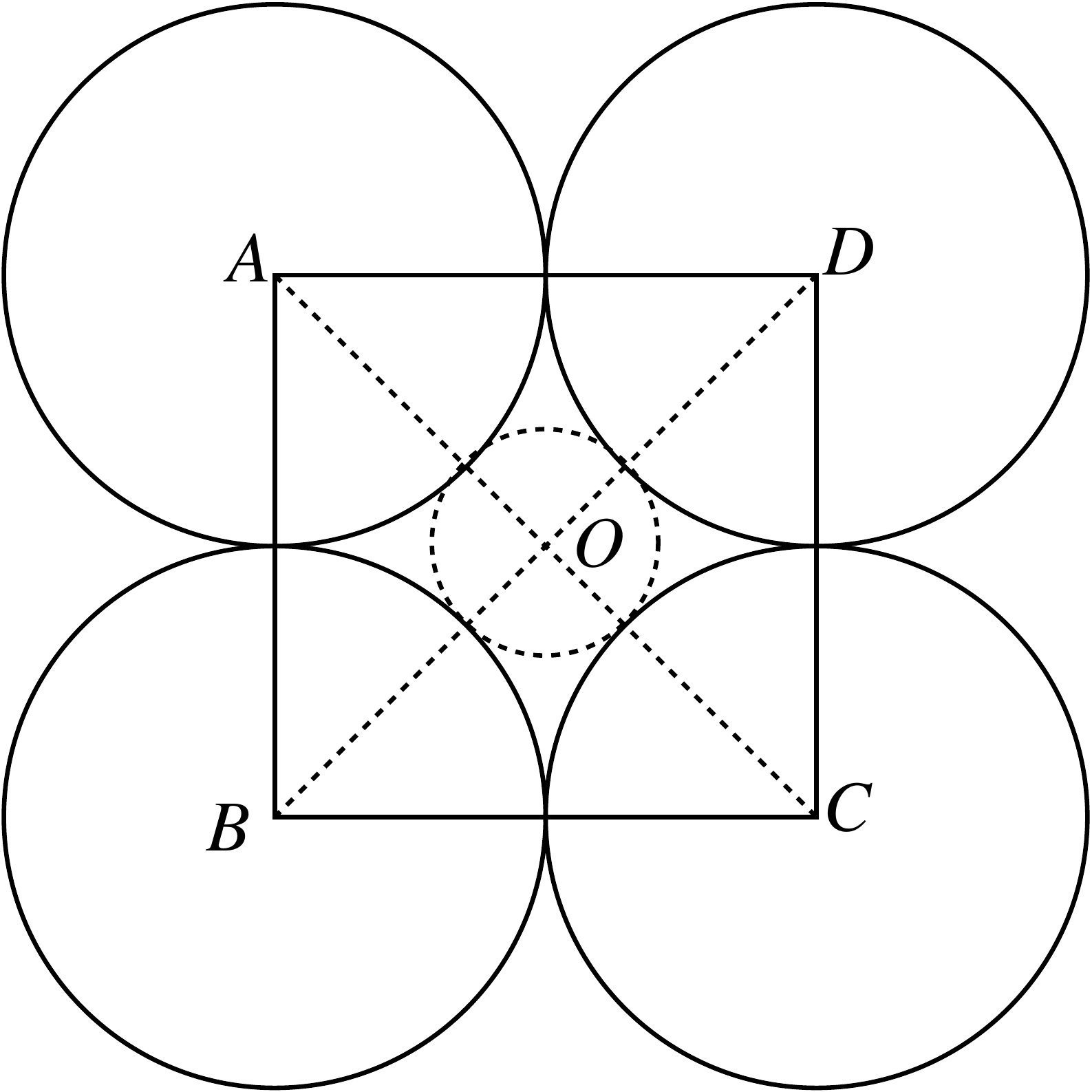}\label{square}}\hspace{12mm}
\subfigure[]{\includegraphics[width=0.3\textwidth, angle=00]{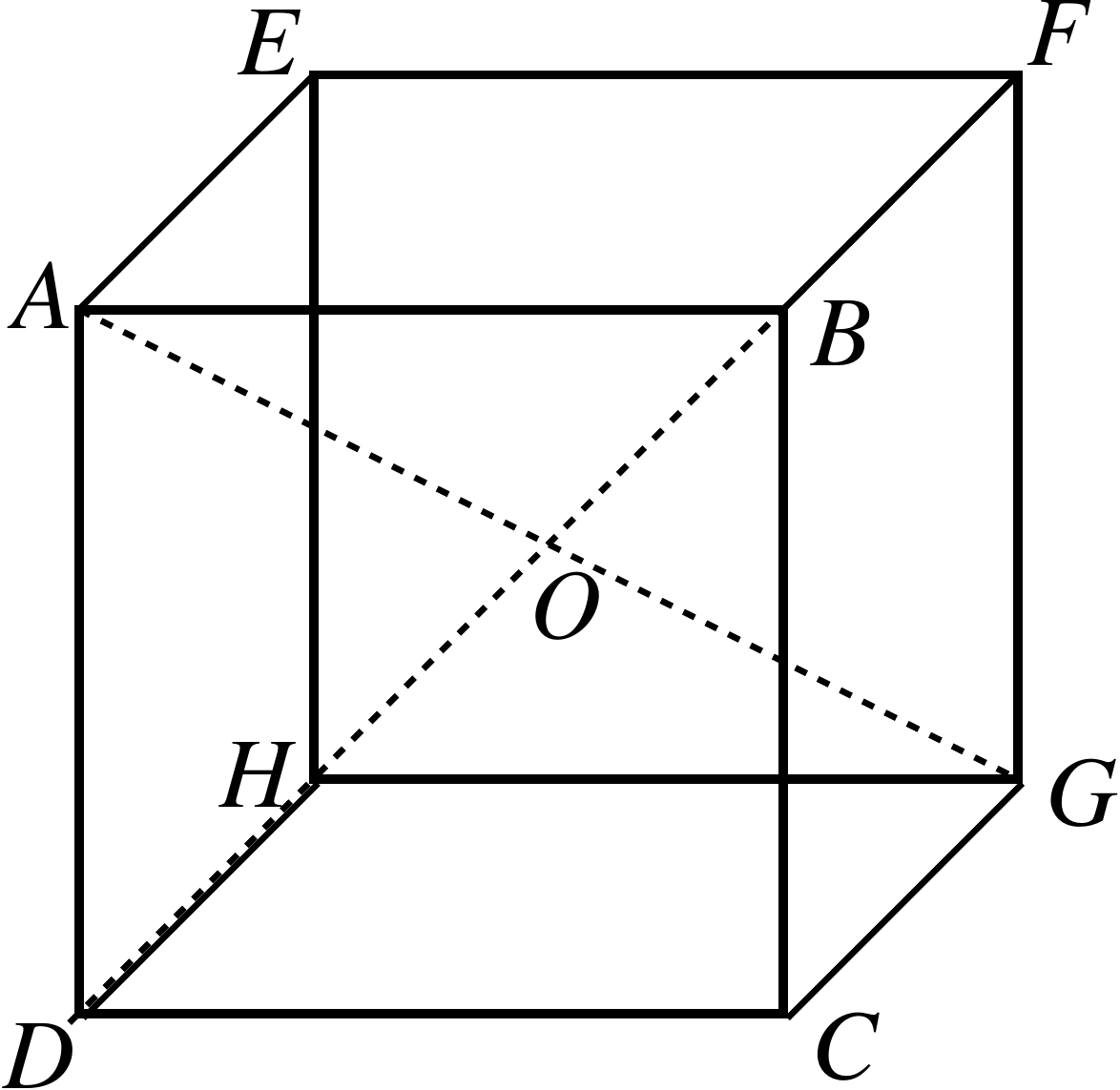}\label{cubic}}
\caption{(a) Triangular (b) tetrahedron (c) square and (d) cubic lattices, where the small particle of radius 
$r_3$, $r_6$, $r_4$ and $r_8$ respectively is residing between bigger particles of radius $r_A$, just touching them. }
\label{lattice}
\end{figure}

In this appendix, we focus on the possible arrangements of particles A and B in the granular assembly in order to characterize some special sizes of voids in the sample. 
At a given density, particles A create voids whose size depends on the geometry. 
Among the many possible arrangements of A, few possibilities are triangular, tetrahedron, square and cubic lattices as shown in Fig.\ \ref{lattice}. 
Particles B can efficiently fill these voids 
at special ratios between the void size and the radius of A. 
%
% Note that this is an very simple attempt to understand the possible arrangements in the assembly, while the full assembly is much more complex. 
%
% \subsection{Triangular lattices} \label{App:tri}

Fig.\ \ref{tri} shows a configuration where three A particles are arranged in a plane forming a triangular lattice and 
one particle B of radius $r_3$ is located in the void, centered at $O$, and touches A particles.
Thus, 
\begin{equation}
\label{eq:tri1}
AP = AO\mathrm{cos(\angle PAO)} = AO\mathrm{cos(30 \degree)},\nonumber
\end{equation}
that means: 
\begin{equation}
\label{eq:tri2}
r_A = \left( r_A + r_3 \right)\frac{\sqrt{3}}{2}.\nonumber 
\end{equation}
From here we can obtain the size ratio such that B efficiently fills the pore throat formed by three A: 
\begin{equation}
\label{eq:triF}
\frac{r_3}{r_A} =  \frac{2}{\sqrt{3}} -1 \approx 0.155.
\end{equation}

Fig.\ \ref{tetra} shows a sample configuration where A particles are arranged in a tetrahedron lattice i.e.,\ a local hexagonal structure, with three of them on a plane while the fourth is out of the plane. 
Assuming the tetrahedron is centered at the origin $O$, that is also the center of particle B of radius $r_6$, while $AB$ is one side of the tetrahedron and connects the centers of two A particles, 
($AB = 2r_A$). The four vertices of the tetrahedron are $r_A\left(\pm 1, 0 , -1/\sqrt{2} \right)$ and $r_A\left(0, \pm1 , 1/\sqrt{2} \right)$.
The tetrahedral angle $\angle AOB$ is $\mathrm{arccos\left(-1/3\right)} \approx 109.47\degree$. 
Using  the law of cosines we get 
\begin{equation}
\label{eq:tetra1}
AB^2 = AO^2 + OB^2 - 2.AO.OB\mathrm{cos(\angle AOB)},\nonumber 
\end{equation}
that in terms of radii becomes:
\begin{align}
 \left( r_A + r_A \right)^2 &= \left( r_A + r_6 \right)^2 + \left( r_A + r_6 \right)^2 - 2\left( r_A + r_6 \right)\left( r_A + r_6 \right)\mathrm{cos}(\mathrm{arccos\left(-1/3\right)}) \nonumber \\
 &= \left( r_A + r_6 \right)^2 \left( 1 + 1 - 2\left(\frac{-1}{3}\right) \right) = \frac{8}{3}\left( r_A + r_6 \right)^2. \nonumber \label{eq:tetra2} 
\end{align}
The tetrahedron void ratio is thus:
\begin{equation}
\label{eq:tetraF}
\frac{r_6}{r_A} =  \sqrt{\frac{3}{2}} -1 \approx 0.225.
\end{equation}

Next in Fig.\ \ref{square}, we show a configuration where A particles are arranged in a planar square lattice and particle B of radius $r_4$ sits in the void of A. 
Using Pythagoras' theorem, the relation between the sides of the lattice is:
\begin{equation}
\label{eq:square1}
AB^2 = AO^2 + OB^2  = 2AO^2, \nonumber
\end{equation}
and introducing the radii
\begin{equation}
\label{eq:square2}
\left( r_A + r_A \right)^2 = 2\left( r_A + r_4 \right)^2, \nonumber
\end{equation}
and the size ratio for efficient packing is
\begin{equation}
\label{eq:squareF}
\frac{r_4}{r_A} =  \sqrt{2} -1 \approx 0.414.
\end{equation}

Finally, Fig.\ \ref{cubic} shows a configuration where A particles are arranged in a body centered lattice with particle B of radius $r_8$ in the center of the cube touching A. 
Using again Pythagoras theorem, we can write
\begin{equation}
\label{eq:cubic1}
AG^2 = AD^2 + DG^2 = AD^2 + CD^2 + CG^2  = 3AD^2, \nonumber
\end{equation}
and 
\begin{equation}
\label{eq:cubic2}
\left( 2r_A + 2r_8 \right)^2 = 3\left( r_A + r_A \right)^2, \nonumber
\end{equation}
that leads to the cubic void size ratio:
\begin{equation}
\label{eq:cubicF}
\frac{r_8}{r_A} =  \sqrt{3} -1 \approx 0.732.
\end{equation}

By comparing the four cases considered here, the triangular lattice produces the smallest size ratio ${r_3}/{r_A}\approx 0.155$ while the cubic lattice ${r_8}/{r_A}\approx 0.732$ creates the largest one. 

For the case of overlapping spheres, the size ratio must be corrected by including 
the average overlap between AA ($\langle\delta_{AA}\rangle$) and AB ($\langle\delta_{AB}\rangle$) interactions.

\setcounter{equation}{0}
\setcounter{figure}{0}
\renewcommand{\theequation}{B.\arabic{equation}}  
\renewcommand{\thefigure}{B.\arabic{figure}}

\renewcommand{\thesubfigure}{\thefigure.(\alph{subfigure})}
\makeatletter
\renewcommand{\p@subfigure}{}
\makeatother

\section{Measurement of $g_3$ and $g_s$ for fabric and bulk modulus} \label{App:AppendixB}

We are interested in relating the isotropic fabric with the mean packing properties (coordination number and volume fraction) via $\Fv = g_3 \nu C$. 
The continuous limit of Eq.\ (\ref{eq:fabriceqtr2}) is given by \cite{shaebani2012influence}:
\begin{equation}
\label{eq:fabriceqtr4app}
\Fv=\frac{N_4}{V}\int_0^\infty V(r) C(r) f(r) dr  = g_3 \nu^* C^*,
\end{equation}
where $C(r)$ is the coordination number of a particle with radius $r$, volume $V(r)=4\pi r^3/3$ and $f(r)$ is the particle size (radii) distribution defined in section\ \ref{sec:moments}.
% $f(r)$ is the distribution of particle radii with probability $f(r)dr$ 
% to find the radius between $r$ and $r+dr$, and with $\int_0^\infty f(r)=1$. 
% $f(r)$ for a bidisperse distribution is given as $f(r) = f_A \Delta\left(r - r_A\right) + f_B\Delta\left(r - r_B\right) $, 
% where $f_A = N_A/\left(N_A + N_B \right)$ and $f_B = N_B/\left(N_A + N_B \right)$ are the number fraction of A and B and $\Delta()$ is the Dirac-delta function. 
% The mean particle radius is $\langle r \rangle = \int_0^\infty rf(r)$ and the mean $n^{\mathrm{th}}$ ordered radius is $\langle r^n \rangle = \int_0^\infty r^nf(r)$. 
The corrected volume fraction in the continuous form is given as: 
\begin{equation}
\label{eq:volfraccontiapp}
\nu^* = \frac{N_4\int_0^\infty V(r)f(r)}{V}.
\end{equation}
Let's assume that a reference $p$-particle with radius $r$ in the system has a contact with a neighboring particle of average radius $\langle r \rangle$. 
The space angle covered by the neighboring particle on the reference particle 
in a three-dimensional packing of sphere is given as \cite{shaebani2012influence, goncu2010constitutive}:
\begin{equation}
\label{eq:omegaapp}
\Omega(r) = 2\pi \left(1 - \sqrt{ 1 - \left(\frac{\langle r \rangle}{r + \langle r \rangle}\right)^2} \right),
\end{equation}
and the linear compacity (or the fraction of the shielded surface) associated with a single interaction is:
\begin{equation}
\label{eq:compacapp1}
c_s(r) = \frac{1}{4 \pi r^2}   \Omega(r) r^2 .
\end{equation}
The total compacity of the reference $p$-particle interacting with its $C(r)$ non-rattler neighboring particle of average radius $\langle r \rangle$ thus becomes:
\begin{equation}
\label{eq:compacapp}
c_s(r) = \frac{1}{4 \pi r^2}   \sum_{p=1}^{C(r)} \Omega(r) r^2 = \frac{\Omega(r) C(r)}{4 \pi}.
\end{equation}
%%%%%%%%%%%%%%%%%
$c_s(r)$ decreases with $r$, starting from 1 when $r/\langle r \rangle \to 0$ and reaches a constant value for $r/\langle r \rangle \ge 1$. 
Refs.\ \cite{shaebani2012influence, goncu2010constitutive} have shown that $c_s(r)$ 
decreases with increasing particle radii and saturates to a constant value $\in[0,1]$ for large sized particles. It is also dependent on the volume fraction of the system. 
Large differences in particle number and size ratio affects the linear compacity $c_s(r)$.
$c_s(r)$ has two bounds: \\
i) \textbf{Upper bound:} the maximum compacity is reached when a small particle is surrounded by two big particles. 
Therefore, for the lower bound, $r/\langle r \rangle \to 0$ leading to $\Omega(r)=2\pi$ and hence $\mathrm{max}\left[c_s(r)\right]=1$. \\
ii) \textbf{Lower bound:} Near the jamming transition (loose states), 
mainly the big particles remain in the system while the smaller particles act as rattlers.
To be mechanically stable, big particles need \textit{six} contacts. 
Using $r/\langle r \rangle \to 1$, we have $\Omega(r)=2\pi \left(1 - \sqrt{3}/2 \right)$ 
and hence $\mathrm{min}\left[c_s(r)\right]= \Lambda = \frac{\Omega(r) C(r)}{4 \pi} = \frac{2\pi \left(1 - \sqrt{3}/2 \right) 6}{4 \pi} = 3\left(1 - \sqrt{3}/2 \right)\approx0.4$, 
generally reached by big particles at low volume fraction, i.e.,\ near the jamming transition. 
%%%%%%%%%%%%%%%%%%%%%%
Using the definition for the average coordination number $C^*$ in the continuous limit:
\begin{equation}
\label{eq:coordrelnapp1}
C^*=\int_0^\infty C(r) f(r) dr = {4 \pi}\int_0^\infty \frac{ c_s(r) f(r) }{\Omega(r)},
\end{equation}
and using Eq.\ (\ref{eq:compacapp}), we get
\begin{equation}
\label{eq:coordrelnapp}
C^*= {4 \pi}\int_0^\infty \frac{ c_s(r) f(r) }{\Omega(r)}.
\end{equation}
Combining Eqs.\ (\ref{eq:volfraccontiapp}), (\ref{eq:compacapp}) and (\ref{eq:coordrelnapp}) in Eq.\ (\ref{eq:fabriceqtr4app}), we have:
\begin{equation}
\label{eq:g_3defn1}
g_3 =\frac{\int_0^\infty r^3 c_s(r) {\Omega(r)}^{-1} f(r) dr}{\int_0^\infty   c_s(r) {\Omega(r)}^{-1} f(r) dr \int_0^\infty   r^3 f(r) dr},
\end{equation}

In a similar fashion, we use a correction term $g_s$ as proposed in Ref.\ \cite{shaebani2012influence} 
to link the bulk modulus $K$ of a granular mixture to the polydispersity as:
\begin{equation}
\label{eq:g_sdefnbi1app}
g_s = \frac{\int_0^\infty   r f(r) dr \int_0^\infty r^2 c_s(r) {\Omega(r)}^{-1}  f(r)  dr}{\int_0^\infty  c_s(r)  {\Omega(r)}^{-1} f(r) dr \int_0^\infty   r^3 f(r) dr}.
\end{equation} 

%%%%%%%%%%%%%%%%

\setcounter{equation}{0}
\setcounter{figure}{0}
\renewcommand{\theequation}{C.\arabic{equation}}  
\renewcommand{\thefigure}{C.\arabic{figure}}

\renewcommand{\thesubfigure}{\thefigure.(\alph{subfigure})}
\makeatletter
\renewcommand{\p@subfigure}{}
\makeatother
\section{Analytical expression for pressure} \label{App:AppendixC}

\begin{figure}[!ht]
\centering
\subfigure[]{\includegraphics[width=0.31\textwidth, angle=0]{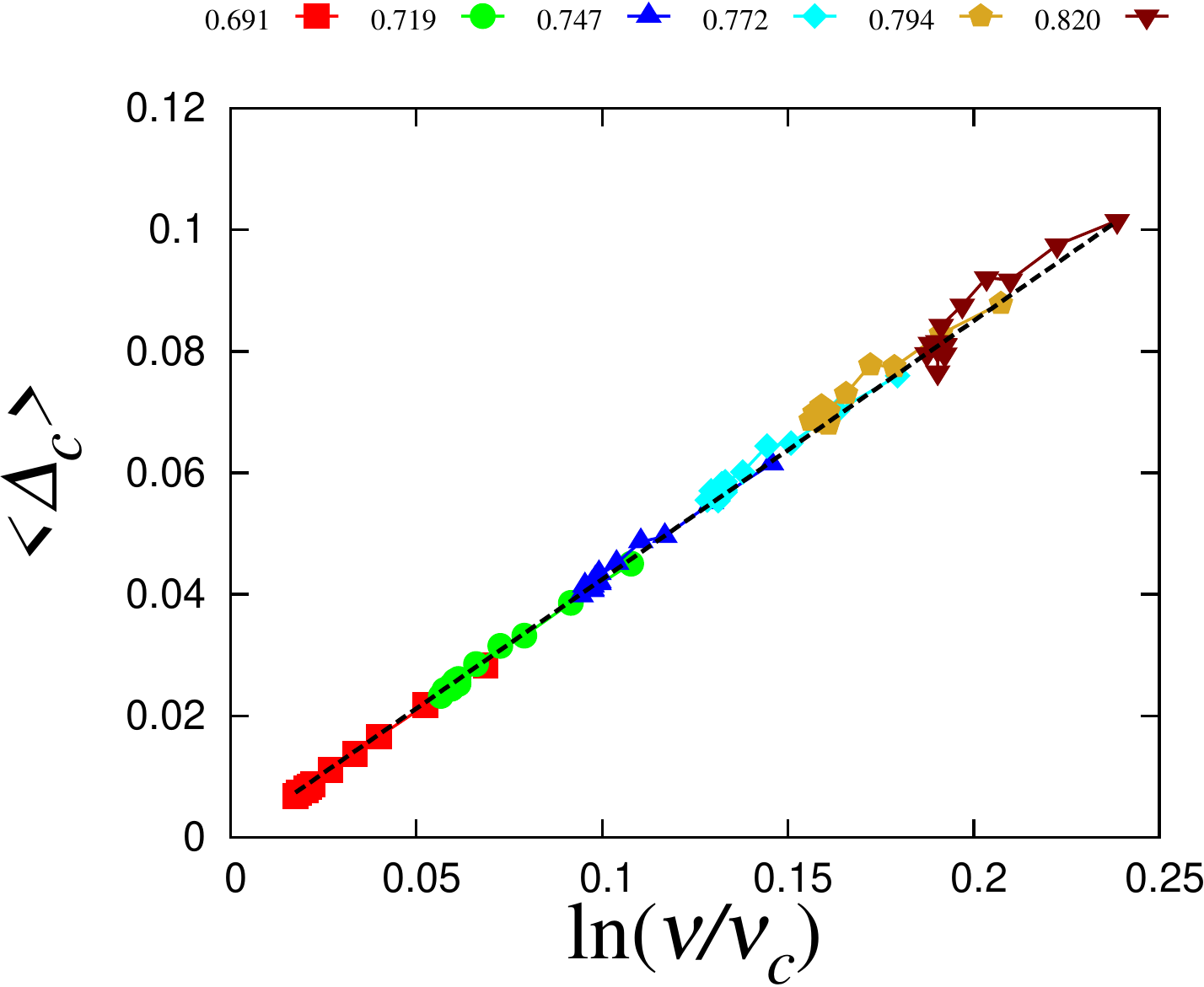}\label{Dcalc_pvskappa_noscale_mu0p0}}
\subfigure[]{\includegraphics[width=0.31\textwidth, angle=0]{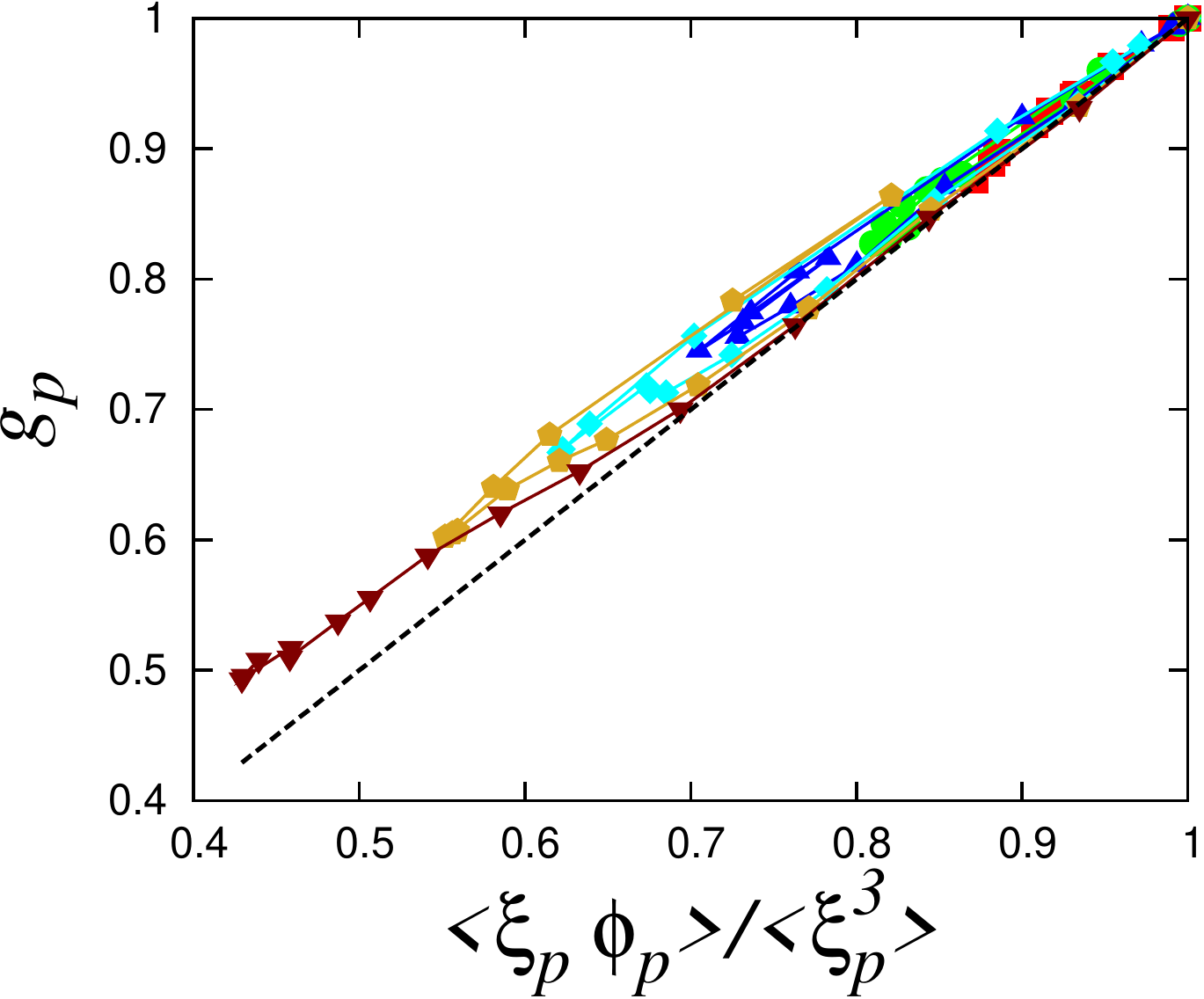}\label{g_pfromcalc_noscale}}\\
%
% \subfigure[]{\includegraphics[width=0.31\textwidth, angle=0]{Figures/g_ppfromcalc_noscale}\label{g_ppfromcalc_noscale}}
%
%
% \subfigure[]{\includegraphics[width=0.4\textwidth, angle=0]{Figures/nonpres_pvskappa_noscale}\label{nonpres_pvskappa_noscale_mu0p0}}
\caption{(a) $\langle\Delta\rangle_c$ against $\mathrm{ln}\left({\nu}/{\nu_c} \right)$ using Eq.\ (\ref{eq:volumefraction}).
(b) $g_p$ measured using Eq.\ (\ref{eq:g_pdefnbi21}) with the assumption that the total force on a particle is proportional to 
square of contacts it has with neighbors and is compared with the analytical expression of $g_p$ in Eq.\ (\ref{eq:pfinaleqn1}).
}
\label{newfigures}
\end{figure}

In order to better understand the final analytical expressions, the stress is rewritten and re-phrased, starting from the traditional definitions. 
Revisiting Eq.\ (\ref{eq:pressnon}), we have:
\begin{eqnarray}
\label{eq:stresseqmodify1}
 p &=& \frac{2 \langle r \rangle}{3 k} \mathrm{tr}(\pmb{\sigma}) \approx \frac{2 \langle r \rangle}{3k}\frac{1}{V}\mathrm{tr}\left(\sum_{c=1}^{M_4}\mathbf{l}^{c}\otimes\mathbf{f}^{c}\right)   =  \frac{{2 \langle r \rangle}}{3V} \sum_{c=1}^{M_4} \mathrm{tr}\left(l_c\mathbf{n}^{c}\otimes\delta_c \mathbf{n}^{c}\right) \nonumber \\
   &=& \frac{{2 \langle r \rangle}}{3V} \sum_{c=1}^{M_4} l_c \delta_c \underbrace{\mathrm{tr}\left( n\otimes \mathbf{n}^{c}\right)}_{\text{=1}}  = \frac{{2 \langle r \rangle}}{3V} \sum_{c=1}^{M_4} l_c \delta_c \nonumber \\
   &=& \frac{{2 \langle r \rangle}}{3V} \sum_{p=1}^{N_4} \left( r_p \sum_{c=1}^{C_p}\delta_c - \frac{1}{2} \sum_{c=1}^{C_p}\delta_c^2 \right),
\end{eqnarray}

where subscript $p$ and $c$ stand for particles and contacts respectively. 
Average overlap per contact is:
\begin{equation}
\label{eq:avgover1}
\langle \delta_c \rangle = \frac{\sum_{p=1}^{N_4}\sum_{c=1}^{C_p}\delta_c}{M_4} = \frac{\sum_{p=1}^{N_4}\sum_{c=1}^{C_p}\delta_c}{N_4 C^*} .
\end{equation}
Similarly, for the average squared overlap, one can write:
\begin{equation}
\label{eq:avgoversquate}
\langle \delta^2_c \rangle = \frac{\sum_{p=1}^{N_4}\sum_{c=1}^{C_p}\delta_c^2}{M_4} = \frac{\sum_{p=1}^{N_4}\sum_{c=1}^{C_p}\delta_c^2}{N_4 C^*} .
\end{equation}

Introducing the average overlap for particle $p$ as: 
\begin{equation}
\label{eq:avgover}
\phi_p =: \frac{\delta_p}{\langle \delta_p \rangle } =\frac{\sum_{c=1}^{C_p}\delta_c}{\left(\sum_{p=1}^{N_4}\sum_{c=1}^{C_p}\delta_c\right)/\sum_{p=1}^{N_4}} = \frac{\sum_{c=1}^{C_p}\delta_c}{\left(\langle \delta_c \rangle M_4\right)/\sum_{p=1}^{N_4}} = \frac{\sum_{c=1}^{C_p}\delta_c}{C^* \langle \delta_c \rangle},
\end{equation}
where $\langle \delta_c \rangle$ is the average overlap per contact. 
Eq.\ (\ref{eq:stresseqmodify1}) then can be written as:
\begin{eqnarray}
\label{eq:stresseqmodify2}
 p &=& \frac{{2 \langle r \rangle}}{3V} \left( \sum_{p=1}^{N_4}  r_p \sum_{c=1}^{C_p}\delta_c - \frac{1}{2} \sum_{p=1}^{N_4} \sum_{c=1}^{C_p}\delta^2_c \right) \nonumber \\
   &=&  \frac{{2 \langle r \rangle}}{3V}  \left(\sum_{p=1}^{N_4} r_p C^* \langle \delta_c \rangle  \phi_p  - \frac{1}{2} \langle \delta^2_c \rangle N_4 C^* \right) \nonumber \\
   &=&  \frac{{2 \langle r \rangle}}{3V}  \left( N_4 C^* \langle \delta_c \rangle \langle r_p \phi_p \rangle - \frac{1}{2} \langle \delta^2_c \rangle N_4 C^* \right) \nonumber \\
   &=&  \frac{{2 \langle r \rangle} N_4 C^* \langle \delta_c \rangle }{3V}  \left(  \langle r_p \phi_p \rangle - \frac{\langle \delta^2_c \rangle}{2 \langle \delta_c \rangle} \right)  \left[\frac{\nu^*}{(4 \pi/3) N_4 \langle r^3 \rangle/V}\right] \nonumber \\
   &=&  \frac{ C^* \nu^*  }{4 \pi} \frac{{ \langle r \rangle} \langle \delta_c \rangle}{\langle r^3 \rangle}\left(  2\langle r_p \phi_p \rangle - \frac{\langle \delta^2_c \rangle}{ \langle \delta_c \rangle} \right)  
\end{eqnarray}
Introducing the normalized particle radius $\xi_p  = r_p/\langle r \rangle$ and overlap $\Delta_c  = \delta_c/\langle r \rangle$ leads to:
\begin{eqnarray}
\label{eq:stresseqmodify3}
 p &=&  \frac{ C^* \nu^*  }{4 \pi} \frac{ \langle \Delta_c \rangle}{\langle \xi_p^3 \rangle}\left(  2\langle \xi_p \phi_p \rangle - \frac{\langle \Delta^2_c \rangle}{ \langle \Delta_c \rangle} \right) \nonumber \\
   &=& \frac{ C^* \nu^*  }{4 \pi} \langle \Delta_c \rangle  \left(  2g_p - b_p \langle \Delta_c \rangle \right) 
\end{eqnarray}
where
\begin{equation}
\label{eq:pfinaleqn1}
g_p = \frac{\langle \xi_p \phi_p \rangle}{\langle \xi_p^3 \rangle}, b_p = \frac{1}{\langle \xi_p^3 \rangle}\frac{\langle \Delta^2_c \rangle}{ \langle \Delta_c \rangle^2}
\end{equation}
The normalized average overlap $\langle \Delta_c \rangle$ is logarithmically related to the volume fraction of the present state via as also presented in Refs.\ \cite{imole2013hydrostatic, goncu2010constitutive}
\begin{equation}
\label{eq:volumefraction}
\langle \Delta_c \rangle = D (-\varepsilon_\mathrm{v}) = D \mathrm{ln}\left( \frac{\nu}{\nu_c} \right). 
\end{equation}
Fig.\ \ref{Dcalc_pvskappa_noscale_mu0p0} shows the measured average overlap per contact $\langle\Delta\rangle_c$ against $\mathrm{ln}\left({\nu}/{\nu_c} \right)$ 
and the slope of the linear line is $D=0.425$, in consistent with the measured value in Ref.\ \cite{imole2013hydrostatic}. 
Therefore, Eq.\ (\ref{eq:stresseqmodify3}) can be written in the same form as given in Ref.\ \cite{goncu2010constitutive}:
\begin{equation}
p=p_0 \frac{ \nu^* C^*  }{\nu_c} (-\varepsilon_\mathrm{v})  \left [ 1-\gamma_p (-\varepsilon_\mathrm{v}) \right ]
\label{eq:pstar}
\end{equation}
where $p_0 = \nu_c g_p D/ 2\pi$ and $\gamma_p = bD /2g_p$. The unknowns are $g_p$ and $b_p$. 
Assuming that total force on a particle is proportional to square of contacts it has with neighbors \cite{shaebani2012influence}, 
$\delta_p \propto C_p^2$, hence $\phi_p = C_p^2/\langle C_p^2 \rangle$, where $C(r) = 4\pi \frac{c_s(r)}{\Omega(r)}$ (see appendix \ref{App:AppendixB}). 
Therefore, for a continuous distribution, $g_p$ is given as: 
\begin{equation}
\label{eq:g_pdefnbi21}
g_p = \frac{\langle r \rangle^2}{ \langle r^3 \rangle}   \frac{ \int_0^\infty r c_s(r)^{2} {\Omega(r)}^{-2} f(r) dr}{\int_0^\infty   c_s(r)^{2} {\Omega(r)}^{-2} f(r) dr},
\end{equation}
which for a bidisperse size distribution is: 
\begin{equation}
\label{eq:g_pdefnbi2}
g_p = \frac{\langle r \rangle^2}{ \langle r^3 \rangle}  \frac{r_A  \Omega_A^{-2}f_A  + r_B^3 \chi^2\Omega_B^{-2}f_B }{\Omega_A^{-2} f_A +  \chi^2 \Omega_B^{-2} f_B   }, 
% = \frac{\langle r \rangle^2 \langle r \rangle_g}{\langle r^3 \rangle}  ,
\end{equation}
where $r_A$ and $r_B$ are the radius of A and B with number fraction $f_A$ and $f_B$ respectively. 
Note that the second term in Eq.\ (\ref{eq:stresseqmodify3}) is very small (maximum 10\% for dense volume fractions) and is a subject of future research. 
$g_p$ measured from the simulations is plotted in Fig.\ \ref{g_pfromcalc_noscale} against Eq.\ (\ref{eq:g_pdefnbi2}) and the results are in close agreement.

%%%%%%%%%%%%%%%%%%%%%%%%%%%
% \end{linenumbers}

\newpage
% \bibliographystyle{plainnat}
% \bibliographystyle{abbrv}
% \bibliographystyle{elsarticle-num}

% \bibliography{referencesA}

\begin{thebibliography}{10}
\expandafter\ifx\csname url\endcsname\relax
  \def\url#1{\texttt{#1}}\fi
\expandafter\ifx\csname urlprefix\endcsname\relax\def\urlprefix{URL }\fi
\expandafter\ifx\csname href\endcsname\relax
  \def\href#1#2{#2} \def\path#1{#1}\fi

\bibitem{thevanayagam2002undrained}
S.~Thevanayagam, T.~Shenthan, S.~Mohan, J.~Liang, {Undrained fragility of clean
  sands, silty sands, and sandy silts}, Journal of geotechnical and
  geoenvironmental engineering 128~(10) (2002) 849--859.

\bibitem{rahman2011equivalent}
M.~M. Rahman, S.~R. Lo, M.~A. Baki, {Equivalent granular state parameter and
  undrained behaviour of sand-fines mixtures}, Acta Geotechnica 6~(4) (2011)
  183--194.

\bibitem{rahman2012initial}
M.~M. Rahman, M.~Cubrinovski, S.~R. Lo, {Initial shear modulus of sandy soils
  and equivalent granular void ratio}, Geomechanics and Geoengineering 7~(3)
  (2012) 219--226.

\bibitem{ni2004contribution}
Q.~Ni, T.~S. Tan, G.~R. Dasari, D.~W. Hight, {Contribution of fines to the
  compressive strength of mixed soils}, G{\'{e}}otechnique 54~(9) (2004)
  561--569.

\bibitem{lade1998effects}
P.~V. Lade, C.~D. Liggio, J.~A. Yamamuro, {Effects of non-plastic fines on
  minimum and maximum void ratios of sand}, Geotechnical Testing Journal 21
  (1998) 336--347.

\bibitem{salgado2000shear}
R.~Salgado, P.~Bandini, A.~Karim, {Shear strength and stiffness of silty sand},
  Journal of Geotechnical and Geoenvironmental Engineering 126~(5) (2000)
  451--462.

\bibitem{Yin2014micromechanics}
Z.-Y. Yin, J.~Zhao, P.-Y. Hicher, {A micromechanics-based model for sand-silt
  mixtures}, International Journal of Solids and Structures 51~(6) (2014)
  1350--1363.

\bibitem{belkhatir2012experimental}
M.~Belkhatir, A.~Arab, N.~Della, T.~Schanz, {Experimental Study of Undrained
  Shear Strength of Silty Sand: Effect of Fines and Gradation}, Geotechnical
  and Geological Engineering 30~(5) (2012) 1103--1118.

\bibitem{chang2011micromechanical}
C.~S. Chang, Z.-Y. Yin, {Micromechanical modeling for behavior of silty sand
  with influence of fine content}, International Journal of Solids and
  Structures 48~(19) (2011) 2655--2667.

\bibitem{yan2011effect}
W.~M. Yan, J.~Dong, {Effect of particle grading on the response of an idealized
  granular assemblage}, International Journal of Geomechanics 11~(4) (2011)
  276--285.

\bibitem{kumar2014effects}
N.~Kumar, O.~I. Imole, V.~Magnanimo, S.~Luding, {Effects of polydispersity on
  the micro-macro behavior of granular assemblies under different deformation
  paths}, Particuology 12 (2014) 64--79.

\bibitem{martin2004isostatic}
C.~L. Martin, D.~Bouvard, {Isostatic compaction of bimodal powder mixtures and
  composites}, International Journal of Mechanical Sciences 46~(6) (2004)
  907--927.

\bibitem{minh2014strong}
N.~H. Minh, Y.~P. Cheng, C.~Thornton, {Strong force networks in granular
  mixtures}, Granular Matter 16~(1) (2014) 69--78.

\bibitem{ogarko2012equation}
V.~Ogarko, S.~Luding, {Equation of state and jamming density for equivalent bi-
  and polydisperse, smooth, hard sphere systems}, Journal of Chemical Physics
  136~(12).

\bibitem{langroudi2010transmission}
M.~K. Langroudi, J.~Sun, S.~Sundaresan, G.~I. Tardos, {Transmission of stresses
  in static and sheared granular beds: The influence of particle size, shearing
  rate, layer thickness and sensor size}, Powder Technology 203~(1) (2010)
  23--32.

\bibitem{ueda2011effect}
T.~Ueda, T.~Matsushima, Y.~Yamada, {Effect of particle size ratio and volume
  fraction on shear strength of binary granular mixture}, Granular Matter
  13~(6) (2011) 731--742.

\bibitem{shaebani2012influence}
M.~R. Shaebani, M.~Madadi, S.~Luding, D.~E. Wolf, {Influence of polydispersity
  on micromechanics of granular materials}, Phys. Rev. E 85~(1) (2012) 011301.

\bibitem{cundall1979discrete}
P.~A. Cundall, O.~D.~L. Strack, {A discrete numerical model for granular
  assemblies}, G\'{e}otechnique 29~(1) (1979) 47--65.

\bibitem{luding2005shear}
S.~Luding, {Shear flow modeling of cohesive and frictional fine powder}, Powder
  Technology 158~(1-3) (2005) 45--50.

\bibitem{kruyt2010micromechanical}
N.~P. Kruyt, I.~Agnolin, S.~Luding, L.~Rothenburg, {Micromechanical study of
  elastic moduli of loose granular materials}, Journal of the Mechanics and
  Physics of Solids 58~(9) (2010) 1286--1301.

\bibitem{duran2010micromechanical}
O.~Dur\'{a}n, N.~P. Kruyt, S.~Luding, {Micro-mechanical analysis of deformation
  characteristics of three-dimensional granular materials}, International
  Journal of Solids and Structures 47~(17) (2010) 2234--2245.

\bibitem{sun2011constitutive}
J.~Sun, S.~Sundaresan, {A constitutive model with microstructure evolution for
  flow of rate-independent granular materials}, Journal of Fluid Mechanics 682
  (2011) 590--616.

\bibitem{alonsomarroquin2005role}
F.~Alonso-Marroquin, S.~Luding, H.~J. Herrmann, I.~Vardoulakis, {Role of
  anisotropy in the elastoplastic response of a polygonal packing}, Phys. Rev.
  E 71~(5).

\bibitem{thornton2010quasistatic}
C.~Thornton, {Quasi-static simulations of compact polydisperse particle
  systems}, Particuology 8~(2) (2010) 119--126.

\bibitem{thornton2010evolution}
C.~Thornton, L.~Zhang, {On the evolution of stress and microstructure during
  general 3D deviatoric straining of granular media}, G\'{e}otechnique 60~(5)
  (2010) 333--341.

\bibitem{luding2008cohesive}
S.~Luding, {Cohesive, frictional powders: contact models for tension}, Granular
  Matter 10~(4) (2008) 235--246.

\bibitem{singh2014effect}
A.~Singh, V.~Magnanimo, K.~Saitoh, S.~Luding, {Effect of cohesion on shear
  banding in quasistatic granular materials}, Phys. Rev. E 90 (2014) 022202.

\bibitem{imole2013hydrostatic}
O.~I. Imole, N.~Kumar, V.~Magnanimo, S.~Luding, {Hydrostatic and Shear Behavior
  of Frictionless Granular Assemblies Under Different Deformation Conditions},
  KONA Powder and Particle Journal 30 (2013) 84--108.

\bibitem{majmudar2007jamming}
T.~S. Majmudar, M.~Sperl, S.~Luding, R.~P. Behringer, {Jamming Transition in
  Granular Systems}, Phys. Rev. Lett. 98~(5).

\bibitem{ohern2002random}
C.~S. O'Hern, S.~A. Langer, A.~J. Liu, S.~R. Nagel, {Random packings of
  frictionless particles}, Phys. Rev. Lett. 88~(7).

\bibitem{makse2000packing}
H.~A. Makse, D.~L. Johnson, L.~M. Schwartz, {Packing of compressible granular
  materials}, Phys. Rev. Lett. 84~(18) (2000) 4160--4163.

\bibitem{van2010jamming}
M.~van Hecke, {Jamming of soft particles: geometry, mechanics, scaling and
  isostaticity}, Journal of Physics: Condensed Matter 22:033101~(3).

\bibitem{goncu2010constitutive}
F.~G\"{o}nc\"{u}, O.~Dur\'{a}n, S.~Luding, {Constitutive relations for the
  isotropic deformation of frictionless packings of polydisperse spheres}, C.
  R. M\'{e}canique 338~(10-11) (2010) 570--586.

\bibitem{madadi2004fabric}
M.~Madadi, O.~Tsoungui, M.~Latzel, S.~Luding, {On the fabric tensor of
  polydisperse granular materials in 2D}, International Journal of Solids and
  Structures 41~(9-10) (2004) 2580.

\bibitem{luding2005anisotropy}
S.~Luding, {Anisotropy in cohesive, frictional granular media}, Journal of
  Physics Condensed Matter 17~(24) (2005) S2623--S2640.

\bibitem{kumar2013evolution}
N.~Kumar, O.~I. Imole, V.~Magnanimo, S.~Luding, {Evolution of the Effective
  Moduli for Anisotropic Granular Materials during Shear}, in: S.~Luding, A.~Yu
  (Eds.), Powders \& Grains 2013, Balkema, Sydney, Australia, 2013, pp.
  1238--1241.

\bibitem{kumar2014macroscopic}
N.~Kumar, S.~Luding, V.~Magnanimo, {Macroscopic model with anisotropy based on
  micro–macro information}, Acta Mechanica 225~(8) (2014) 2319--2343.

\end{thebibliography}
% 

\end{document}